\documentclass[runningheads]{llncs}
\usepackage{geometry}
\usepackage[utf8]{inputenc}
\usepackage[T1]{fontenc}
\usepackage{stmaryrd}
\usepackage{amssymb}
\usepackage[pdftex]{graphicx}
\usepackage{verbatim}
\usepackage{amsmath}
\usepackage{times}
\usepackage{xspace}
\usepackage{tikz}
\usetikzlibrary{arrows,automata,fit,arrows,decorations.pathreplacing,calc,positioning,decorations.text}
\usepackage{algorithm}
\usepackage{algpseudocode}
\usepackage{xcolor}
\usepackage{multirow}
\usepackage{multicol}
\usepackage{hhline}
\usepackage{tablefootnote}
\usepackage{mathtools}
\usepackage{gensymb}
\usepackage{url}

\usepackage[compatibility=false]{caption}
\usepackage{subcaption}
\usepackage{flexisym}
\usepackage{amsmath}
\usepackage{graphicx}
\usepackage{xcolor,colortbl}
\usepackage{mathpartir}
\usepackage{booktabs}  
\usepackage{eqparbox}  
\usepackage{booktabs}
\usepackage[]{hyperref}

\usepackage{lipsum}
\usepackage{wrapfig}
\usepackage{ccpres}
\usepackage[colorinlistoftodos,prependcaption,textsize=footnotesize]{todonotes}

\newcommand{\red}[1]{{#1}}
\newcommand{\redd}[1]{{#1}}

\newcommand{\ie}{\text{i.e.,}\xspace}
\newcommand{\eg}{{e.g.,}\xspace}

\newcommand{\secref}[1]{Section~\ref{#1}}

\newcommand{\figref}[1]{Figure~\ref{#1}}
\newcommand{\mfigref}[1]{Fig.~\ref{#1}}

\newcommand{\lowoh}{< 0.1\%}

\newcommand{\Data}{\mathrm{Data}}
\newcommand{\ignore}[1]{}
\newcommand{\true}{\ensuremath{\mathtt{true}}}
\newcommand{\false}{\ensuremath{\mathtt{false}}}
\newcommand{\goesto}[1][]{\stackrel{#1}{\longrightarrow}}

\DeclareMathOperator{\Tr}{Tr}
\DeclareMathOperator{\interactions}{interactions}
\DeclareMathOperator{\RGT}{RGT} 
\DeclareMathOperator{\Acc}{acc}
\DeclareMathOperator{\D}{discriminant}
\DeclareMathOperator{\upd}{upd}
\DeclareMathOperator{\last}{last}
\DeclareMathOperator{\W}{W}

\DeclareMathOperator{\length}{length}
\DeclareMathOperator{\pref}{pref}
\DeclareMathOperator{\free}{\mathit{free}}
\DeclareMathOperator{\ready}{\mathit{hold}}
\DeclareMathOperator{\delivered}{\mathit{delivered}}
\DeclareMathOperator{\done}{\mathit{done}}
\newcommand{\taskgenerator}{\mathit{Generator}}
\newcommand{\worker}{\mathit{Worker}}
\newcommand{\monitor}{\mathit{Monitor}}
\DeclareMathOperator{\ex}{\mathit{ex}}
\DeclareMathOperator{\exec}{\mathit{exec}}
\DeclareMathOperator{\finish}{\mathit{finish}}
\DeclareMathOperator{\maintenance}{\mathit{reset}}
\DeclareMathOperator{\newtask}{\mathit{newtask}}
\DeclareMathOperator{\deliver}{\mathit{deliver}}
\DeclareMathOperator{\stable}{stable}
\DeclareMathOperator{\equ}{equ}
\DeclareMathOperator{\Map}{map}
\DeclareMathOperator{\upto}{..}
\DeclareMathOperator{\quant}{\centerdot}

\newcommand{\longtrans}[2]{\ {\ensuremath{\xrightarrow{#2}}_{#1}}\ }

\begin{document}
\title{
Concurrency-Preserving and Sound
Monitoring of Multi-Threaded
Component-Based Systems\\
{\normalsize Theory, Algorithms, Implementation, and Evaluation}}
\author{
Hosein Nazarpour, Yli\`es Falcone, Saddek Bensalem, Marius Bozga
}
%
\institute{
Univ. Grenoble Alpes, Inria, CNRS, VERIMAG, LIG, Grenoble, France\\
\texttt{Firstname.Lastname@imag.fr}
}
\titlerunning{Concurrency-Preserving and Sound Monitoring of Multi-Threaded CBS}
\authorrunning{H. Nazarpour, Y. Falcone, S. Bensalem, M. Bozga}
\maketitle

\begin{abstract}
This paper addresses the monitoring of logic-independent linear-time user-provided properties in multi-threaded component-based systems.
We consider intrinsically independent components that can be executed concurrently with a centralized coordination for multiparty interactions.
In this context, the problem that arises is that a global state of the system is not available to the monitor.
A naive solution to this problem would be to plug in a monitor which would force the system to synchronize in order to obtain the sequence of global states at runtime.
Such a solution would defeat the whole purpose of having concurrent components. 
Instead, we reconstruct on-the-fly the global states by accumulating the partial states traversed by the system at runtime.
We define transformations of components that preserve their semantics and concurrency and, at the same time, allow to monitor global-state properties.
Moreover, we present RVMT-BIP, a prototype tool implementing the transformations for monitoring multi-threaded systems described in the BIP (Behavior, Interaction, Priority) framework, an expressive framework for the formal construction of heterogeneous systems.
Our experiments on several multi-threaded BIP systems show that RVMT-BIP induces a cheap runtime overhead.
\end{abstract}
%
\section{Introduction}
%
Component-based design is the process leading from given requirements and a set of predefined components to a system meeting the requirements.
Building systems from components is essential in any engineering discipline.
Components are abstract building blocks encapsulating behavior. 
They can be composed in order to build composite components. Their composition should be rigorously defined so that it is possible to infer the behavior of composite components from the behavior of their constituents as well as global properties from the properties of individual components.

The problem of building component-based systems (CBSs) can be defined as follows.
Given a set of components $\{B_{1},\ldots, B_{n}\}$ and a property of their product state space $\varphi$, find multiparty interactions $\gamma$  (i.e., ``glue" code) such that the coordinated behavior $\gamma (B_{1}, \ldots, B_{n} )$ meets the property $\varphi$.
It is, however, generally not possible to ensure or verify the desired property $\varphi$ using static verification techniques such as model-checking or static analysis, either because of the state-explosion problem or because $\varphi$ can only be decided with information available at runtime (\eg from the user or the environment).
In this paper, we are interested in complementary verification techniques for CBSs such as runtime verification.
In~\cite{FalconeJNBB15}, we introduced runtime verification of sequential CBSs against properties referring to the global states of the system, which, in particular, implies that properties can not be ``projected" and checked on individual components.
From an input composite system $\gamma\left(B_1, \ldots, B_n\right)$ and a regular linear-time property, a component monitor $M$ and a new set of interactions $\gamma'$ are synthesized to build a new composite system $\gamma'\left(B_1, \ldots, B_n, M\right)$ where the property is checked at runtime.
The underlying model of CBSs relies on multiparty interactions which consist of actions that are jointly executed by certain components, either sequentially or concurrently.
In the sequential setting, components are coordinated by a single centralized controller and joint actions are atomic.
Components notify the controller of their current states.
Then, the controller computes the possible interactions, selects one, and then sequentially executes the actions of each component involved in the interaction.
When components finish their executions, they notify the controller of their new states, and the aforementioned steps are repeated.
For performance reasons, it is desirable to parallelize the execution of components.
In the multi-threaded setting, each component executes on a thread and a controller is in charge of coordination.
Parallelizing the execution of $\gamma\left(B_1, \ldots, B_n\right)$ yields a bisimilar~\cite{mil95} component (\cite{basu2008distributed}) where each synchronized action $a$ occurring on $B_i$ is broken down into $\beta_i$ and $a'$ where $\beta_i$ represents an internal computation of $B_i$ and $a'$ is a synchronization action.
Between $\beta_i$ and $a'$, a new \textit{busy location} is added.
Consequently, the components can perform their interaction independently after synchronization, and the joint actions become non atomic.
After starting an interaction, and before this interaction completes (meaning that certain components are still performing internal computations), the controller can start another interaction between ready components.
The problem that arises in the multi-threaded setting is that a global steady state of the system (where all components are ready to perform an interaction) may never exist at runtime.
\redd{Note that we do not target distributed but multi-threaded systems in which components execute with a centralized controller, there is a global clock and communication is instantaneous and atomic.}
We define a method to monitor CBSs against any linear-time property referring to global states.
Our method preserves the concurrency and semantics of the monitored system.
It transforms the system so that global states can be reconstructed by accumulating partial states at runtime.
The execution trace of a \redd{multi-threaded} CBS is a sequence of partial states. 
For an execution trace of a \redd{multi-threaded} CBS, we define the notion of \textit{witness} trace, which is intuitively the unique trace of global states corresponding to the trace of the \redd{multi-threaded} CBS if this CBS was executed on a single thread.
For this purpose, we define transformations allowing one to add a new component building the witness trace.\par
We prove that the transformed and initial systems are bisimilar: the obtained reconstructed sequence of global states from a parallel execution is as the sequence of global states obtained when the multi-threaded CBS is executed with a single thread.

We introduce RVMT-BIP, a tool integrated in the BIP tool suite.\footnote{RVMT-BIP is available for download at~\cite{rvmt}.}
BIP (Behavior, Interaction, Priority) framework is a powerful and expressive component framework for the formal construction of heterogeneous systems.
BIP offers two powerful mechanisms for composing components by using \emph{multiparty interactions} and \emph{priorities}. The combination of interactions and priorities is expressive enough to express usual composition operators of other languages as shown in~\cite{bliudze07}.
A system model is layered. 
The lowest layer contains atomic components whose behavior is described by state machines with data and functions described in the C language. 
As in process algebras, atomic components can communicate by using ports. 
The second layer contains interactions which are relations between communication ports of individual components.
Priorities are used to express scheduling policies by selecting among the enabled interactions of the layer underneath.
RVMT-BIP takes as input a BIP CBS and a monitor description which expresses a property $\varphi$, and outputs a new BIP system whose behavior is monitored against $\varphi$ while running concurrently.
%
%
\figref{fig:over-view} presents an overview of our approach.
Recall that according to~\cite{basu2008distributed}, a BIP system with global-state semantics $S_g$ (sequential model), is (weakly) bisimilar with the corresponding partial-state model $S_p$ (concurrent model).
This is formalized as $S_g \sim S_p$ ($\sim$ is formally defined in \secref{sec:prelim}).
Moreover, $S_p$ generally runs faster than $S_g$ because of its parallelism.
Thus, if a trace of $S_g$, \ie $\sigma_g$, satisfies $\varphi$, then the corresponding trace of $S_p$, \ie $\sigma_p$, satisfies $\varphi$ as well.
The technique in~\cite{FalconeJNBB15} \emph{could} serve as a monitoring solution.
In short, \cite{FalconeJNBB15} instruments the components and synthesizes additional interactions in such a way that, whenever the system performs an interaction, the monitor receives the current global state of the system.
Hence, based on~\cite{FalconeJNBB15}, a couple of naive solutions to monitor $S_p$ would be (i) to monitor $S_g$ and run $S_p$, which would incur unpredictable delays in detecting verdicts or (ii) plug the monitor (as in~\cite{FalconeJNBB15}) into $S_p$, which would force the (concurrent) components to synchronize for the monitor to take a snapshot of the global state of the system.
Such solutions would completely defeat the purpose of using multi-threaded models.
Instead, we here propose a transformation technique to build another system $S_{\it pg}$ out of $S_p$ such that (i) $S_{\it pg}$ and $S_p$ are bisimilar (hence $S_g$ and $S_{\it pg}$ are bisimilar), (ii) $S_{\it pg}$ is as concurrent as $S_p$ and preserves the performance gained from multi-threaded execution and (iii) $S_{\it pg}$ produces a witness trace, that is the \emph{unique} trace that allows to check the property $\varphi$.
Our method does not introduce any delay in the detection of verdicts since it always reconstructs the maximal (information-wise) prefix of the witness trace (Theorem~\ref{theorem_witness}).
Moreover, we show that our method is correct in the sense that it always produces the correct witness trace (Theorem~\ref{theorem_Correctness}).
\pgfdeclarelayer{background}
\pgfdeclarelayer{foreground}
\pgfsetlayers{background,main,foreground}
\begin{figure}
    \centering
\resizebox{\textwidth}{!}{             
\tikzstyle{block} = [rectangle, draw, fill=gray!20, 
    text centered]   
\tikzstyle{block2} = [rectangle, draw,fill=yellow,  
    text width=5em, text centered, minimum height=2.5em]       
\tikzstyle{block3} = [rectangle, draw,fill=green!40,  
    text width=5em, text centered, rounded corners, minimum height=7em]     
\tikzstyle{line} = [draw, -latex']  
\def\myarm{1cm}
\def\myangle{0}
\tikzset{
  arm/.default=1cm,
  arm/.code={\def\myarm{#1}},
  angle/.default=0,
  angle/.code={\def\myangle{#1}} 
}   
\tikzset{
    myncbar/.style = {to path={
        let
            \p1=($(\tikztotarget)+(\myangle:\myarm)$)
        in
            -- ++(\myangle:\myarm) coordinate (tmp)
            -- ($(\tikztotarget)!(tmp)!(\p1)$)
            -- (\tikztotarget)\tikztonodes
    }}
} 
\begin{tikzpicture}[node distance = 1.5cm, auto]   
 \node [block3,node distance = 1cm,text width=12em, yshift=-0.5cm] (Tg) {};
 \node [yshift=-1.4cm]  {Execution Trace};
 \node [block,text width=10em,node distance = 1cm, yshift=-0.65cm]  (Sg) {Global-state semantics};
 \node [block2, text width=10em]  (BIPg) {BIP system $S_g$};

 \node [block3,node distance = 1cm,text width=12em, xshift=9cm, yshift=-0.5cm] (Tp) {};
 \node [xshift=9cm,yshift=-1.4cm]  {Execution Trace};
 \node [block,text width=10em,node distance = 1cm,xshift=9cm, yshift=-0.65cm]  (Sp) {Partial-state semantics};
 \node [block2, text width=10em,xshift=9cm]  (BIPp) {BIP system $S_p$};
 
 \node [block3,node distance = 1cm,text width=12em, xshift=8cm, yshift=-4.5cm] (Tw) {};
 \node [xshift=8cm,yshift=-5.4cm]  {Witness Trace};
 \node [block,text width=10em,node distance = 1cm,xshift=8cm, yshift=-4.65cm]  (Sw) {Partial-state semantics};
 \node [block2, text width=10em,xshift=8cm, yshift=-4cm]  (BIPw) {BIP system $S_{pg}$};
 
 \path [line] (BIPp)  to [myncbar,angle=0,arm=2,xshift=1cm] (BIPw); 
 \node [xshift=12.3cm,yshift=-2cm,rotate=-90]  {Transformation w.r.t $\varphi$};

 \node [xshift=4.5cm, yshift=0.2cm]  {Transformation in~\cite{basu2008distributed}};
 \node [xshift=4.5cm, yshift=-0.3cm]  {($\beta$ transitions)};
 
 \path [line] (BIPg)  -- (BIPp); 
 \node [block2, text width=5em,yshift=-4.5cm,fill=gray!40]  (M) {Runtime\\ Monitor};
 \node [xshift=-0.2cm,yshift=-2.9cm,rotate=-90]  {Global states};
 \node [xshift=4.6cm, yshift=-4.2cm]  {Global states};
 \path [line, dashed] (Tg)  -- (M); 
 \path [line, dashed] (Tw)  -- (M);
 
 \node [block2, text width=9em,xshift=-7cm,yshift=-4.5cm,fill=blue!5]  (P) {Property $\varphi$\\ on global states}; 
 
 \path [line] (P)  -- (M);
 \node [xshift=-3.1cm, yshift=-4.2cm]  {Monitor synthesis \cite{FalconeJNBB15}};
\def\myshift#1{\raisebox{1ex}}
\draw [->,>=latex,line width=2pt,red!60,postaction={decorate,decoration={text along path,text align=center,text={|\sffamily\myshift|Refers to}}}]      (P) to [bend left=35] (BIPg);
 \node (MV) at ([yshift=-1.2cm]M) {Verdicts};
 \path [line] (M)  -- (MV);
 
     \begin{pgfonlayer}{background}
        \draw[rounded corners=2em,line width=5em,blue!20,cap=round,xshift=1cm]
   (10,-0.2)--(11.5,-0.2)--(11.5,-3.5)--(3.5,-3.5)--(3.5,-5.5)--(11.5,-5.5)--(11.5,-2.5)-- (11.1,-4.7)--(3.5,-4.7);
    \end{pgfonlayer}
    \node [xshift=7.5cm,yshift=-2.5cm]  {This paper's contribution};
                  
\end{tikzpicture}}
\caption{Approach overview}
\label{fig:over-view}
\end{figure}
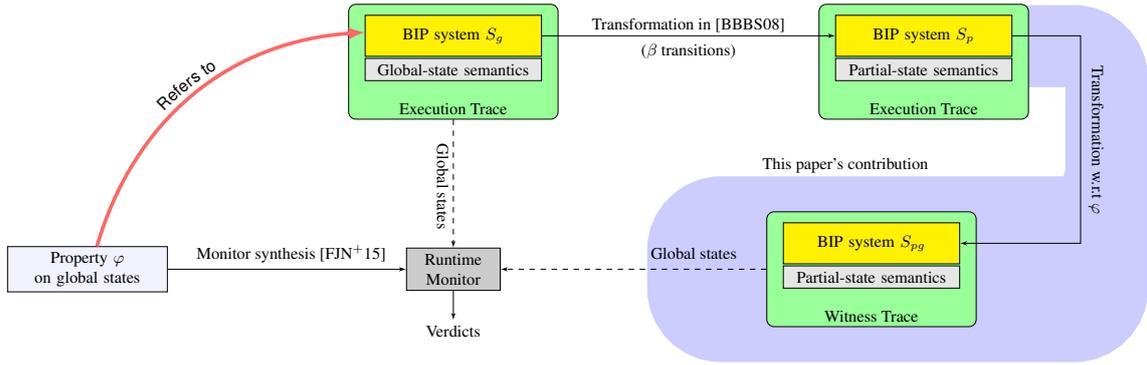
\begin{remark}[On the monitored properties]
Note that our approach allows one to monitor \emph{any} linear-time property.
Moreover, how the property is defined is irrelevant as one can use the approaches in~\cite{BauerLS10,FalconeFM12} to synthesize a monitor which emits verdicts in a 4-valued domain.
Our approach directly uses the definition of a monitor as input and is thus compatible with the various approaches compatible with the ones in~\cite{BauerLS10,FalconeFM12}. 
\end{remark}

This paper extends a previous contribution~\cite{NazarpourFBBC16} that appeared in the 12$^{\textrm{th}}$ International Conference on integrated Formal Methods, with the following additional contributions:
\begin{itemize}
\item 
we propose detailed and rigorous proofs of the propositions and theorems related to the soundness of our monitoring approach;
\item 
we improve the presentation and readability of \cite{NazarpourFBBC16} by (i) formalizing some concepts that remained informal in the conference version, (ii) providing more detailed explanations in each section, and (iii) illustrating the concepts with additional examples;
\item
we present the actual algorithms used in the instrumented system to reconstruct global states;
\item
we further validate our approach against additional case studies and report on additional experimental data;
\item we propose a deeper study of related work.
\end{itemize}
\paragraph{Running example.}
We use a task system, called Task, to illustrate our approach throughout the paper.
The system consists of a task generator (\textit{Generator}) along with 3 task executors (\textit{Workers}) that can run in parallel.
Each newly generated task is processed whenever two cooperating workers are available. 
A desirable property of system Task is the homogeneous distribution of the tasks among the workers.
\paragraph{Outline.}
The remainder of this paper is organized as follows.
\secref{sec:prelim} introduces some preliminary concepts.
\secref{sec:cbs} overviews CBS design and semantics.
In \secref{sec:mon}, we define a theoretical framework for the monitoring of multi-threaded CBSs.
In \secref{sec:inst}, we present the transformation of a multi-threaded CBS model for introducing monitors.
\secref{sec:implem} describes RVMT-BIP, an implementation of the approach and its evaluation on several examples. 
\secref{sec:rw} presents related work.
\secref{sec:conc} concludes and presents future work.
Complete proofs related to the correctness of the approach are given in Appendix~\ref{sec:proofs}.
%
%
\section{Preliminaries and Notations}
\label{sec:prelim}
%
We introduce some preliminary concepts and notations.
\paragraph{Functions.}
For two domains of elements $E$ and $F$, we note $E\rightarrow F$  the set of functions from $E$ to $F$. 
For two functions $v\in  X \rightarrow Y$ and $v'\in X' \rightarrow Y'$, the function obtained by overriding $v$ images by $v'$ images is denoted by $v \backslash v'$, where $v \backslash v' \in X \cup X' \rightarrow Y \cup Y'$, and is defined as follows:
\[
v \backslash v'(x) = \left\{
  \begin{array}{ll}
v'(x)& \text{if } x \in X',\\
v(x) & \text{otherwise}.
\end{array} \right.
\]
\paragraph{Sequences.}
Given a set of elements $E$, $e_1\cdot e_2\cdots e_n$ is a sequence or a list of length $n$ over $E$, where $\forall i\in [1 \upto n] \quant e_i\in E$.
Sequences of assignments are delimited by square brackets for clarity.
The empty sequence is denoted by $\epsilon$ or $[~]$, depending on the context.
The set of (finite) sequences over E is denoted by $E^*$.
$E^+$ is defined as $E^* \setminus \{\epsilon\}$.
The length of a sequence $s$ is denoted by $\length(s)$.
We define $s(i)$ as the $\redd{i^{\rm th}}$ element of $s$ and $s(i\cdots j)$ as the factor of $s$ from the $\redd{i^{\rm th}}$ to the $\redd{j^{\rm th}}$ element.
We also denote by $\pref(s)$, the set of \textit{prefixes} of $s$ such that $\pref(s)=\{s(1\cdots k) \mid k\leq \length(s)\}$.
Operator $\pref$ is naturally extended to sets of sequences.
Function $\max_\preceq$ (resp. $\min_\preceq$) returns the maximal (resp. minimal) sequence w.r.t. prefix ordering of a set of sequences.
We define function  $\last:E^+\rightarrow E$ such that $\last(e_1\cdot e_2\cdots e_n)=e_n$.
\paragraph{Map operator: applying a function to a sequence.}
For a sequence $e=e_1\cdot e_2\cdots e_n$ of elements over $E$ of some length $n \in \mathbb{N}$, and a function $f : E \rightarrow F$, $\Map \ f \ e$ is the sequence of elements of $F$ defined as $f(e_1)\cdot f(e_2)\cdots f(e_n)$ where $\forall i\in[1 \upto n]\quant f(e_i)\in F$.
\paragraph{Labeled transition systems.}
Labeled Transition System (LTS) are used to define the semantics of component-based systems.
An LTS is defined over an alphabet $\Sigma$ and is a 3-tuple $(\mathrm{Sta},\mathrm{Lab}, \mathrm{Trans})$ where $\mathrm{Sta}$ is a non-empty set of states, $\mathrm{Lab}$ is a set of labels, and $\mathrm{Trans}\subseteq \mathrm{Sta}\times \mathrm{Lab} \times \mathrm{Sta}$ is the transition relation.
A transition $(q,e,q') \in \mathrm{Trans}$ means that the LTS can move from state $q$ to state $q'$ by consuming label $e$. 
We abbreviate $(q,e,q') \in \mathrm{Trans}$ by $q\longtrans{\mathrm{Trans}}{e} q'$ or by $q \longtrans{}{e} q'$ when clear from context. 
Moreover, relation $\mathrm{Trans}$ is extended to its reflexive and transitive closure in the usual way and we allow for regular expressions over $\mathrm{Lab}$ to label moves between states: if $\mathit{expr}$ is a regular expression over $\mathrm{Lab}$ (i.e., $\mathit{expr}$ denotes a subset of $\mathrm{Lab}^*$), $q \longtrans{}{\mathit{expr}} q'$ means that there exists one sequence of labels in $\mathrm{Lab}$ matching $\mathit{expr}$ such that the system can move from $q$ to $q'$.
\paragraph{Observational equivalence and bi-simulation.}
The \emph{observational equivalence} of two transition systems is based on the usual definition of weak bisimilarity~\cite{mil95}, where $\theta$-transitions are considered to be unobservable.
Given two transition systems $S_1 = (\mathrm{Sta}_1, \mathrm{Lab} \cup \{ \theta \},\rightarrow_{\mathrm{Trans}_1})$ and $S_2 = (\mathrm{Sta}_2 , \mathrm{Lab} \cup \{ \theta \}, \rightarrow_{\mathrm{Trans}_2})$, system $S_1$ \emph{weakly simulates} system $S_2$, if there exists a relation $R \subseteq \mathrm{Sta}_{1} \times \mathrm{Sta}_{2}$ such that the two following conditions hold:
\begin{enumerate}
 \item $\forall ( q_1 , q_2 ) \in R, \forall a \in \mathrm{Lab} \quant q_1 \goesto[a]_{\mathrm{Trans}_1} q_1' \implies \exists q_2' \in \mathrm{Sta}_2 \quant \left((q_1' , q_2') \in R \wedge q_2 \longtrans{\mathrm{Trans}_2}{\theta^*\cdot a \cdot \theta^*} q_2'\right)$, and
 \item  $\forall (q_1,q_2) \in R \quant \left(\exists q_1' \in \mathrm{Sta}_1 \quant q_1 \goesto[\theta]_{\mathrm{Trans}_1} q_1'\right) \implies \exists q_2' \in \mathrm{Sta}_2 \quant \left((q_1',q_2') \in R \wedge q_2 \goesto[\theta^*]_{\mathrm{Trans}_2} q_2' \right)$.
\end{enumerate}
Equation \textit{1.} says that if a state $q_1$ simulates a state $q_2$ and if it is possible to perform $a$ from $q_1$ to end in a state $q_1'$, then there exists a state $q_2'$ simulated by $q_1'$ such that it is possible to go from $q_2$ to $q_2'$ by performing some unobservable actions, the action $a$, and then some unobservable actions.
Equation \textit{2.} says that if a state $q_1$ simulates a state $q_2$ and it is possible to perform an unobservable action from $q_1$ to reach a state $q_1'$, then it is possible to reach a state $q_2'$ by a sequence of unobservable actions such that $q_1'$ simulates $q_2'$.
In that case, we say that the relation $R$ is a weak simulation over $S_1$ and $S_2$ or equivalently that the states of $S_1$ are (weakly) similar to the states of $S_2$. Similarly, a weak bi-simulation over $S_1$ and $S_2$ is a relation $R$ such that $R$ and $R^{-1} = \{ (q_2,q_1) \in \mathrm{Sta}_2 \times \mathrm{Sta}_1 \mid (q_1,q_2)\in R\}$ are both weak simulations. In this latter case, we say that $S_1$ and $S_2$ are \emph{observationally equivalent} and we write $S_1 \sim S_2$ to express this formally.
%
\section{Component-Based Systems with Multiparty Interactions}
\label{sec:cbs}
%
An action of a CBS is an interaction {\ie} a coordinated operation between certain atomic components.
Atomic components are transition systems with a set of ports labeling individual transitions.
Ports are used by components to communicate.
Composite components are obtained from atomic components by specifying interactions.
\paragraph{Atomic Components.}
An atomic component is endowed with a finite set of local variables $X$ taking values in a set $\Data$.
Atomic components synchronize and exchange data with other components through \textit{ports}.
\begin{definition}[Port] 
A port $p[x_p]$, where $x_p \subseteq X$, is defined by a port identifier $p$ and some data variables in a set $x_p$.
\end{definition}
Variables attached to ports are purposed to transfer values between interacting components (see also Definition~\ref{def:interaction} for interactions).
The variables attached to the port are also used to determine whether a communication through this port can take place (see below).
\begin{definition}[Atomic component]
\label{def:atomic_component}
An atomic component is defined as a tuple $(P, L, T,$ $ X)$ where $P$ is the set of ports, $L$ is the set of (control) locations, $T\subseteq L \times P \times {\cal G}(X) \times {\cal F}^*(X) \times L$ is the set of transitions, and $X$ is the set of variables.
${\cal G}(X)$ denotes the set of Boolean expressions over $X$ and ${\cal F}(X)$ the set of assignments of expressions over $X$ to variables in $X$.
For each transition $\tau = (l, p, g_{\tau}, f_{\tau}, l') \in T$, $g_{\tau}$ is a Boolean expression over $X$ (the guard of $\tau$), $f_{\tau}\in \{ x:= f^x(X)\mid x\in X \wedge f^x \in {\cal F}(X) \}^*$: the computation step of $\tau$, a sequence of assignments to variables.

The semantics of the atomic component is an LTS $(Q, P, \rightarrow)$ where $Q= L\times (X\rightarrow \mathit{\Data})$ is the set of states, and $\rightarrow= \{ ((l,$ $v),$ $p(v_p),$ $(l',$ $v'))\in Q\times P\times Q\mid \exists \tau= (l, p, g_{\tau}, f_{\tau}, l') \in T \quant  g_{\tau}(v) \wedge v'=f_{\tau}(v \backslash v_p)\}$ is the transition relation.
\end{definition}
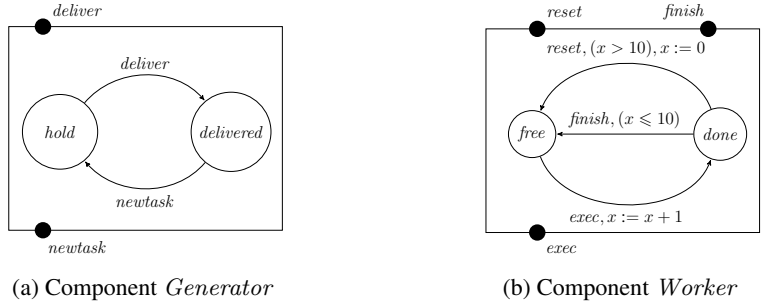
\begin{figure}[t]
    \centering
    \begin{subfigure}[b]{0.4\textwidth}
    \centering
    \Large
   \scalebox{.45} {   
\begin{tikzpicture}[->,>=stealth',shorten >=1pt,auto,node distance=5cm, semithick]
  \tikzstyle{every state}=[fill=none,draw=black,text=black]  

  \filldraw[fill=none,draw=black](0,0.5)--(0,6.5)--(8,6.5)--(8,0.5)--(0,0.5)--cycle; 
  \node at (1.5,3.4) [state] (A)        [minimum size=2.2cm]              {$\ready$};
  \node            [state] (B)  [right of =A,minimum size=2.2cm]          {$\delivered$};
  \path [in=-50,out=-130]  (B) edge  [below]       node[text centered] {$\newtask$} (A);
  \path [in=130,out=50]    (A) edge  [above]       node[text centered] {$\deliver$} (B); 
  \fill (1,0.5) circle (0.24cm);\node at (2,0){$\newtask$};
  \fill (1,6.5) circle (0.24cm);\node at (2,7) {$\deliver$};
   

\end{tikzpicture}}
\caption{Component $\taskgenerator$} 
\label{fig:Task generator}
    \end{subfigure}
    \begin{subfigure}[b]{0.4\textwidth}
       \centering
           \Large
   \scalebox{.45} {   
\begin{tikzpicture}[->,>=stealth',shorten >=1pt,auto,node distance=5.5cm, semithick]
  \tikzstyle{every state}=[fill=none,draw=black,text=black]  

  \filldraw[fill=none,draw=black](-0.5,10)--(-0.5,16)--(7.5,16)--(7.5,10)--(-0.5,10)--cycle; 
  \node at (0.85,12.9) [state] (A)        [minimum size=0.5cm]         {$\free$};
  \node            [state] (B)  [right of =A,minimum size=0.5cm]       {$\done$};
  \path [out=-70,in=-110]  (A) edge  [below]       node[text centered] {$\exec,x$ :$=x+1$} (B);
  \path [out=180,in=0]     (B) edge  [above ]      node[text centered] {$\finish,(x\leqslant 10)$} (A);
  \path [out=110,in=70]    (B) edge  [above ]      node[text centered] {$\maintenance,(x>10),x$ :$=0$} (A); 
  \fill (1,10) circle (0.24cm);\node at (1.7,9.5) {$\exec$};
  \fill (1,16) circle (0.24cm);\node at (1.8,16.5) {$\maintenance$};
  \fill (6,16) circle (0.24cm);\node at (5.3,16.5) {$\finish$};
  

\end{tikzpicture}}
\caption{Component $\worker$}
\label{fig:Workers}
       \end{subfigure}
       \caption{Atomic components of system Task} 
\label{fig:task atoms}
\end{figure}
A state is a pair $(l, v) \in Q$, where $l \in L$, $v \in X \rightarrow \mathit{\Data}$ is a valuation of the variables in $X$.
The evolution of states $(l,$ $v) \goesto[p(v_{p})] (l',$ $v')$, where $v_{p}$ is a valuation of the variables $x_p$ attached to port $p$, is possible if there exists a transition $(l,$ $p[x_p], g_\tau, f_\tau, l')$, such that $g_\tau(v)=\true$.
As a result, the valuation $v$ of variables is modified to $v'=f_\tau(v \backslash v_p)$.

We use the dot notation to denote the elements of atomic components. \eg for an atomic component $B$, $B.P$ denotes the set of ports of the atomic component $B$, $B.L$ denotes its set of locations, etc.
\begin{example}[Atomic component]
Figure~\ref{fig:task atoms} shows the atomic components of system Task.
\begin{itemize}
\item
Figure~\ref{fig:Task generator} depicts a model of component $\taskgenerator$\footnote{For the sake of simpler notation, the variables attached to the ports are not shown.} defined as follows:
\begin{itemize}
\item
${\taskgenerator}.P=\{\deliver[\emptyset],\  \newtask[\emptyset]\}$,
\item
${\taskgenerator}.L=\{\ready$ $,$ $\delivered\}$,
\item
${\taskgenerator}.T=\{(\ready, \deliver,\true, [~],\delivered)$, $(\delivered$ $, \newtask,$ $\true, [~], \ready)\}$,
\item
${\taskgenerator}.X=\emptyset$.
\end{itemize}
\item
Figure~\ref{fig:Workers} depicts a model of component $\worker$ defined as follows:
\begin{itemize}
\item
 $\worker.P=\{ \exec[\emptyset], \finish[\emptyset], \maintenance[\emptyset]\ \}$,
 \item
$\worker.L $ $ = \{ \free,\done \}$,
\item
$\worker.T =\{(\free$, $\exec, \true, [x:=x+1], \done)$, $(\done, \finish, (x\leqslant 10), [~],$ $ \free),(\done, \maintenance,$ $(x>10), [x:=0], \free)\}$,
\item
$\worker.X$ $=\{x\}$.
\end{itemize}
\end{itemize}
\end{example}
\begin{definition}[Interaction]
\label{def:interaction}
An interaction $a$ is a tuple $({\cal P}_a, F_a)$, where ${\cal P}_a = \{p_i[x_i]\mid p_i\in B_i.P\}_{i \in I}$ is the set of ports such that $\forall i\in I \quant {\cal P}_a \cap B_i.P = \{p_i\}$ and $F_a$ is a sequence of assignments to the variables in $\cup_{i \in I} x_i$.
\end{definition}
When clear from context, in the following examples, an interaction $(\{p[x_p]\}, F_a)$ consisting of only one port $p$ is denoted by $p$.
\begin{definition}[Composite component]
\label{def:composite_component}
A composite component $\gamma(B_1,\ldots,B_n)$ is defined from a set of atomic components $\{B_i\}_{i=1}^n$ and a set of interactions $\gamma$.

A state $q$ of a composite component $\gamma(B_1,\ldots,B_n)$ is an n-tuple $q=(q_1,\ldots,q_n)$, where $q_i=(l_i,v_i)$ is a state of atomic component $B_i$.
The semantics of the composite component is an LTS $(Q,\gamma,\goesto)$, where $Q= B_1.Q\times \ldots\times B_n.Q$ is the set of states, $\gamma$ is the set of all possible interactions and $\goesto$ is the least set of transitions satisfying the following rule:
\begin{mathpar}
\inferrule*
{
   a=(\{p_i[x_i]\}_{i\in I}, F_a) \in \gamma \and 
       \forall i\in I \quant \ q_i \goesto[p_i(v_i)]_i q'_i  \wedge v_i=F_{a_i}(v(X)) \and
    \forall i\not\in I \quant \ q_i = q'_i
}
{
    (q_1,\dots,q_n) \goesto[a] (q'_1,\dots,q'_n)
}
\end{mathpar}
$X$ is the set of variables attached to the ports of $a$, $v$ is the global valuation, and $F_{a_i}$ is the restriction of $F$ to the variables of $p_i$.
\end{definition}
The semantic rule in Definition~\ref{def:composite_component} says that a composite component moves from state $(q_1,\ldots,q_n)$ to a state $(q'_1,\ldots,q'_n)$ through some interaction $a$ if there exists an interaction $a \in \gamma$ of the form $(\{p_i[x_i]\}_{i\in I}, F_a)$, i.e., involving component of index in a set $I \subseteq [1\upto n]$.
The components involved in interaction $a$ (i.e., components with index in set $I$) evolve according to their transition relation $\goesto{}_i$ (as per Definition~\ref{def:atomic_component}): they move from state $q_i$ to state $q_i'$ by executing port $p_i$ with valuation $v_i$ obtained after executing the assignments $F_{a_i}$ related to the variables of port $p_i$ (obtained from the sequence of assignments $F_a$ of interaction $a$).
The components not involved in interaction $a$ (i.e., components with index not in set $I$) remain in the same state.

A trace is a sequence of states and interactions $(q_0 \cdot a_1 \cdot q_1\cdots a_s \cdot q_s)$ such that:
$q_0 = \mathit{Init} \wedge\big(\forall i\in [1\upto s] \quant  q_i\in Q\wedge  a_i\in \gamma \wedge q_{i-1}\stackrel{a_i}{\longrightarrow}q_i \big)$, where $\mathit{Init}\in Q$ is the initial state. 
Given a trace $(q_0 \cdot a_1 \cdot q_1\cdots a_s \cdot q_s)$, the sequence of interactions is defined as $\interactions(q_0 \cdot a_1 \cdot q_1\cdots a_s \cdot q_s) = a_1 \cdots a_s$.
The set of traces of composite component $B$ is denoted by $\Tr(B)$.
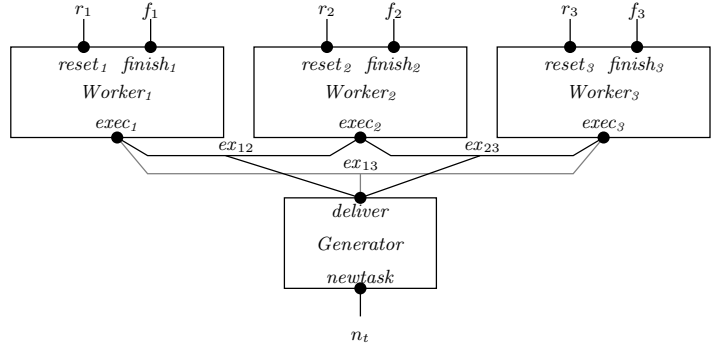
\begin{figure}[t]
    \centering
   \scalebox{0.8} {   
\begin{tikzpicture}[->,>=stealth',shorten >=1pt,auto,node distance=5cm, semithick]
  \tikzstyle{every state}=[fill=none,draw=black,text=black]  
\draw[-] (0.75,2)--(0.25,1.7)--(-2.75,1.7)--(-3.25,2);
\draw[-] (4.75,2)--(4.25,1.7)--(1.25,1.7)--(0.75,2);  
\draw[-,gray] (4.75,2)--(4.25,1.4)--(-2.75,1.4)--(-3.25,2);
\draw[-,gray] (0.75,1)--(0.75,1.43); \node  at (0.75,1.55) {\textbf{$\ex_{13}$}};
\draw[-] (0.75,1)--(2.75,1.71);      \node  at (2.75,1.85) {\textbf{$\ex_{23}$}};
\draw[-] (0.75,1)--(-1.5,1.71);      \node  at (-1.3,1.85) {\textbf{$\ex_{12}$}};
\draw[-] (0.75,-0.5)--(0.75,-1);     \node  at (0.75,-1.3)  {\textbf{$n_t$}};
\draw[-] (1.3,3.5)--(1.3,4);         \node  at (1.3,4.1)   {\textbf{$f_2$}};
\draw[-] (0.2,3.5)--(0.2,4);         \node  at (0.2,4.1)   {\textbf{$r_2$}};
\draw[-] (5.3,3.5)--(5.3,4);         \node  at (5.3,4.1)   {\textbf{$f_3$}};
\draw[-] (4.2,3.5)--(4.2,4);         \node  at (4.2,4.1)   {\textbf{$r_3$}};
\draw[-] (-2.7,3.5)--(-2.7,4);       \node  at (-2.7,4.1)  {\textbf{$f_1$}};
\draw[-] (-3.8,3.5)--(-3.8,4);       \node  at (-3.8,4.1)  {\textbf{$r_1$}};
  \filldraw[fill=none,draw=black](-0.5,-0.5)--(-0.5,1)--(2,1)--(2,-0.5)--(-0.5,-0.5)--cycle; 
  \node at (0.75,0.25)   {$\mathit{Generator}$};  
  \fill (0.75,1) circle (0.1cm); \node at (0.75,-0.3)   {{\small $\mathit{newtask}$}}; 
  \fill (0.75,-0.5) circle (0.1cm); \node at (0.75,0.8)   {{\small $\mathit{deliver}$}};
  \filldraw[fill=none,draw=black](-1,2)--(-1,3.5)--(2.5,3.5)--(2.5,2)--(-1,2)--cycle;  
  \node at (0.75,2.75)    {$\mathit{Worker_2}$};
  \fill (0.75,2) circle (0.1cm);\node at (0.75,2.2) {{\small $\mathit{exec_2}$}}; 
  \fill (1.3,3.5) circle (0.1cm);\node at (1.3,3.2)   {{\small $\mathit{finish_2}$}};
  \fill (0.2,3.5) circle (0.1cm);\node at (0.2,3.2)  {{\small $\mathit{\maintenance_2}$}};
  \filldraw[fill=none,draw=black,xshift=-4cm](-1,2)--(-1,3.5)--(2.5,3.5)--(2.5,2)--(-1,2)--cycle;   
  \node [xshift=-4cm] at (0.75,2.75)   {$\mathit{Worker_1}$};
  \fill [xshift=-4cm](0.75,2) circle (0.1cm);\node [xshift=-4cm] at (0.75,2.2) {{\small $\mathit{exec_1}$}}; 
  \fill [xshift=-4cm](1.3,3.5) circle (0.1cm);\node [xshift=-4cm] at (1.3,3.2)   {{\small $\mathit{finish_1}$}};
  \fill [xshift=-4cm](0.2,3.5) circle (0.1cm);\node [xshift=-4cm] at (0.2,3.2)  {{\small $\mathit{\maintenance_1}$}};
  \filldraw[fill=none,draw=black,xshift=4cm](-1,2)--(-1,3.5)--(2.5,3.5)--(2.5,2)--(-1,2)--cycle;  
  \node [xshift=4cm] at (0.75,2.75)   {$\mathit{Worker_3}$};
  \fill [xshift=4cm](0.75,2) circle (0.1cm);\node [xshift=4cm] at (0.75,2.2) {{\small $\mathit{exec_3}$}}; 
  \fill [xshift=4cm](1.3,3.5) circle (0.1cm);\node [xshift=4cm] at (1.3,3.2)   {{\small $\mathit{finish_3}$}};
  \fill [xshift=4cm](0.2,3.5) circle (0.1cm);\node [xshift=4cm] at (0.2,3.2)  {{\small $\mathit{\maintenance_3}$}};
\end{tikzpicture}}
\caption{Composite component of system Task} 
\label{fig:running_composite}
\end{figure}
\begin{example}[Interaction, composite component]
Figure~\ref{fig:running_composite} depicts the composite component $\gamma(\worker_1,$ $\worker_2,$ $\worker_3,$ $\taskgenerator)$ of system Task, where each $\mathit{\worker_i}$ is identical to the component in \mfigref{fig:Workers} and $\taskgenerator$ is the component depicted in \mfigref{fig:Task generator}.
The set of interactions is $\gamma=\{\ex_{12},$ $\ex_{13},$ $\ex_{23},$ $r_1,$ $r_2,$ $ r_3,$ $f_1,$ $f_2,$ $f_3,$ $n_t\}$.
We have 
$\ex_{12}=(\{\deliver,$ $\exec_1,$ $\exec_2\},[~])$
, $\ex_{23}=(\{\deliver,$ $\exec_2,$ $ \exec_3\},[~])$, $\ex_{13}=(\{\deliver,$ $\exec_1,$ $\exec_3\},[~])$,
$r_1=(\{\maintenance_1\},[~])$, $r_2=(\{\maintenance_2\},[~])$, $r_3=(\{\maintenance_3\},[~])$, $f_1=(\{\finish_1\},[~])$, $f_2=(\{\finish_2\},[~])$, $f_3=(\{\finish_3\},[~])$, and $n_t=(\{\newtask\},[~])$.

One of the possible traces\footnote{For the sake of simpler notation, we denote a state by its location (and omit the valuation of variables).} of system Task is: $(\free,$ $\free,$ $\free,$ $\ready) \cdot \ex_{12}\cdot(\done,$ $\done,$ $\free,$ $\delivered) \cdot n_t\cdot(\done,$ $\done,$ $\free,$ $\ready)$ such that from the initial state $(\free,$ $\free,$ $\free,$ $\ready)$, where workers are at location $\free$ and task generator is ready to deliver a task, interaction $\ex_{12}$ is fired and $\worker_1$ and $\worker_2$ move to location $\done$ and $\taskgenerator$ moves to location $\delivered$.
Then, a new task is generated by the execution of interaction $n_t$ so that $\taskgenerator$ moves to location $\ready$.
\end{example}
Two composite components are bi-similar if the LTSs of their semantics are bi-similar.
%
%
\section{Monitoring Multi-Threaded CBSs with Partial-State Semantics}
\label{sec:mon}
%
The general semantics defined in the previous section is referred to as the global-state semantics of CBSs because each state of the system is defined in terms of the local states of components, and, all local states are defined.
In this section, we consider what we refer to as the partial-state semantics where the states of a system may contain undefined local states because of the concurrent execution of components.
%
\subsection{Partial-State Semantics}
\label{sec:partial_state_sem}
%
To model concurrent behavior, we associate a partial-state model to each atomic component.
In global-state semantics, one does not distinguish the beginning of an interaction (or a transition) from its completion.
That is, the interactions and transitions of a system execute atomically and sequentially.
Partial states and the corresponding internal transitions are needed for modeling non-atomic executions.
Atomic components with partial states behave as atomic components except that each transition is decomposed into a sequence of two transitions: a visible transition followed by an internal $\beta$-labeled transitions (aka busy transition).
Between these two transitions, a so-called \textit{busy location} is added.
Intuitively, busy transitions are notifications indicating the completion of internal computations. 
Below, we define the transformation of a component with global-state semantics to a component with partial-state semantics (extending the definition in~\cite{basu2008distributed} with variables, guards, and computation steps on transitions).
\begin{definition}[Atomic component with partial states]
\label{def:atomic-partial}
The partial-state-semantics version of atomic component $B=(P,L, T,X)$ is $B^\perp=(P\cup\{\beta\}$, $L\cup L^\perp $, $T^\perp$, $X)$, where $\beta \notin P$ is a special port, $L^\perp= \{l^\perp_t\mid t\in T\}$ (resp. $L$) is the set of busy locations (resp. ready locations) such that $L^\perp\cap L=\emptyset$ and
$T^\perp=\{(l,p,g_{\tau},[~],l^\perp_{\tau}),(l^\perp_{\tau},\beta, \true, f_{\tau}, l')\mid \exists \tau=(l, p, g_{\tau}, f_{\tau}, l')\in T\}$ is the set of transitions.
\end{definition}
Assuming some available atomic components with partial states $B_1^\perp,\ldots,B_n^\perp$, we construct a composite component with partial states.
\begin{definition}[Composite component with partial states]
$B^\perp=\gamma^\perp(B_{1}^\perp,...,B_{n}^\perp)$ is a composite component where $\gamma^\perp=\gamma\cup \{\{{\beta}_i\}\}_{i=1}^n$, and $\{\{{\beta}_i\}\}_{i=1}^n$ is the set of singleton busy interactions.
\end{definition}
The notions and notation related to traces are lifted to components with partial states in the natural way.
We extend the definition of $\interactions$ (defined in Section~\ref{sec:cbs}) to traces in partial-state semantics such that $\beta$ interactions are filtered out.   
\begin{example}[Composite component with partial states]
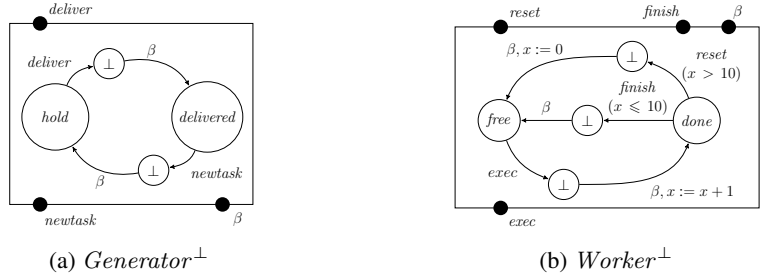
\begin{figure}[t]
    \centering
    \begin{subfigure}[b]{0.4\textwidth}
    \centering
    \Large
   \scalebox{.4} {   
\begin{tikzpicture}[->,>=stealth',shorten >=1pt,auto,node distance=5cm, semithick]
  \tikzstyle{every state}=[fill=none,draw=black,text=black]  

  \filldraw[fill=none,draw=black](0,0.5)--(0,6.5)--(8,6.5)--(8,0.5)--(0,0.5)--cycle; 
 \node at (1.5,3.4) [state] (A)        [minimum size=2.2cm]              {$\ready$};
  \node            [state] (B)  [right of =A,minimum size=2.2cm]          {$\delivered$};
  \node [state] (A') [above right of =A,minimum size=0.5cm,node distance=2.5cm]         {$\perp$};
\node [state] (B') [below left of =B,minimum size=0.5cm,node distance=2.5cm]          {$\perp$};

\path [out=10,in=120](A') edge  [above left ] node[text centered] {$\beta$} (B);
\path [out=190,in=-60](B') edge  [below]  node[text centered] {$\beta$} (A);  
\path [out=-110,in=10] (B) edge  [below right]   node[text centered] { $\newtask$} (B');
\path [out=70,in=190](A) edge  [above left]   node[text centered] { $\deliver$} (A');  
  \fill (1,0.5) circle (0.24cm);\node at (2,0){$\newtask$};
  \fill (1,6.5) circle (0.24cm);\node at (2,7) {$\deliver$};
  \fill (7,0.5) circle (0.24cm);\node at (7.5,0){$\beta$};


\end{tikzpicture}}
\caption{$\taskgenerator^\bot$} 
\label{fig:Task generator_partial}
    \end{subfigure}
    \begin{subfigure}[b]{0.4\textwidth}
       \centering
          \Large
   \scalebox{.4} {   
\begin{tikzpicture}[->,>=stealth',shorten >=1pt,auto,node distance=5cm, semithick]
  \tikzstyle{every state}=[fill=none,draw=black,text=black]  

  \filldraw[fill=none,draw=black](-0.5,10)--(-0.5,16)--(9.5,16)--(9.5,10)--(-0.5,10)--cycle;
\node at (0.95,12.9) [state] (A)[minimum size=0.5cm]         {$\free$};
\node [state] (B)  [right of =A,minimum size=0.5cm,node distance=6.5cm]         {$\done$};
\node [state] (A') [below right of =A,minimum size=0.5cm,node distance=3cm]   {$\perp$};
\node [state] (B') [above left of =B,minimum size=0.5cm,node distance=3cm]    {$\perp$};
\node [state] (C') [left of =B,minimum size=0.5cm,node distance=3.6cm]          {$\perp$};

\path [out=0,in=-110](A') edge  [below right ] node[text centered] {$\beta,x$ :$=x+1$} (B);
\path [out=180,in=0] (C') edge [above] node[text centered] { $\beta$} (A);
\path [out=180,in=70](B') edge  [above]  node[text centered] {$\beta,x$ :$=0\quad\quad $} (A);
\path (B) edge  [above]   node[text centered,text width= 3cm] { $\finish$ $(x\leqslant 10)$} (C');
\path [out=110,in=-20] (B) edge  [above right]   node[text centered,xshift=-0.4cm,text width= 3cm,yshift=-0.4cm] { $\maintenance$ $(x>10)$}(B');
\path [out=-70,in=160](A) edge  [below left]   node[text centered] { $exec$} (A'); 
  \fill (1,10) circle (0.24cm);\node at (1.7,9.5) {$\exec$};
  \fill (1,16) circle (0.24cm);\node at (1.8,16.5) {$\maintenance$};
  \fill (7,16) circle (0.24cm);\node at (6.3,16.5) {$\finish$};
  \fill (8.5,16) circle (0.24cm);\node at (8.8,16.5){$\beta$};


\end{tikzpicture}}
\caption{ {$\worker^\bot$}} 
\label{fig:Workers_partial}
       \end{subfigure}
       \caption{Atomic components of system Task with partial-states} 
\label{fig:task atoms partial}
\end{figure}
The corresponding composite component of system Task with partial-state semantics is $\gamma^\bot(\mathit{\worker_1^\perp}$, $\mathit{\worker_2^\perp}$, $\mathit{\worker_3^\perp}$, $\taskgenerator^\perp)$, where each $\worker_i^\perp$ for $i\in[1\upto 3]$ is identical to the component in \mfigref{fig:Workers_partial} and $\taskgenerator^\perp$ is the component in \mfigref{fig:Task generator_partial}.
To simplify the depiction of these components, we represent each busy location $l^\perp$ as $\perp$.
The set of interactions is $\gamma^\perp=\{\mathit{ex_{12}}$, $\mathit{ex_{13}}$, $ \mathit{ex_{23}}$, $\mathit{r_1}$, $\mathit{r_2}$, $\mathit{r_3}$, $\mathit{f_1}$, $\mathit{f_2}$, $\mathit{f_3}$, $\mathit{n_t}\} \cup \{\{\beta_1\}$, $\{\beta_2\}$, $\{\beta_3\}$, $\{\beta_4\}\}$.
One possible trace of system Task with partial-state semantics is: 
$(\free$, $\free$, $\free$, $\ready)\cdot ex_{12}\cdot(\perp$, $\perp$, $\free$, $\perp)\cdot\beta_4\cdot(\perp$, $\perp$, $\free$, $\delivered)\cdot n_t\cdot(\perp$, $\perp$, $\free$, $\perp)$.
\end{example}
It is possible to show that the partial-state system is a correct implementation of the global-state system, that is, the two systems are (weakly) bisimilar (cf. \cite{basu2008distributed}, Theorem 1).
A weak bisimulation relation $R$ is defined between the set of states of the model in global-state semantics (\ie $Q$) and the set of states of its partial-state model (\ie $Q^\bot$), such that $R=\{(q,r)\in Q\times Q^\bot \mid r\goesto[\beta^*] q\}$. Any global state in partial-state semantics model is equivalent to the corresponding global state in global-state semantics model, and any partial state in partial-state semantics model is equivalent to the successor global state obtained after stabilizing the system by executing busy interactions (which take place independently).

In the sequel, we consider a CBS with global-state semantics $B$ and its partial-states semantics version $B^\perp$.
Intuitively, from any trace of $B^\perp$, we want to reconstruct on-the-fly the corresponding trace in $B$ and evaluate a property which is defined over global states of $B$.
\begin{remark}
	We note that transforming a CBS with global-state semantics into a CBS with partial-state semantics resembles a special case of splitting semantics carried out for process algebraic systems \cite{van97,hoare78}, i.e., splitting each action into two atomic sub-actions.
	Indeed, we shall see that the method presented in this paper re-constructs a trace in the interleaving semantics from a trace in the concurrent semantics based on splitting.
\end{remark}
%
\subsection{Witness Relation and Witness Trace}
\begin{figure}[t]
    \centering
\tikzstyle{line} = [dashed,draw,-]
\tikzstyle{trans} = [ultra thick,draw,->,>=stealth']
\tikzstyle{beta} = [thick,draw,->,>=stealth']
\resizebox{\textwidth}{!}{
   \Large
\begin{tikzpicture}[every state/.style={minimum size=0.9cm}] 
    \node at (0,0)    (A) {$(\free,\free,\free,\ready)$};
    \node [node distance = 3cm, below of=A] (B) {$(\free,\free,\free,\ready)$};
    \node [node distance = 6cm, right of=A]  (C) {$(\perp,\perp,\free,\perp)$};  
    \node [node distance = 5.5cm, right of=C]  (D) {$(\perp,\perp,\free,\delivered)$}; 
    \node [node distance = 5.5cm, right of=D]  (E) {$(\perp,\perp,\free,\perp)$}; 
    \node [node distance = 3cm, below of=D,minimum size=0.2cm] (Q) {$(\done,\done,\free,\delivered)$};   
    \node [node distance = 10cm, right of=Q,minimum size=0.2cm] (R) {$(\done,\done,\free,\ready)$}; 

    (-0.5,-3.8) coordinate (t_k_unten) -- (24,-3.8) coordinate (t_k_opt_unten); 

    (-0.5,0.8) coordinate (t_k_unten) -- (24,0.8) coordinate (t_k_opt_unten);
         
    \path [line] (A) -- node[] {$R$} (B);
    \path [trans](A) -- node[above] {$\ex_{12}$} (C);
    \path [beta] (C) -- node[above] {$\beta_4$} (D);
    \path [trans] (D) -- node[above] {$n_t$} (E);
    \path [trans] (Q) -- node[above] {$n_t$} (R);
    \path [trans](B) -- node[above] {$\ex_{12}$} (Q);   
    \path [line] (C) -- node[] {$R$} (Q);
    \path [line] (D) -- node[] {$R$} (Q);
    \path [line] (E) -- node[] {$R$} (R);

\end{tikzpicture}
}
\caption{An example of witness trace in system Task}
\label{fig:witness_task}
\end{figure}
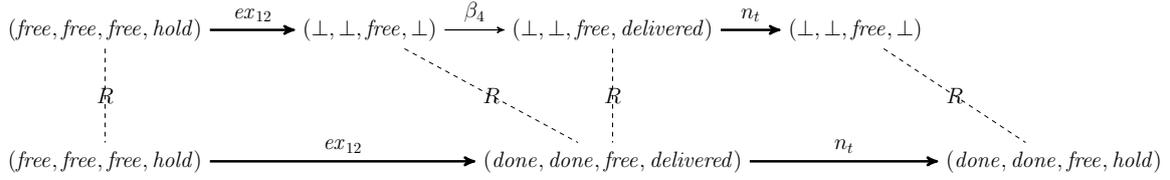

%
We define the notion of \textit{witness} relation between traces in global-state semantics and traces in partial-state semantics, based on the bisimulation between $B$ and $B^\perp$.
Any trace of $B^\perp$ is related to a trace of $B$, i.e., its \emph{witness}.
The witness trace allows to monitor the system in partial-state semantics (thus benefiting from the parallelism) against properties referring to the global behavior of the system.
\begin{definition}[Witness relation and witness trace]
\label{def:witness}
Given the bisimulation $R$ between $B$ and $B^\perp$, the witness relation $\W\subseteq \Tr(B)\times \Tr(B^\perp)$ is the smallest set that contains $(\mathit{Init},\mathit{Init})$ and satisfies the following rules:
\begin{itemize}
\item $(\sigma_1\cdot a\cdot q_1,\sigma_2\cdot a\cdot q_2)\in \W$, if $a\in \gamma$ and $(q_1,q_2)\in R$,
\item $(\sigma_1,\sigma_2\cdot\beta\cdot q_2)\in \W$, if $(\last(\sigma_1),q_2)\in R$;
\end{itemize}
whenever $(\sigma_1,\sigma_2)\in \W$.

If $(\sigma_1,\sigma_2)\in \W$, we say that $\sigma_1$ is a \emph{witness trace} of $\sigma_2$.
\end{definition}
Suppose that the witness relation relates a trace in partial-state semantics $\sigma_2$ to a trace in global-state semantics $\sigma_1$.
The states obtained after executing the same interaction in the two systems are bisimilar.
Moreover, any move through a busy interaction in $B^\perp$ preserves the bisimulation between the state of $\sigma_2$ followed by the busy interaction in $B^\perp$ and the last state of $\sigma_1$ in $B$.
%
%
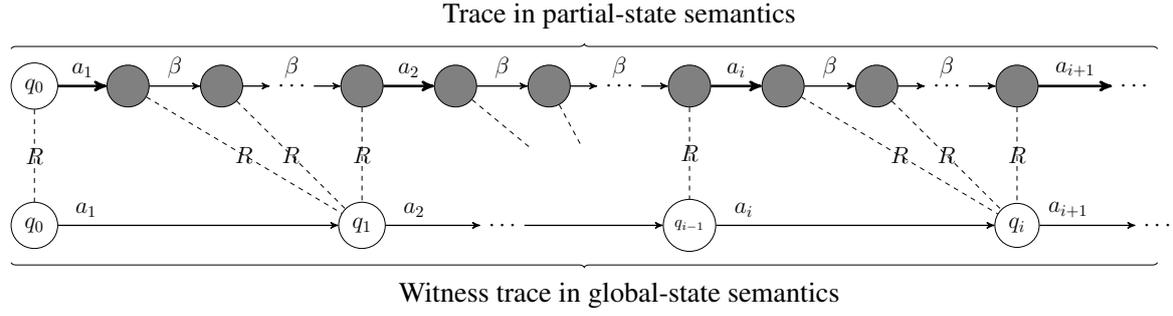
\begin{figure}[t]
    \centering
\tikzstyle{line} = [dashed,draw,-]
\tikzstyle{trans} = [ultra thick,draw,->,>=stealth']
\tikzstyle{beta} = [thick,draw,->,>=stealth']
\resizebox{\textwidth}{!}{
   \Large
\begin{tikzpicture}[every state/.style={minimum size=0.9cm}] 
    \node at (0,0)[state]  [minimum size=0.2cm]  (A) {$q_0$};
    \node [state][node distance = 3cm, below of=A,minimum size=0.2cm] (B) {$q_0$};
    \node [state][node distance = 2cm, right of=A,fill=gray]  (C) {};  
    \node [state][node distance = 2cm, right of=C,fill=gray]  (D) {}; 
    \node [node distance = 1.5cm, right of=D]                 (E) {$\cdots$}; 
    \node [state][node distance = 1.5cm, right of=E,fill=gray](F) {};
    \node [state][node distance = 2cm, right of=F,fill=gray]  (G) {};
    \node [state][node distance = 2cm, right of=G,fill=gray]  (H) {};
    \node [node distance = 1.5cm, right of=H]                 (I) {$\cdots$};
    \node [state][node distance = 1.5cm, right of=I,fill=gray](J) {};
    
    \node [state][node distance = 2cm, right of=J,fill=gray]  (K) {};
    \node [state][node distance = 2cm, right of=K,fill=gray]  (L) {};
    \node [node distance = 1.5cm, right of=L]                 (M) {$\cdots$};
    \node [state][node distance = 1.5cm, right of=M,fill=gray](N) {};
    \node [node distance = 2.5cm, right of=N]                 (O) {$\cdots$};

    \node [state][node distance = 3cm, below of=F,minimum size=0.2cm] (Q) {$q_1$};   
    \node [node distance = 3cm, right of=Q,minimum size=0.2cm]        (R) {$\cdots$}; 
    \node [state][node distance = 3cm, below of=J,minimum size=0.2cm] (S) {\normalsize {$q_{i-1}$}}; 
    \node [state][node distance = 3cm, below of=N,minimum size=0.2cm] (T) {$q_i$};
    \node [node distance = 3cm, right of=T,minimum size=0.2cm]        (U) {$\cdots$};  
    \node [node distance = 0.4cm, above of=E,minimum size=0.2cm]      (EE) {$\beta$}; 
    \node [node distance = 0.4cm, above of=I,minimum size=0.2cm]      (II) {$\beta$}; 
    \node [node distance = 0.4cm, above of=M,minimum size=0.2cm]      (MM) {$\beta$}; 
    \node at (10.8,-1.5) [minimum size=0.2cm]  (GG) {};
    \node at (11.8,-1.5) [minimum size=0.2cm]  (HH) {};
    \node [node distance = 1.5cm, above of=I,minimum size=0.2cm]      (II) {{\LARGE Trace in partial-state semantics}};
    \node [node distance = 4.5cm, below of=I,minimum size=0.2cm]      (III) {{\LARGE Witness trace in global-state semantics}};

\draw[decorate,decoration={brace,amplitude=3pt,mirror}] 
    (-0.5,-3.8) coordinate (t_k_unten) -- (24,-3.8) coordinate (t_k_opt_unten); 

\draw[decorate,decoration={brace,amplitude=3pt}] 
    (-0.5,0.8) coordinate (t_k_unten) -- (24,0.8) coordinate (t_k_opt_unten);

    \path [line] (A) -- node[] {$R$} (B);
    \path [trans](A) -- node[above] {$a_1$} (C);
    \path [beta] (C) -- node[above] {$\beta$} (D);
    \path [beta] (D) -- node[above] {} (E);
    \path [beta] (E) -- node[above] {} (F);
    \path [trans](F) -- node[above] {$a_2$} (G);
    \path [beta] (G) -- node[above] {$\beta$} (H);
    \path [beta] (H) -- node[above] {} (I);
    \path [beta] (I) -- node[above] {} (J);
    \path [trans](J) -- node[above] {$a_i$} (K);
    \path [beta] (K) -- node[above] {$\beta$} (L);
    \path [beta] (L) -- node[above] {} (M);
    \path [beta] (M) -- node[above] {} (N);
    \path [trans](N) -- node[above] {$a_{i+1}$} (O);
    \path [beta] (B) -- node[above,pos=0.1] {$a_1$} (Q);
    \path [line] (F) -- node[] {$R$} (Q);
    \path [line] (C) -- node[] {$R$} (Q);
    \path [line] (D) -- node[] {$R$} (Q);
    \path [beta] (Q) -- node[above,pos=0.3] {$a_2$} (R);
    \path [beta] (R) -- node[above] {} (S);
    \path [beta] (S) -- node[above,pos=0.1] {$a_i$} (T);
    \path [beta] (T) -- node[above,pos=0.3] {$a_{i+1}$} (U);
    
    \path [line] (G) -- node[] {} (GG);
    \path [line] (H) -- node[] {} (HH);
    
    \path [line] (J) -- node[] {$R$} (S);
    \path [line] (K) -- node[] {$R$} (T);
    \path [line] (L) -- node[] {$R$} (T);
    \path [line] (N) -- node[] {$R$} (T);
\end{tikzpicture}
}
\caption{\red{Witness trace built using weak bisimulation ($R$)}}
\label{fig:witness}
\end{figure}
\begin{example}[Witness relation]
Figure~\ref{fig:witness} illustrates the witness relation.
State $q_0$ is the initial state of $B$ and $B^\perp$.
In the trace of $B^\perp$, gray circles after each interaction represent partial states which are bisimilar to the global state that comes after the corresponding trace of $B$.
\end{example}
\begin{example}[Witness trace]
Let us consider $\sigma_2$ as a trace of system Task with partial-state semantics depicted in \mfigref{fig:witness_task} where $\sigma_2=(\free$, $\free$, $\free$, $\ready)\cdot \ex_{12}\cdot(\perp$, $\perp$, $\free$, $\perp)\cdot\beta_4\cdot(\perp$, $\perp$, $\free$, $\delivered)\cdot n_t\cdot(\perp$, $\perp$, $\free$, $\perp)$.
The witness trace corresponding to trace $\sigma_2$ is $(\free$, $\free$, $\free$, $\ready)\cdot ex_{12}\cdot(\done$, $\done$, $\free$, $\delivered)\cdot n_t\cdot(\done$, $\done$, $\free$, $\ready)$.
\end{example}
The following property states that any trace in partial-state semantics and its witness trace have the same sequence of interactions.
\begin{property}
\label{property:same_sequence}
$\forall (\sigma_1,\sigma_2)\in \W \quant  \interactions(\sigma_1) = \interactions (\sigma_2)$.
\end{property}
\begin{proof}
The proof is done by induction on the length of the sequence of interactions and follows from the definitions of the witness relation and witness trace.
The proof of this property can be found in Appendix~\ref{proof:same_sequence} (p.~\pageref{proof:same_sequence}).
\end{proof}
The next property states that any trace in the partial-state semantics has a unique witness trace in the global-state semantics.
\begin{property}
\label{property:unique_witness}
$\forall \sigma_2\in \Tr(B^\perp), \exists! \sigma_1 \in \Tr(B)  \quant  (\sigma_1, \sigma_2) \in \W $.
\end{property}
\begin{proof}
This proof is done by contradiction. The proof of this property is given in Appendix~\ref{proof:unique_witness} (p.~\pageref{proof:unique_witness}).
\end{proof}

We note $\W(\sigma_2)=\sigma_1$ when $(\sigma_1,\sigma_2)\in \W$.

Note that, when running a system in partial-state semantics, the global state of the witness trace after an interaction $a$ is not known until all the components involved in $a$ have reached their ready locations after the execution of $a$.
Nevertheless, even in non-deterministic systems, after a deterministic execution, this global state is uniquely defined and consequently there is always a unique witness trace (that is, non-determinism is resolved at runtime).
%
\subsection{Construction of the Witness Trace}
%
Given a trace in partial-state semantics, the witness trace is computed using function $\RGT$ (Reconstructor of Global Trace).
The global states (of the trace in the global-state semantics) are reconstructed from partial states.
We define a function to reconstruct global states from partial states.
\begin{definition}[Function $\RGT$ - Reconstructor of Global States]
\label{def:rgt}
Function $\RGT:\Tr(B^\perp) \longrightarrow \pref(\Tr(B))$ is defined as:
\[
\RGT(\sigma)=\D(\Acc(\sigma)),
\]
where:
\begin{itemize}
\item[--] $\Acc:\Tr(B^\perp) \longrightarrow Q \cdot (\gamma\cdot Q)^* \cdot (\gamma\cdot (Q^\perp \backslash Q))^*$ is defined as:
\begin{itemize}
\item[$\bullet$] $\Acc(\mathit{Init})=\mathit{Init}$,
\item[$\bullet$] $\Acc(\sigma\cdot a\cdot q)=\Acc(\sigma)\cdot a\cdot q \quad \quad \quad \! \! \quad\quad\qquad \qquad\quad$ for $a\in \gamma,$
\item[$\bullet$] $\Acc(\sigma\cdot\beta\cdot q)=\Map \ [ x \mapsto \upd(q,x)] \ (\Acc(\sigma)) \qquad\,$ for $\beta\in \{\{\beta_i\}\}^n_{i=1};$
\end{itemize}
\item[--] $\D:Q \cdot (\gamma\cdot Q)^* \cdot (\gamma\cdot (Q^\perp \backslash Q))^* \longrightarrow \pref(\Tr(B))$ is defined as:
\begin{itemize}
\item[] $\D(\sigma)=\max_\preceq(\{\sigma'\in \pref(\sigma)\mid \last(\sigma')\in Q\})$
\end{itemize}
\end{itemize}
with $\upd: Q^\perp \times (Q^\perp\cup \gamma)\longrightarrow Q^\perp\cup \gamma$ defined as:
\begin{itemize}
\item[--] $\upd((q_1,\ldots,q_n),a)=a$, for $a\in \gamma$,
\item[--] $\upd\bigg((q_1,\ldots,q_n),(q'_1,\ldots ,q'_n)\bigg)=(q''_1,\ldots ,q''_n)$,\\
where $ \forall k\in [1\upto n] \quant  q''_k=\begin{dcases}
   q_k  &\text{ if } (q_k\notin Q^\perp_k) \wedge (q'_k\in Q^\perp_k) \\
   q'_k &\text{ otherwise.}
   \end{dcases}$
\end{itemize}
\end{definition}
Function $\RGT$ uses helper functions $\Acc$ and $\D$.
First, function $\Acc$ is an \emph{accumulator} function which takes as input a trace in partial-state semantics $\sigma$, removes $\beta$ interactions and the partial states after $\beta$.
Function $\Acc$ uses the (information in the) partial state after $\beta$ interactions in order to update the partial states using function $\upd$.
Then, function $\D$ returns the longest prefix of the result of $\Acc$ corresponding to a trace in global-state semantics.

Note that, because of the inductive definition of function $\Acc$, the input trace can be processed step by step by function $\RGT$ and allows to generate the witness incrementally.
Moreover, such definition allows to apply the function $\RGT$ to a running system by monitoring execution of interactions and partial states of components.
Finally, we note that function $\RGT$ is monotonic (w.r.t. prefix ordering on sequences).

Such an online computation is illustrated in the following example.
\begin{example}[Applying function RGT]
Table~\ref{Table:RGT} illustrates Definition~\ref{def:rgt} on one trace of system Task with initial state $(\free,\free,\free,\ready)$ followed by interactions $\ex_{12}$, $\beta_4$, $n_t$, $\beta_2$, and $\beta_1$.
We comment on certain steps illustrated in Table~\ref{Table:RGT}.
At step 0, the outputs of functions $\Acc$ and $\D$ are equal to the initial state.
At step 1, the execution of interaction $\ex_{12}$ adds two elements $\ex_{12}\cdot(\perp,\perp,\free,\perp)$ to traces $\sigma$ and $\Acc(\sigma)$.
At step 2, the state after $\beta_4$ has fresh information on component $\taskgenerator$ which is used to update the existing partial states, so that $(\perp,\perp,$ $\free,\perp)$ is updated to $(\perp,\perp,$ $\free,\delivered)$.
At step 5, $\worker_1$ becomes ready after $\beta_1$, and the partial state $(\perp,$ $\done,\free,\delivered)$ in the intermediate step is updated to the global state $(\done,\done,\free,\delivered)$, therefore it appears in the output trace.
\end{example}
\begin{table}[!t]
\centering
\caption{Values of function $\RGT$ for a sample input} 
\resizebox{\textwidth}{!}{
\begin{tabular}{|c|c|c|c|}
\hline
Step & \begin{tabular}[c]{@{}c@{}} Input trace in partial semantics \\$\sigma$\end{tabular} & \begin{tabular}[c]{@{}c@{}} Intermediate step \\$\Acc(\sigma)$\end{tabular} & \begin{tabular}[c]{@{}c@{}} Output trace in global semantics \\$\RGT(\sigma)$ \end{tabular} \\ \hline
\rowcolor[HTML]{EFEFEF} 
0 & $(\free,\free,\free,\ready)$ & $(\free,\free,\free,\ready)$ & $(\free,\free,\free,\ready)$ \\ 
1 &\begin{tabular}[c]{@{}c@{}}$(\free,\free,\free,\ready)\cdot\ex_{12}\cdot$\\$(\perp,\perp,\free,\perp)$\end{tabular}  & \begin{tabular}[c]{@{}c@{}}$(\free,\free,\free,\ready)\cdot\ex_{12}\cdot$\\$(\perp,\perp,\free,\perp)$\end{tabular} & \begin{tabular}[c]{@{}c@{}}$(\free,\free,\free,\ready)\cdot\ex_{12}$\end{tabular} \\
\rowcolor[HTML]{EFEFEF} 
2 & \begin{tabular}[c]{@{}c@{}}$(\free,\free,\free,\ready)\cdot\ex_{12}\cdot$\\$(\perp,\perp,\free,\perp)\cdot\beta_4\cdot$\\$(\perp,\perp,\free,\delivered)$\end{tabular}  & \begin{tabular}[c]{@{}c@{}}$(\free,\free,\free,\ready)\cdot\ex_{12}\cdot$\\$(\perp,\perp,\free,\delivered)$\end{tabular} & \begin{tabular}[c]{@{}c@{}}$(\free,\free,\free,\ready)\cdot\ex_{12}$\end{tabular} \\
3 & \begin{tabular}[c]{@{}c@{}}$(\free,\free,\free,\ready)\cdot\ex_{12}\cdot$\\$(\perp,\perp,\free,\perp)\cdot\beta_4\cdot$\\$(\perp,\perp,\free,\delivered)\cdot n_t\cdot$\\$(\perp,\perp,\free,\perp)$\end{tabular} & \begin{tabular}[c]{@{}c@{}}$(\free,\free,\free,\ready)\cdot\ex_{12}\cdot$\\$(\perp,\perp,\free,\delivered)\cdot n_t\cdot$\\$(\perp,\perp,\free,\perp)$\end{tabular} & \begin{tabular}[c]{@{}c@{}}$(\free,\free,\free,\ready)\cdot\ex_{12}$\end{tabular} \\
\rowcolor[HTML]{EFEFEF} 
4 & \begin{tabular}[c]{@{}c@{}}$(\free,\free,\free,\ready)\cdot\ex_{12}\cdot$\\$(\perp,\perp,\free,\perp)\cdot\beta_4\cdot$\\$(\perp,\perp,\free,\delivered)\cdot n_t\cdot$\\$(\perp,\perp,\free,\perp)\cdot\beta_2\cdot$\\$(\perp,\done,\free,\perp)$\end{tabular} & \begin{tabular}[c]{@{}c@{}}$(\free,\free,\free,\ready)\cdot\ex_{12}\cdot$\\$(\perp,\done,\free,\delivered)\cdot n_t\cdot$\\$(\perp,\done,\free,\perp)$\end{tabular} &  \begin{tabular}[c]{@{}c@{}}$(\free,\free,\free,\ready)\cdot\ex_{12}$\end{tabular}\\
5 & \begin{tabular}[c]{@{}c@{}}$(\free,\free,\free,\ready)\cdot\ex_{12}\cdot$\\$(\perp,\perp,\free,\perp)\cdot\beta_4\cdot$\\$(\perp,\perp,\free,\delivered)\cdot n_t\cdot$\\$(\perp,\perp,\free,\perp)\cdot\beta_2\cdot$\\$(\perp,\done,\free,\perp)\cdot\beta_1\cdot$\\$(\done,\done,\free,\perp)$\end{tabular}& \begin{tabular}[c]{@{}c@{}}$(\free,\free,\free,\ready)\cdot\ex_{12}\cdot$\\ $(\done,\done,\free,\delivered)\cdot n_t\cdot$\\$(\done,\done,\free,\perp)$\end{tabular} & \begin{tabular}[c]{@{}c@{}}$(\free,\free,\free,\ready)\cdot\ex_{12}\cdot$\\ $(\done,\done,\free,\delivered)\cdot n_t$\end{tabular} \\
\hline
\end{tabular}}
\label{Table:RGT}
\end{table}
%
%
\subsection{Properties of Global-trace Reconstruction}
%
We state some properties of global-trace reconstruction based on function $\RGT$, namely the soundness and maximality (information-wise) of the reconstructed global trace.
To do so, we first start by stating some intermediate lemmas on the computation performed by function $\RGT$. 
\begin{lemma}
\label{lemma:length-acc}
$\forall (\sigma_1,\sigma_2)\in W \quant |\Acc(\sigma_2)|=|\sigma_1|=2s+1$, where $s=|\interactions (\sigma_1)|$,
 $\Acc$ is the accumulator used in the definition of function $\RGT$ (Definition~\ref{def:rgt}), and function $\interactions$ (defined in \secref{sec:partial_state_sem}) returns the sequence of interactions in a trace (removing $\beta$).
\end{lemma}
Lemma~\ref{lemma:length-acc} states that, for a given trace in partial-state semantics $\sigma_2$, the length of $\Acc(\sigma_2)$ is equal to the length of the witness of $\sigma_2$ (\ie $\sigma_1$).
\begin{lemma}
\label{lemma:acc}
$\forall \sigma\in \Tr(B^\perp) \quant $ let $\Acc(\sigma)=(q_0\cdot a_1 \cdot q_1 \cdots a_s\cdot q_s)$ in 
\begin{itemize}
\item[] $\exists k\in[1\upto s] \quant  q_k\in Q\implies\forall z\in[1\upto k] \quant  q_z\in Q \wedge q_{z-1} \stackrel{a_z}{\longrightarrow}q_z$.
\end{itemize}
\end{lemma}
Lemma~\ref{lemma:acc} states that, for a given trace in partial-state semantics $\sigma$, if there exists a global state $q_k\in Q, k\in[1 \upto s]$ in sequence $\Acc(\sigma)$, then all the states occurring before $q_k$ in $\Acc(\sigma)$ are global states.

The next proposition states that the sequence of global states produced by function $\RGT$ (which is the composition of functions $\D$ and $\Acc$) follows the global-state semantics.
\begin{proposition}
\label{proposition:length-discriminant}
$\forall \sigma\in \Tr(B^\perp) \quant $
\[
\begin{array}{l}
\quad |\D(\Acc(\sigma))|\leq |\Acc(\sigma)| \\
\wedge \D(\Acc(\sigma))=q_0\cdot a_1 \cdot q_1 \cdots a_d\cdot q_d \implies \forall i\in[1 \upto d] \quant q_{i-1} \stackrel{a_i}{\longrightarrow}q_i,
\end{array}
\]
where $\Acc$ (resp. $\D$) is the accumulator (resp. discriminant) function used in the definition of function $\RGT$ (Definition~\ref{def:rgt}) such that $\RGT(\sigma)=\D(\Acc(\sigma))$.
\end{proposition}
Proposition~\ref{proposition:length-discriminant} states that, for any trace in partial-state semantics $\sigma$, 1) the length of the output trace of function $\RGT$ (i.e., $\D(\Acc(\sigma))$) is lower than or equal to the length of the output of function $\Acc$ (i.e., $\Acc(\sigma)$), and 2) the output trace of function $\RGT$ is a trace in global-state semantics.

Moreover, the last element of a given trace in partial-state semantics $\sigma$ is always the same as the last element of output of $\Acc(\sigma)$, as stated by the following lemma.
\begin{lemma}
\label{lemma:last-acc}
$\forall \sigma\in \Tr(B^\perp) \quant  \last(\Acc(\sigma))=\last(\sigma)$.
\end{lemma}
Finally, any trace in partial-state semantics $\sigma$ and its image through function $\Acc$ have the same sequence of interactions, as stated by the following lemma. 
\begin{lemma}
\label{lemma:interactions-acc}
$\forall \sigma\in \Tr(B^\perp) \quant  \interactions(\Acc(\sigma))=\interactions(\sigma)$.
\end{lemma}
Based on the above lemmas, we have the following theorem which states the \emph{soundness} and \emph{maximality} of the reconstructed global trace.
That is, applying function $\RGT$ on a trace in partial-state semantics produces the longest possible prefix of the corresponding witness trace with respect to the current trace of the partial-state semantics model.
\begin{theorem}[On the reconstructed global trace with function $\RGT$]
\label{theorem_witness}
$\forall \sigma\in \Tr(B^\perp) \quant $
\[
\begin{array}{l}
\qquad \quad \last(\sigma) \in Q \implies \RGT(\sigma)=\W(\sigma)\\
\qquad \wedge \last(\sigma) \notin Q \implies
\RGT(\sigma)=\W(\sigma') \cdot a, \text{with}\\
\sigma' = \min_{\preceq} \{ \sigma_p \in \Tr(B^\perp) \mid \exists a\in\gamma, \exists \sigma'' \in \Tr(B^\perp)  \quant  \sigma=\sigma_p \cdot a \cdot \sigma'' \wedge \exists i\in[1 \upto n] \quant  \\
\qquad\qquad\qquad\qquad\qquad\qquad (B_i.P\cap a \neq \emptyset) \wedge(\forall j\in[1 \upto \length(\sigma'')] \quant \beta_i \neq\sigma''(j))\}
\end{array}
\]
\end{theorem}
Theorem~\ref{theorem_witness} distinguishes two cases:
\begin{itemize}
\item
When the last state of a system is a global state ($\last(\sigma)\in Q$), none of the components are in a busy location.
Moreover, function $\RGT$ has sufficient information to build the corresponding witness trace ($\RGT(\sigma)=\W(\sigma)$).
\item
When the last state of a system is a partial state, at least one component is in a busy location and function $\RGT$ can not build a complete witness trace because it lacks information on the current state of such components.
It is possible to decompose the input sequence $\sigma$ into two parts $\sigma'$ and $\sigma''$ separated by an interaction $a$.
The separation is made on the interaction $a$ occurring in trace $\sigma$ such that, for the interactions occurring after $a$ (i.e., in $\sigma''$), at least one component involved in $a$ has not executed any $\beta$ transition (which means that this component is still in a busy location).
Note that it may be possible to split $\sigma$ in several manners with the above description.
In such a case, function $\RGT$ computes the witness for the smallest sequence $\sigma'$ (w.r.t. prefix ordering) as above because it is the only sequence for which it has information regarding global states.
Note also that such splitting of $\sigma$ is always possible as $\last(\sigma) \notin Q$ implies that $\sigma$ is not empty, and $\sigma'$ can be chosen to be $\epsilon$.
\end{itemize}
In both cases, because of its inductive definition and monotonicity, $\RGT$ returns the maximal prefix of the corresponding witness trace that can be built with the information contained in the partial states observed so far. 

The above explanation can be extended to a full proof which is given in Appendix~\ref{proof:RGT-witness} (p.~\pageref{proof:RGT-witness}).
\begin{example}[Illustration of Theorem~\ref{theorem_witness}]
We illustrate the correctness of Theorem~\ref{theorem_witness} based on the execution trace in Table~\ref{Table:RGT}.
At step $0$, since the last element in the trace is the initial state we can see that the output of function $\RGT$ is equal to the witness trace which is the initial state as well.
At step $5$, the output of function $\RGT$ is a sequence which consists of the witness of sequence $(\free,\free,\free,\ready)\cdot\ex_{12}\cdot(\perp,\perp,\free,\perp) \cdot \beta_4 \cdot( \perp, \perp, \free,\delivered)$ ({\ie} $(\free,\free,\free,\ready)\cdot\ex_{12}\cdot(\done,\done,\free,\delivered)$) followed by $n_t$.
At this step, function $\RGT$ can not process partial states following interaction $n_t$, because the component involved in $n_t$ is still busy.
\end{example}

%
\section{Model Transformation}
\label{sec:inst}
%
We propose a model transformation of a composite component $B^\perp=\gamma^\perp(B^\perp_1,\ldots,B^\perp_n)$ such that it can produce the witness trace on-the-fly.
The transformed system can be plugged to a runtime monitor as described in~\cite{FalconeJNBB15}.
Our model transformation consists of three steps: 1) instrumentation of atomic components (\secref{sec:inst-atom}), 2) construction of a new component (RGT) which implements Definition~\ref{def:rgt} (\secref{sec:rgt-atom}), 3) modification of interactions in $\gamma^\perp$ such that (i) component RGT can interact with the other components in the system and (ii) new interactions connect RGT to a runtime monitor (\secref{sec:connections}).
\newcommand{\lsquarebracket}{[}
\newcommand{\rsquarebracket}{]}
\begin{figure}
    \centering
   \scalebox{.8} {             
\tikzstyle{block} = [rectangle, draw, fill=gray!20, 
    text width=5em, text centered, rounded corners, minimum height=2.5em]   
\tikzstyle{block2} = [rectangle, draw,  
    text width=5em, text centered, minimum height=2.5em]           
\tikzstyle{line} = [draw, -latex']      
\begin{tikzpicture}[node distance = 1.5cm, auto]   
 \node [block2, text width=15em]  (BIP) {Initial Model With Partial States};
 \node [block, below left of=BIP,text width=9em,node distance = 2cm, xshift=-2.5cm](NI){Building \\ RGT Atom}; 
 \node [block, below right of=BIP,text width=9em,node distance = 2cm, xshift=2.5cm](AT){Modification of Interactions}; 
 \node [block, below of=BIP,text width=9em,node distance = 1.42cm](AI){Atomic Components \\ Instrumentation}; 
 \node [block2,below of=BIP,text width=15em,node distance = 3cm](F){Model With Partial States That Produces the Witness Trace};
 
     \path [line] (BIP)  -| (AT);
     \path [line] (BIP)  -| (NI);
     \path [line] (BIP)  -- (AI);
     \path [line] (NI)    |- (F);
     \path [line] (AT)    |- (F);
     \path [line] (AI)    -- (F);
\end{tikzpicture}}
\caption{Model transformation}
\label{fig:Model transformation} 
\end{figure}
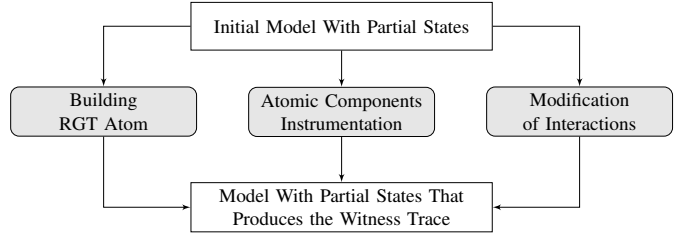
%
\subsection{Instrumentation of Atomic Components}
\label{sec:inst-atom}
%
Given an atomic component with partial-state semantics as per Definition~\ref{def:atomic-partial}, we instrument this atomic component such that it is able to transfer its state through port $\beta$.
The state of an instrumented component is delivered each time the component moves out from a busy location. 
In the following instrumentation, the state of a component is represented by the values of variables and the current location.
\begin{definition}[Instrumenting an atomic component]
Given an atomic component in partial-state semantics $B^\perp=(P\cup\{\beta\},$ $L\cup L^\perp,$ $T^\perp,$ $X)$ with initial location $l_0\in L$, we define a new component $B^r=(P^r, L\cup L^\perp, T^r, X^r)$ where:
\begin{itemize}
\item $X^r = X \cup \{\mathit{loc}\}$, $\mathit{loc}$ is initialized to $l_0$;
\item $P^r=P\cup\{\beta^r\}$, with $\beta^r=\beta[X^r];$
\item $T^r=\{(l,p,g_{\tau},[~],l^\perp_{\tau}), (l^\perp_{\tau},\beta,\true,f_{\tau};[\mathit{loc}:=\text{$l'$}],l') \mid \{(l,p,g_{\tau},[~],l^\perp_{\tau}),$ $(l^\perp_{\tau},\beta,$ $\true,f_{\tau},l')\} \subseteq T^\perp \}$.
\end{itemize}
\label{def:inst-atom}
\end{definition}
In $X^r$, $loc$ is a variable containing the current location.
$X^r$ is exported through port $\beta$.
An assignment is added to the computation step of each transition to record the location.
\begin{example}[Instrumenting an atomic component]
\begin{figure}[t]
    \centering
    \begin{subfigure}[b]{0.45\textwidth}
    \centering
    \Large
   \scalebox{.4} {   
\begin{tikzpicture}[->,>=stealth',shorten >=1pt,auto,node distance=5.5cm, semithick]
  \tikzstyle{every state}=[fill=none,draw=black,text=black]  

  \filldraw[fill=none,draw=black](0,0)--(0,8)--(9,8)--(9,0)--(0,0)--cycle; 
 \node at (1.7,4) [state] (A)        [minimum size=2.5cm]              {$\ready$};
  \node            [state] (B)  [right of =A,minimum size=2.5cm]          {$\delivered$};
  \node [state] (A') [above right of =A,minimum size=0.5cm,node distance=2.8cm]         {$\perp$};
\node [state] (B') [below left of =B,minimum size=0.5cm,node distance=2.8cm]          {$\perp$};

\path [out=10,in=120](A') edge  [above right] node[text centered,text width=3.5cm,xshift=-0.8cm] { $loc$:$= \delivered$ $\quad\beta\quad\quad$ } (B);
\path [out=190,in=-60](B') edge  [below]  node[text centered,text width=3.5cm] {$\quad\beta\quad\quad$  $loc$:$= \ready$} (A);  
\path [out=-120,in=10] (B) edge  [below right]   node[text centered] {$\newtask$} (B');
\path [out=70,in=190](A) edge  [above left]   node[text centered] {$\deliver$} (A');  
  \fill (1,0) circle (0.24cm);\node at (2,-0.5){$\newtask$};
  \fill (1,8) circle (0.24cm);\node at (2,8.5) {$\deliver$};
  \fill (7,0) circle (0.24cm);\node at (7.5,-0.5){$\beta$};

  \node [draw,dashed] at (7.8,0.8) {$loc$}; 
  \draw[dashed,-] (7,0)--(7.8,0.4);


\end{tikzpicture}}
\caption{Instrumented component $\taskgenerator^r$} 
\label{fig:Task generator inst}
    \end{subfigure}
    \begin{subfigure}[b]{0.45\textwidth}
       \centering
          \Large
   \scalebox{.4} {   
\begin{tikzpicture}[->,>=stealth',shorten >=1pt,auto,node distance=5cm, semithick]
  \tikzstyle{every state}=[fill=none,draw=black,text=black]  

  \filldraw[fill=none,draw=black](0,0)--(0,8)--(12,8)--(12,0)--(0,0)--cycle;
\node at (1.5,3.9) [state] (A)[minimum size=0.5cm]         {$\free$};
\node [state] (B)  [right of =A,minimum size=0.5cm,node distance=8cm]         {$\done$};
\node [state] (A') [below right of =A,minimum size=0.5cm,node distance=3.2cm]   {$\perp$};
\node [state] (B') [above left of =B,minimum size=0.5cm,node distance=3.2cm]    {$\perp$};
\node [state] (C') [left of =B,minimum size=0.5cm,node distance=3.5cm]          {$\perp$};

\path [out=0,in=-110](A') edge  [below right ] node[text centered,text width=3.5cm,yshift=0.2cm,xshift=0.3cm] {$\beta$ , $x$ :$=x+1$ $loc$:$= \done$} (B);
\path [out=180,in=0] (C') edge [above] node[text centered,text width=3cm] { $\quad\beta\quad$  $loc$:$= \free$} (A);
\path [out=180,in=70](B') edge  [above]  node[text centered,text width=4.5cm,yshift=-0.1cm] {$\quad\beta$ , $x$ :$=0\quad $  $loc$:$= \free$} (A);
\path (B) edge  [above]   node[text centered,text width=3cm] { $\finish$ $(x\leqslant 10)$ } (C');
\path [out=110,in=-20] (B) edge  [above right]   node[text centered,xshift=-0.5cm,yshift=-0.6cm,text width=3cm] {$(x> 10)$ $\maintenance$}(B');
\path [out=-70,in=160](A) edge  [below left]   node[text centered] { $\exec$} (A'); 
  \fill (1,0) circle (0.24cm);\node at (1.7,-0.65) {$\exec$};
  \fill (1,8) circle (0.24cm);\node at (1.8,8.5) {$\maintenance$};
  \fill (7,8) circle (0.24cm);\node at (6.3,8.5) {$\finish$};
  \fill (9.5,8) circle (0.24cm);\node at (9.8,8.5){$\beta$};

  \node [draw,dashed] at (11,7) {$loc$}; 
  \node [draw,dashed,text height=0.4cm] at (9.4,7) {$x$};
  \draw[dashed,-] (9.5,8)--(9.4,7.3);
  \draw[dashed,-] (9.5,8)--(11,7.3);
    

\end{tikzpicture}}
\caption{Instrumented component $\worker^r$} 
\label{fig:Workers_partial_inst}
       \end{subfigure}
       \caption{Instrumented atomic components of system Task} 
\label{fig:task atoms inst}
\end{figure}
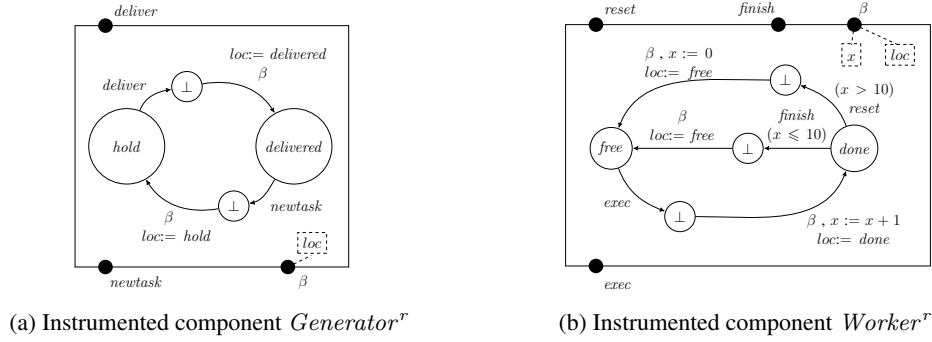
Figure~\ref{fig:task atoms inst} shows the instrumented version of atomic components in system Task (depicted in \mfigref{fig:task atoms partial}).
\begin{itemize}
\item
Figure~\ref{fig:Task generator inst} depicts component task generator, where $\taskgenerator^r.P^r=\{\deliver[\emptyset],\newtask[\emptyset],$ $\beta[\{\mathit{loc}\}]\}$,$\quad $  $\taskgenerator^r.T^r=\{(\ready, \deliver, \true, [~], \bot), (\bot, \beta,$ $\true, [\mathit{loc}:=\delivered],\delivered), (\delivered, \newtask,$ $\true, [~], \bot)$, $(\bot, \beta,\true,$ $[\mathit{loc}:=\ready], \ready)\}$, $\taskgenerator^r.X^r=\{\mathit{loc}\}$.
\item
Figure~\ref{fig:Workers_partial_inst} depicts a worker component, where
$\worker^r.P^r = \{\exec[\emptyset],   \finish[\emptyset],  \maintenance[\emptyset],  \beta[\{x,\mathit{loc}\}]\}$,$\quad\quad $ $\worker^r.T^r = \{(\free, \exec, \true, [~], \bot),  (\bot, \beta, \true, [x:=x+1; \mathit{loc}:=\done], \done),  (\done, \finish,\\ (x\leqslant 10), [~], \bot)$, $(\bot, \beta,\true, \mathit{loc}:=\free], \free)$, $(\done, \maintenance, (x>10), [~], \bot)$, $(\bot, \beta, \true, [x:=0;\\ \mathit{loc}:=\free], \free)\}$, $\worker^\bot.X^r=\{x,\mathit{loc}\}$.
\end{itemize}
\end{example}
%
\subsection{Creating a New Atomic Component to Reconstruct Global States}
\label{sec:rgt-atom}
%
Let us consider a composite component $B^\perp=\gamma^\perp(B^\perp_1,\ldots,B^\perp_n)$ with partial-state semantics, such that:
\begin{itemize}
\item $\mathit{Init} = (q^0_1,\ldots,q^0_n)$ is the initial state,
\item
$\gamma$ is the set of interactions in the corresponding composite component with global-state semantics with $\gamma =\gamma^\perp\setminus \{\{\beta_i\}\}_{i=1}^n$, and
\item
the corresponding instrumented atomic components $B_1^r,\ldots,B_n^r$ have been obtained through Definition~\ref{def:inst-atom} such that $B^r_i$ is the instrumented version of $B_i^\perp$.
\end{itemize}
We define a new atomic component, called RGT, which is in charge of accumulating the global states of the system $B^{\bot}$.
Component RGT is an operational implementation, as a component of function RGT (Definition~\ref{def:rgt}).
At runtime, we represent a global state as a tuple consisting of the valuation of variables and the location for each atomic component.
After a new interaction gets fired, component RGT builds a new tuple using the current states of components.
Component RGT builds a sequence with the generated tuples.
The stored tuples are updated each time the state of a component is updated.
Following Definition~\ref{def:inst-atom}, atomic components transfer their states through port $\beta$ each time they move from a busy location to a ready location.
RGT reconstructs global states from these received partial states and delivers them through the dedicated ports.
\begin{definition}[RGT atom]
\label{def:RGT-atom}
Component $\RGT$ is defined as $(P$, $L$, $T$, $X)$ where:
\begin{itemize}
\item[--] $X=\bigcup_{i\in[1 \upto n]}\{B^r_i.X^r\} \bigcup_{i\in[1 \upto n]}\{B^r_i.X^r_c\} \cup\{\mathit{gs}_a\mid a\in\gamma\} \cup\{(z_1,\ldots,z_n)\}\cup\{V,v,m\}$, where $B^r_i.X^r_c$ is a set containing a copy of the variables in $B^r_i.X^r$.
\item[--] $P=\bigcup_{i\in[1 \upto n]}\{\beta_i[B^r_i.X^r]\}\cup \{p_a[\emptyset]\mid a \in\gamma\}\cup \{p_a'[\bigcup_{i\in[1 \upto n]}\{B^r_i.X_c\}]\mid a\in \gamma\}$.
\item[--] $L=\{l\}$ is a set with one control location.
\item[--] $T=T_{\rm{new}}\cup T_{\rm{upd}}\cup T_{\rm{out}}$ is the set of transitions, where:
\begin{itemize}
\item $T_{\rm{new}}=\{(l,p_a,\bigwedge_{a \in \gamma} (\neg \mathit{gs}_a),\mathtt{new}(a),l)\mid a\in\gamma \}$,
\item $T_{\rm{upd}}=\{(l,\beta_i, \bigwedge_{a \in \gamma} (\neg \mathit{gs}_a), \mathtt{upd}(i),l) \mid i\in [1 \upto n]\}$,
\item $T_{\rm{out}}=\{(l,p'_a,\mathit{gs}_a,\mathtt{get},l)\mid a\in \gamma\}$.
\end{itemize}
\end{itemize}
\end{definition}
$X$ is a set of variables that contains the following variables:
\begin{itemize}
\item the variables in $B^r_i.X^r$ for each instrumented atomic component $B^r_i$;
\item  a Boolean variable $\mathit{gs}_a$ that holds $\true$ whenever a global state corresponding to interaction $a$ is reconstructed;
\item a tuple $(z_1,\ldots,z_n)$ of Boolean variables initialized to $\false$;
\item an $(n+1)$-tuple $v=(v_1,\ldots, v_n, v_{n+1})$.
\end{itemize}
For each $i\in[1 \upto n]$, $z_i$ is $\true$ when component $i$ is in a busy location and $\false$ otherwise.
For $i\in[1 \upto n]$, $v_i$ is a state of $B_i^r$ and $v_{n+1}\in\gamma$.
$V$ is a sequence of $(n+1)$-tuples initialized to $(q^0_1,\ldots,q^0_n,-)$.
$m$ is an integer variable initialized to $1$.

$P$ is a set of ports.
\begin{itemize}
\item For each atomic component $B^r_i$ for $i\in[1 \upto n]$, RGT has a corresponding port $\beta_i$. States of components are exported to RGT through this port. 
\item For each interaction $a\in\gamma$, RGT has two corresponding ports $p_a$ and $p'_a$.
Port $p_a$ is added to interaction $a$ (later in Definition~\ref{def:composit_transformation}) in order to notify RGT when a new interaction is fired. A reconstructed global state which is related to the execution of interaction $a$, is exported to a runtime monitor through  port $p'_a$.
\end{itemize} 

RGT has three types of transitions:
\begin{itemize}
\item The transitions labeled by port $p_a$, for $a\in\gamma$, are in $T_{\rm{new}}$.
When no reconstructed global state can be delivered (that is, the Boolean variables in $\{\mathit{gs}_a\mid a\in\gamma\}$ are $\false$), the transitions occur when the corresponding interaction $a$ is fired.
\item The transitions labeled by port $\beta_i$, for $i\in [1 \upto n]$, are in $T_{\rm{upd}}$.
When no reconstructed global state can be delivered, to obtain the state of component $B^\bot_i$, these transitions occur at the same time  transition $\beta$ occurs in component $B^\bot_i$.
\item The transition labeled by port $p'_a$ for $a\in\gamma$ are in $T_{\rm{get}}$. If RGT has a reconstructed global state corresponding to the global state of the system after executing interaction $a\in\gamma$, these transitions deliver the reconstructed global state to a runtime monitor.
\end{itemize}

$\RGT$ uses three algorithms.

Algorithm $\mathtt{new}$
(see Algorithm~\ref{alg:NEW}) 
implements the case of function $\Acc$ that corresponds to the occurrence of a new interaction $a\in \gamma$ (Definition~\ref{def:rgt}).
It takes $a\in\gamma$ as input and then: 1) sets $z_i$ to {\true} if component $i$ is involved in interaction $a$, for $i\in[1 \upto n]$; 2) fills the elements of the $(n+1)$-tuple $v$ with the states of components after the execution of the new interaction $a$ in such a way that the $i^{\rm th}$ element of $v$ corresponds to the state of component $B^\bot_i$.
Moreover, the state of busy components is $\mathtt{null}$.
The $(n+1)^{\rm th}$ element of $v$ is dedicated to interaction $a$, as a record specifying that tuple $v$ is related to the execution of $a$; 3) appends $v$ to $V$.
\begin{algorithm}
\caption{$\mathtt{new}(a)$}
\label{alg:NEW}
\begin{algorithmic}[1]
    \For {$i = 1 \to n$}
       	\If {$B_i.P \cap a \neq \emptyset$}
       	  \Comment{Check if component $B_i$ is involved in interaction $a$.}
      	 	\State {$z_i := $ \true}
      	 	\Comment{In case component $B_i$ is busy, $z_i$ is $\true$.}
           	\State {$v_i := $ $\mathtt{null}$}
           	\Comment{The $i^{\rm th}$ element of tuple $v$ is represented by $v_i$.}
        \Else
          	\State {$v_i := B_i^r.X^r$}
          	\Comment{$v_i$ receives the state of $B^r_i$.}
        \EndIf
    \EndFor
    \State {$v_{n+1} := a$}
   			 \Comment{Last element of $v$ receives interaction $a$.}
    \State {$V := V\cdot v$}
              \Comment{$v$ is added to $V$.}
\end{algorithmic}
\end{algorithm}\\
Algorithm $\mathtt{upd}$ (see Algorithm~\ref{alg:UPD})  implements the case of function $\Acc$ which corresponds to the occurrence of transition $\beta$ of atomic component $B^\bot_i$ for $i\in[1 \upto n]$.
According to Definition~\ref{def:inst-atom}, the current state of the instrumented atomic component $B^r_i$ for $i\in[1 \upto n]$ is exported through port $\beta$ of $B^r_i$.
Algorithm $\mathtt{upd}$ takes the current state of $B^r_i$ and looks into each element of $V$ and replaces $\mathtt{null}$ values which correspond to $B^r_i$ with the current state of $B^r_i$.
Finally, algorithm $\mathtt{upd}$ invokes algorithm $\mathtt{check}$ to check the elements of $V$.
If any tuple of $V$, associated to $a \in \gamma$, becomes a global state and has no $\mathtt{null}$ element, then the corresponding Boolean variable $\mathit{gs}_a$ is set to $\true$.
\begin{algorithm}[]
\caption{$\mathtt{upd}(i)$}
\label{alg:UPD}
\begin{algorithmic}[1]
  \State{$z_i := \false$}
    \For {$j = 1 \to \length(V)$}
            \If {$V(j)_i$ == $\mathtt{null}$}
                       	\Comment{ The $i^{\rm th}$ element of the $j^{\rm th}$ tuple in $V$ is represented by $V(j)_i$.}
            	\State {$V(j)_i := B_i^r.X^r$}
            	\Comment{Update the $\mathtt{null}$ states.}
            \EndIf
    \EndFor
\State \Call{$\mathtt{check()}$}{}
        \Comment{Check the elements of sequence $V$ (cf. Algorithm.~\ref{alg:check})}
\end{algorithmic}
\end{algorithm}
\begin{algorithm}[]
\caption{$\mathtt{check()}$}
\label{alg:check}
\begin{algorithmic}[1]
	\For {$i = m \to \length(V)$}
	\Comment{Check those tuples of $V$ which have not been delivered to the monitor.}
		\If {$\neg \mathit{gs}_{(V(i)_{n+1})}$}
			\State {$b_{\rm{tmp}} := \true$}
			\Comment{Make a temporary boolean variable initialized to $\true$.}
			\For {$j = 1 \to n$}     
				\State {$b_{\rm{tmp}} := b_{\rm{tmp}} \wedge (V(i)_j\neq \mathtt{null})$}
				\Comment{{$b_{\rm{tmp}}$ remains $\true$ until a $\mathtt{null}$ is found in the $i^{\rm th}$ tuple of $V$.}}
			\EndFor    
			\State {$\mathit{gs}_{(V(i)_{n+1})} := b_{\rm{tmp}}$} 
   \Comment{Update the value of Boolean $\mathit{gs}$ associated to $V(i)_{n+1}$.}
		\EndIf
	\EndFor
\end{algorithmic}
\end{algorithm}\\
Algorithm $\mathtt{get}$ (see Algorithm~\ref{alg:get}) is called whenever component $\RGT$ has a reconstructed global state to deliver.
Algorithm $\mathtt{get}$ takes the $m^{\rm th}$ tuple in $V$ and copies its values into $\{B^r_i.X^r_c\}_{i=1}^n$ and then increments $m$.
Finally, algorithm $\mathtt{get}$ calls algorithm $\mathtt{check}$ in order to update the value of the Boolean variables $\mathit{gs}_a$ for $a \in \gamma$, because there are possibly several reconstructed global states associated to an interaction $a \in \gamma$.
In this case, after delivering one of those reconstructed global states and resetting $\mathit{gs}_a$ to $\false$, one must again set variable $\mathit{gs}_a$ to $\true$ for the rest of the reconstructed global states associated to interaction $a$.
\begin{algorithm}
\caption{$\mathtt{get()}$}
\label{alg:get}
\begin{algorithmic}[1]
    \For {$i = 1 \to n$}
            \State {$B_i^r.X^r_c := V(m)_i$}
             \Comment{Copy the $m^{\rm th}$ tuple of $V$.}
    \EndFor    
    \State {$\mathit{gs}_{(V(m)_{n+1})} := \false$}
    \Comment{Reset the corresponding $\mathit{gs}_a$ of the $V(m)$.}
    \State {$m := m+1$}
    \Comment{Increment $m$.}
    \State \Call{$\mathtt{check()}$}{}
            \Comment{Check the elements of sequence $V$ (cf. Algorithm.~\ref{alg:check})}
\end{algorithmic}
\end{algorithm}
Note, to facilitate the presentation of proofs in Appendix~\ref{sec:proofs}, component $\RGT$ is defined in such a way that it does not discard the reconstructed global states of the system after delivering them to the monitor.
In our actual implementation of $\RGT$, these states are discarded because they are not useful after being delivered to the monitor.
At runtime, $\RGT.V$ contains the sequence of global states associated with the witness trace (as stated later by Proposition~\ref{proposition_RGT}).
\begin{example}[Component $\RGT$]
\begin{figure}[t]
    \centering
    \Large
   \scalebox{.55} {   
\begin{tikzpicture}[->,>=stealth',shorten >=1pt,auto,node distance=5.5cm, semithick]
  \tikzstyle{every state}=[fill=none,draw=black,text=black]  

  \filldraw[fill=none,draw=black](0,0)--(0,8)--(21.5,8)--(21.5,0)--(0,0)--cycle; 
 \node at (7.8,3) [state] (A)        [minimum size=1.5cm]              {$l$};


  \path [out=-70,in=10] (A) edge [loop]   node[text centered,right,text width=4cm,xshift=-0.3cm] {$p_a$, $\mathtt{new}(p_a)$\\ $(\bigwedge_{a \in \gamma} (\neg \mathit{gs}_a))$\\ for $a\in\gamma$} ();
  \path [out=50,in=130] (A) edge [loop]   node[text centered,above,text width=4cm] {$p'_a$, $\mathtt{get}$ $(\mathit{gs}_a == \true)$\\ for $a\in\gamma$} ();
  \path [out=170,in=250] (A) edge [loop]   node[text centered,left,text width=5.5cm] {$\beta_i$, $\mathtt{upd}(i)$ for $i\in[1 \upto 4]$\\ $(\bigwedge_{a \in \gamma} (\neg \mathit{gs}_a))$} ();
  
  \fill (1,8) circle (0.24cm);\node at (1,8.7) {$p'_{n_t}$};
  \fill (2.5,8) circle (0.24cm);\node at (2.5,8.7) {$p'_{ex_{12}}$};
  \fill (4,8) circle (0.24cm);\node at (4,8.7) {$p'_{r_1}$}; 
  \fill (5.5,8) circle (0.24cm);\node at (5.5,8.7) {$p'_{f_1}$}; 
  \fill (8.5,8) circle (0.24cm);\node at (8.5,8.7) {$p'_{r_2}$};
  \fill (10,8) circle (0.24cm);\node at (10,8.7) {$p'_{f_2}$};
  \fill (13,8) circle (0.24cm);\node at (13,8.7) {$p'_{r_3}$};
  \fill (14.5,8) circle (0.24cm);\node at (14.5,8.7) {$p'_{f_3}$};
  \fill (17.5,8) circle (0.24cm);\node at (17.5,8.7) {$p'_{ex_{23}}$};
  \fill (19,8) circle (0.24cm);\node at (19,8.7) {$p'_{ex_{13}}$};

  \fill (1,0) circle (0.24cm);\node at (1,-0.7) {$p_{n_t}$};
  \fill (2.5,0) circle (0.24cm);\node at (2.5,-0.7) {$p_{ex_{12}}$};
  \fill (4,0) circle (0.24cm);\node at (4,-0.7) {$p_{r_1}$}; 
  \fill (5.5,0) circle (0.24cm);\node at (5.5,-0.7) {$p_{f_1}$}; 
  \fill (7,0) circle (0.24cm);\node at (7,-0.7) {$\beta_1$};
  \fill (8.5,0) circle (0.24cm);\node at (8.5,-0.7) {$p_{r_2}$};
  \fill (10,0) circle (0.24cm);\node at (10,-0.7) {$p_{f_2}$};
  \fill (11.5,0) circle (0.24cm);\node at (11.5,-0.7) {$\beta_2$};
  \fill (13,0) circle (0.24cm);\node at (13,-0.7) {$p_{r_3}$};
  \fill (14.5,0) circle (0.24cm);\node at (14.5,-0.7) {$p_{f_3}$};  
  \fill (16,0) circle (0.24cm);\node at (16,-0.7) {$\beta_3$};
  \fill (17.5,0) circle (0.24cm);\node at (17.5,-0.7) {$p_{ex_{23}}$};
  \fill (19,0) circle (0.24cm);\node at (19,-0.7) {$p_{ex_{13}}$};
  \fill (20.5,0) circle (0.24cm);\node at (20.5,-0.7) {$\beta_4$};

\node at (16,4) {$\gamma=\{\mathit{ex_{12}},\mathit{ex_{13}},\mathit{ex_{23}},\mathit{r_1},\mathit{r_2},\mathit{r_3},\mathit{f_1},\mathit{f_2},\mathit{f_3},\mathit{n_t}\}$};


\end{tikzpicture}}
\caption{Component RGT \redd{for} system Task} 
\label{fig:RGT}
\end{figure}
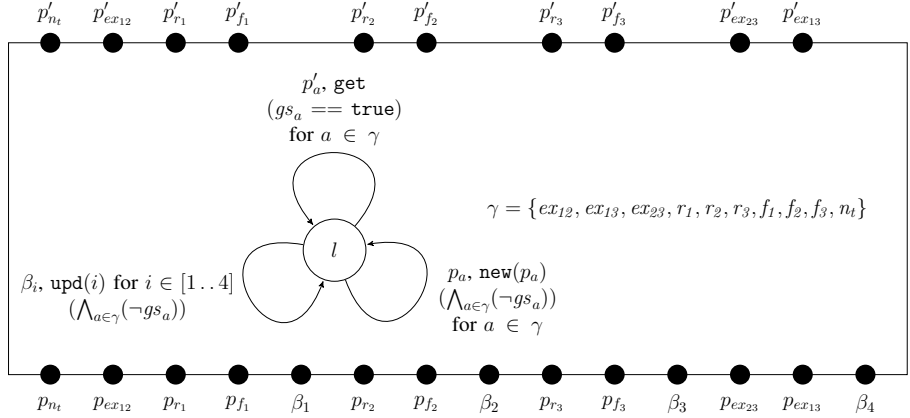
Figure~\ref{fig:RGT} depicts the component $\RGT$ for system Task.
For readability, only one instance of each type of transitions is shown.
The execution of a new interaction $a \in \{\mathit{ex_{12}},\mathit{ex_{13}},\mathit{ex_{23}},\mathit{r_1},\mathit{r_2},\mathit{r_3},\mathit{f_1},\mathit{f_2},$ $\mathit{f_3},\mathit{n_t}\}$ in  system Task is synchronized with the execution of transition $p_a$ of the component $\RGT$ which applies the algorithm $\mathtt{new}$.
Each busy interaction in the system Task is synchronized with the execution of transition $\beta_i$ ($i \in [1 \upto 4]$ are the indexes of the four components in system Task) which applies the algorithm $\mathtt{upd}$ to update the reconstructed states so far and check whether or not a new global state is reconstructed.
Transition $\beta_i$, $i \in [1 \upto 4]$, is guarded by $\bigwedge_{a \in \gamma} (\neg \mathit{gs}_a)$ which ensures the delivery of the new reconstructed global state through the ports $p_{a\in \gamma}$ as soon as they are reconstructed.
At runtime, $\RGT$ produces the sequence of global states in the right-most column of Table~\ref{Table:RGT}.
\end{example}
%
\subsection{Connections}
\label{sec:connections}
%
After building component $\RGT$ (see Definition~\ref{def:RGT-atom}), and instrumenting atomic components (see Definition~\ref{def:inst-atom}), we modify all interactions and define new interactions to build a new transformed composite component. 
To let $\RGT$ accumulate states of the system, first we transform all the existing interactions by adding a new port to communicate with component $\RGT$, then we create new interactions that allow $\RGT$ to deliver the reconstructed global states of the system to a runtime monitor.

Given a composite component $B^\perp=\gamma^\perp(B^\perp_{1},\ldots,B^\perp_{n})$ with corresponding component $\RGT$  and instrumented components $B^r = (P\cup\{\beta^r\},$ $L\cup L^\perp,$ $T^r,$ $X^r)$ such that $B^r=B_i^r\in\{B_1^r, \ldots, B_n^r\}$, we define a new composite component.
\begin{definition}[Composite component transformation]
\label{def:composit_transformation}
For a composite component $B^\perp=\gamma^\perp(B^\perp_{1},\ldots,$ $B^\perp_{n})$, we introduce a corresponding transformed component $B^r=\gamma^r(B_1^r,\ldots,B_n^r,RGT)$ such that $\gamma^r=a^r_\gamma \cup a^r_\beta\cup a^m$ where:
\begin{itemize}
\item $a^r_\gamma$ and $a^r_\beta$ are the sets of transformed interactions such that:

$\forall a \in\gamma^\perp \quant a^r= \begin{dcases}   
a\cup \{RGT.p_a\} &\text{ if } a\in\gamma\\
a\cup \{RGT.\beta_i\}  &\text{ otherwise }(a\in\{\{\beta_i\}\}_{i\in[1 \upto n]})\\
   \end{dcases}$
   
$a^r_\gamma=\{a^r\mid a\in \gamma\}$, $a^r_\beta=\{a^r\mid a\in \{\{\beta_i\}\}_{i\in[1 \upto n]}\}$
\item $a^m$ is a set of new interactions such that:

$a^m=\{a'\mid a \in \gamma\}$ where 
$\forall a\in\gamma \quant  a'=\{RGT.p'_a\}$ is a corresponding unary interaction.
\end{itemize}
\end{definition}
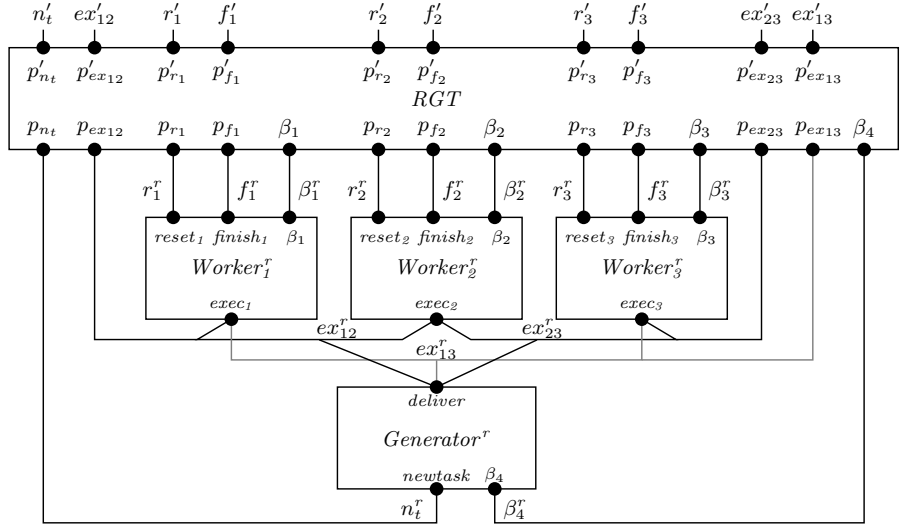
\begin{figure}[t]
    \centering
   \scalebox{0.9} {   
\begin{tikzpicture}[->,>=stealth',shorten >=1pt,auto,node distance=5cm, semithick]
  \tikzstyle{every state}=[fill=none,draw=black,text=black]  
\draw[-] (0.75,2)--(0.25,1.7)--(-2.75,1.7)--(-2.25,2);
\draw[-] (3.75,2)--(4.25,1.7)--(1.25,1.7)--(0.75,2);  
\draw[-,gray] (3.75,2)--(3.75,1.4)--(-2.25,1.4)--(-2.25,2);
\draw[-,gray] (0.75,1)--(0.75,1.43);\node  at (0.75,1.56) {\textbf{$ex^r_{13}$}};
\draw[-] (0.75,1)--(2.25,1.71);\node  at (2.3,1.86) {\textbf{$ex^r_{23}$}};
\draw[-] (0.75,1)--(-1,1.71);  \node  at (-0.7,1.86) {\textbf{$ex^r_{12}$}};
\draw[-] (0.75,-0.5)--(0.75,-1)--(-5,-1)--(-5,4.5);\node  at (0.4,-0.8) {\textbf{$n^r_t$}};
\draw[-] (0.7,3.5)--(0.7,4.5); \node  at (1,3.9) {\textbf{$f^r_2$}};
\draw[-] (-0.1,3.5)--(-0.1,4.5); \node  at (-0.4,3.9) {\textbf{$r^r_2$}};
\draw[-] (3.7,3.5)--(3.7,4.5); \node  at (4,3.9) {\textbf{$f^r_3$}};
\draw[-] (2.9,3.5)--(2.9,4.5); \node  at (2.6,3.9) {\textbf{$r^r_3$}};
\draw[-] (-2.3,3.5)--(-2.3,4.5); \node  at (-2,3.9) {\textbf{$f^r_1$}};
\draw[-] (-3.1,3.5)--(-3.1,4.5); \node  at (-3.4,3.9) {\textbf{$r^r_1$}};
\draw[-] (1.6,3.5)--(1.6,4.5); \node  at (1.9,3.9) {\textbf{$\beta^r_2$}};
\draw[-] (-1.4,3.5)--(-1.4,4.5); \node  at (-1.1,3.9) {\textbf{$\beta^r_1$}};
\draw[-] (4.6,3.5)--(4.6,4.5); \node  at (4.9,3.9) {\textbf{$\beta^r_3$}};
\draw[-] (1.6,-0.5)--(1.6,-1)--(7,-1)--(7,4.5); \node  at (1.9,-0.8) {\textbf{$\beta^r_4$}};
\draw[-] (-2.75,1.7)--(-4.25,1.7)--(-4.25,4.5);
\draw[-] (4.25,1.7)--(5.5,1.7)--(5.5,4.5);
\draw[-,gray] (3.75,1.4)--(6.25,1.4)--(6.25,4.5);
  \filldraw[fill=none,draw=black](-0.7,-0.5)--(-0.7,1)--(2.2,1)--(2.2,-0.5)--(-0.7,-0.5)--cycle; 
  \node at (0.75,0.25)    {$\mathit{Generator^r}$};   
  \fill (0.75,1) circle (0.1cm);  \node at (0.75,0.8) {{\scriptsize $\mathit{deliver}$}};
  \fill (0.75,-0.5) circle (0.1cm); \node at (0.75,-0.3) {{\scriptsize $\mathit{newtask}$}}; 
  \fill (1.6,-0.5) circle (0.1cm); \node at (1.6,-0.3)  {{\scriptsize $\beta_4$}};

  \filldraw[fill=none,draw=black](-0.5,2)--(-0.5,3.5)--(2,3.5)--(2,2)--(-0.5,2)--cycle;  
  \node at (0.75,2.75)    {$\mathit{Worker^r_2}$};
  \fill (0.75,2) circle (0.1cm);  \node at (0.75,2.2) {{\scriptsize $\mathit{exec_2}$}}; 
  \fill (0.7,3.5) circle (0.1cm); \node at (0.9,3.2)  {{\scriptsize $\mathit{finish_2}$}};
  \fill (-0.1,3.5) circle (0.1cm);\node at (0,3.2){{\scriptsize $\mathit{\maintenance_2}$}};
  \fill (1.6,3.5) circle (0.1cm); \node at (1.7,3.2)  {{\scriptsize $\beta_2$}};
  
  \filldraw[fill=none,draw=black,xshift=-3cm](-0.5,2)--(-0.5,3.5)--(2,3.5)--(2,2)--(-0.5,2)--cycle;  
  \node [xshift=-3cm] at (0.75,2.75)    {$\mathit{Worker^r_1}$};
  \fill [xshift=-3cm](0.75,2) circle (0.1cm);
  \fill [xshift=-3cm](0.7,3.5) circle (0.1cm); 
  \fill [xshift=-3cm](-0.1,3.5) circle (0.1cm); 
  \fill [xshift=-3cm](1.6,3.5) circle (0.1cm); 
  
  \node [xshift=-3cm]at (0.75,2.2) {{\scriptsize $\mathit{exec_1}$}}; 
  \node [xshift=-3cm]at (0,3.2)  {{\scriptsize $\mathit{\maintenance_1}$}};
  \node [xshift=-3cm]at (0.9,3.2)  {{\scriptsize $\mathit{finish_1}$}};
  \node [xshift=-3cm]at (1.7,3.2)  {{\scriptsize $\beta_1$}};
  \filldraw[fill=none,draw=black,xshift=3cm](-0.5,2)--(-0.5,3.5)--(2,3.5)--(2,2)--(-0.5,2)--cycle;   
  \node [xshift=3cm] at (0.75,2.75)    {$\mathit{Worker^r_3}$};
  \fill [xshift=3cm](0.75,2) circle (0.1cm);
  \fill [xshift=3cm](0.7,3.5) circle (0.1cm); 
  \fill [xshift=3cm](-0.1,3.5) circle (0.1cm);
  \fill [xshift=3cm](1.6,3.5) circle (0.1cm);   
  
  \node [xshift=3cm]at (0.75,2.2) {{\scriptsize $\mathit{exec_3}$}}; 
  \node [xshift=3cm]at (0,3.2)  {{\scriptsize $\mathit{\maintenance_3}$}};
  \node [xshift=3cm]at (0.9,3.2)  {{\scriptsize $\mathit{finish_3}$}};
  \node [xshift=3cm]at (1.7,3.2)  {{\scriptsize $\beta_3$}};
  \filldraw[fill=none,draw=black](-5.5,4.5)--(-5.5,6)--(7.5,6)--(7.5,4.5)--(-5,4.5)--cycle;  
  \node  at (0.75,5.25)  {$\mathit{RGT}$};
  
  \fill (7,4.5) circle (0.1cm);
  \fill (6.25,4.5) circle (0.1cm);
  \fill (5.5,4.5) circle (0.1cm);
  \fill (4.6,4.5) circle (0.1cm);
  \fill (3.7,4.5) circle (0.1cm);  
  \fill (2.9,4.5) circle (0.1cm);
  \fill (1.6,4.5) circle (0.1cm);  
  \fill (0.7,4.5) circle (0.1cm);
  \fill (-0.1,4.5) circle (0.1cm); 
  \fill (-1.4,4.5) circle (0.1cm);
  \fill (-2.3,4.5) circle (0.1cm);
  \fill (-3.1,4.5) circle (0.1cm);
  \fill (-4.25,4.5) circle (0.1cm);
  \fill (-5,4.5) circle (0.1cm);

  \fill (6.25,6) circle (0.1cm);\draw[-] (6.25,6)--(6.25,6.3); \node at (6.25,6.5) {\textbf{$ex'_{13}$}};  
  \fill (5.5,6) circle (0.1cm);\draw[-] (5.5,6)--(5.5,6.3); \node at (5.5,6.5) {\textbf{$ex'_{23}$}};
  \fill (3.7,6) circle (0.1cm);\draw[-] (3.7,6)--(3.7,6.3); \node at (3.7,6.5) {\textbf{$f'_3$}};
  \fill (2.9,6) circle (0.1cm);\draw[-] (2.9,6)--(2.9,6.3); \node at (2.9,6.5) {\textbf{$r'_3$}};
  \fill (0.7,6) circle (0.1cm);\draw[-] (0.7,6)--(0.7,6.3); \node at (0.7,6.5) {\textbf{$f'_2$}};
  \fill (-0.1,6) circle (0.1cm);\draw[-] (-0.1,6)--(-0.1,6.3); \node at (-0.1,6.5) {\textbf{$r'_2$}};
  \fill (-2.3,6) circle (0.1cm);\draw[-] (-2.3,6)--(-2.3,6.3); \node at (-2.3,6.5) {\textbf{$f'_1$}};
  \fill (-3.1,6) circle (0.1cm);\draw[-] (-3.1,6)--(-3.1,6.3); \node at (-3.1,6.5) {\textbf{$r'_1$}};
  \fill (-4.25,6) circle (0.1cm);\draw[-] (-4.25,6)--(-4.25,6.3); \node at (-4.25,6.5) {\textbf{$ex'_{12}$}};
  \fill (-5,6) circle (0.1cm);\draw[-] (-5,6)--(-5,6.3); \node at (-5,6.5) {\textbf{$n'_t$}};
    
\node at (-5,4.8) {$p_{n_t}$};
\node at (-4.15,4.8) {$p_{ex_{12}}$};
\node at (-3.1,4.8) {$p_{r_1}$}; 
\node at (-2.3,4.8) {$p_{f_1}$}; 
\node at (-1.4,4.8) {$\beta_1$};
\node at (-0.1,4.8) {$p_{r_2}$};
\node at (0.7,4.8) {$p_{f_2}$};
\node at (1.6,4.8) {$\beta_2$};
\node at (2.9,4.8) {$p_{r_3}$};
\node at (3.7,4.8) {$p_{f_3}$};  
\node at (4.6,4.8) {$\beta_3$};
\node at (5.5,4.8) {$p_{ex_{23}}$};
\node at (6.35,4.8) {$p_{ex_{13}}$};
\node at (7,4.8) {$\beta_4$};

\node at (-5,5.65) {$p'_{n_t}$};
\node at (-4.15,5.65) {$p'_{ex_{12}}$};
\node at (-3.1,5.65) {$p'_{r_1}$}; 
\node at (-2.3,5.65) {$p'_{f_1}$}; 
\node at (-0.1,5.65) {$p'_{r_2}$};
\node at (0.7,5.65) {$p'_{f_2}$};
\node at (2.9,5.65) {$p'_{r_3}$};
\node at (3.7,5.65) {$p'_{f_3}$};
\node at (5.5,5.65) {$p'_{ex_{23}}$};
\node at (6.35,5.65) {$p'_{ex_{13}}$};

\end{tikzpicture}}
\caption{Composite component of system Task obtained by applying the transformation in Definition~\ref{def:composit_transformation}} 
\label{fig:running_rgt}
\end{figure}
For each interaction $a\in\gamma^\perp$, we associate a transformed interaction $a^r$ which is the modified version of interaction $a$ such that a corresponding port of component $\RGT$ is added to $a$.
Instrumenting interaction $a\in\gamma$ does not modify sequence of assignment $F_a$, whereas instrumenting busy interactions $a\in \{\{\beta_i\}\}_{i=1}^n$ adds assignments to transfer attached variables of port $\beta_i$ to the component $\RGT$.
The transformed interactions belong to two subsets, $a^r_\gamma$ and $a^r_\beta$.
The set $a^m$ is the set of all unary interactions $a'$ associated to each existing interaction $a\in\gamma$ in the system.

The set of the states of transformed composite component $B^r$ is $Q^r= B^r_1.Q\times \ldots\times B^r_n.Q \times RGT.Q$.
\begin{example}[Transformed composite component]
Figure~\ref{fig:running_rgt} shows the transformed composite component of system Task.
The goal of building $a'$ for each interaction $a$ is to enable $\RGT$ to connect to a runtime monitor.
Upon the reconstruction of a global state corresponding to interaction $a\in\gamma$, the corresponding interaction $a'$ delivers the reconstructed global state to a runtime monitor.  
\end{example}
%
\subsection{Correctness of the Transformations}
\label{sec:correctness}
%
Combined together, the transformations preserve the semantics of the initial model as stated in the rest of this section.

Intuitively, the component $\RGT$ defined in Definition~\ref{def:RGT-atom} implements function $\RGT$ defined in Definition~\ref{def:rgt}. Reconstructed global states can be transferred through the ports $p'_a$ with $a\in \gamma$. 
If interaction $a$ happens before interaction $b$, then in component $\RGT$, port $p'_{a}$ which contains the reconstructed global state after executing $a$ will be enabled before port $p'_{b}$.
In other words, the total order between executed interactions is preserved.

In the transformed composite component $\gamma^r(B_1^r,\ldots,B_n^r,RGT)$, the notion of equivalence is used to relate the tuples constructed by component $\RGT$ to the states of the initial system in partial-state semantics.
Below, we define the notion of equivalence between an $(n+1)$-tuple $v=(v_1,\ldots, v_n, v_{n+1})$ and a state of the system $q=(q_1,\ldots,q_n)$ such that, for $i\in[1 \upto n]$, $v_i$ is a state of $B_i^r$ and $v_{n+1} \in \gamma$.
\begin{definition}[Equivalence of an $(n+1)$-tuple and a state]
An $(n+1)$-tuple $v=(v_1,\ldots, v_n, v_{n+1})$ is equivalent to a state $q=(q_1,\ldots,q_n)$  if:
\[
\forall i\in[1 \upto n] \quant  v_i = \left\{
  \begin{array}{ll}
q_i & \text{if } q_i \in Q_i,\\
\mathtt{null}  & \text{otherwise}.
\end{array} \right.
\]
When an $(n+1)$-tuple $v$ is equivalent to a state $q$, we denote it by $v\cong q$.
\end{definition}
A tuple $(v_1, \ldots, v_n, v_{n+1})$ and a state $(q_1,\ldots,q_n)$ are equivalent if $v_i=q_i$ for each position $i$ where the state $q_i$ of component $B_i^r$ is also a state of the initial model, and $v_i = \mathtt{null}$ otherwise.
The notion of equivalence is extended to traces and sequences of $(n+1)$-tuples.
A trace $\sigma=q'_0.a_1.q'_1\ldots a_k.q'_k$ and a sequence of $(n+1)$-tuples $V=v(0) \cdot v(1) \ldots v(k)$ are equivalent, denoted $\sigma \cong V$, if $q'_j$ is equivalent to $v(j)$ for all $j\in[0 \upto k]$ and $v(j)_{n+1}= a_j$ for all $j\in[1 \upto k]$.
\begin{proposition}[Correctness of component $\RGT$]
\label{proposition_RGT}
$\forall \sigma\in \Tr(B^\perp) \quant \RGT.V \cong \Acc(\sigma)$.
\end{proposition}
Proposition~\ref{proposition_RGT} states that, for any trace $\sigma$, at any time, variable $\RGT.V$ encodes the witness trace $\Acc(\sigma)$ of the current trace: $\RGT.V$ is a sequence of tuples where each tuple consists of the state and the interaction that led to this state, in the same order as they appear on the witness trace.
\begin{proof}
The proof is done by induction on the length of $\sigma \in \Tr(B^\perp)$, i.e., the trace of the system in partial-state semantics. The proof is given in Appendix~\ref{proof:RGT_correcness} (p.~\pageref{proof:RGT_correcness}).
\end{proof}
For each trace resulting from an execution with partial-state semantics, component $\RGT$ produces a trace of global states which is the witness of this trace in the initial model.
\begin{definition}[State stability]
\label{def:stability}
State $(l, v) \in RGT.Q$ is said to be \emph{stable} when $\forall x \in \{\RGT.\mathit{gs}_a\mid a \in \gamma\}  \quant  v(x) = \false$.
\end{definition}
A state $q$ in the the semantics of atomic component $\RGT$ is said to be stable when all Boolean variables in set $\{\RGT.\mathit{gs}_a\mid a \in \gamma\}$ evaluate to $\false$ with the valuation of variables in state $q$.
In other words, the current state of component $\RGT$ is stable when it has no reconstructed global states to deliver.
We say that the composite component $B^r$ is stable when the state of its associated component $\RGT$ is stable.
\begin{example}[Stable state]
We illustrate Definition~\ref{def:stability} based on the execution trace in Table~\ref{Table:RGT}.
By the evolution of system Task from step $4$ to step $5$, component $\RGT$ reconstructs the global state associated to the execution of $\mathit{ex}_{12}$ and respectively sets boolean variable $\mathit{gs}_{\mathit{ex}_{12}}$ to \true.
Once $\mathit{gs}_{\mathit{ex}_{12}}$ becomes  \true , we say that the state of the component $\RGT$ is not stable.
In component $\RGT$, the execution of transition labeled by port $p'_{\mathit{ex}_{12}}$ delivers the reconstructed global state (\ie $(\done,\done,\free,\delivered)$) to the monitor and sets boolean variable $\mathit{gs}_{\mathit{ex}_{12}}$ to \false.
Consequently, component $\RGT$ becomes stable.
We say that component $\RGT$ is not stable whenever there exists at least one reconstructed global state which has not been delivered to the monitor.
Whenever component $\RGT$ is not stable, we say that the system is not stable as well. 
\end{example}
The following lemma states a property of the algorithms in \secref{sec:rgt-atom} ensuring that whenever component $\RGT$ has reconstructed some global states, it transmits them to the monitor before the system can execute any new partial state can be created.
\begin{lemma}
\label{lemma:reset_befor_set}
In any state of the transformed system, if there is a non-empty set $\mathit{GS} \subseteq  \{\RGT.\mathit{gs}_a \mid a \in \gamma\}$ in which all variables are $\true$, the variables in $\{ \RGT.\mathit{gs}_a \mid a \in \gamma \} \setminus \mathit{GS}$ cannot be set to $\true$ until all variables in $\mathit{GS}$ are reset to $\false$ first.
\end{lemma}
The following lemma states that any state of the composite component $B^r$ can be stabilized by executing interactions in $a^m$.
\begin{lemma}
\label{lemma:stable-state}
We shall prove that for any state $q \in Q^r$, there exists a state $q'\in Q^r$ reached after interactions in $a^m$ (i.e., $q \xrightarrow{({a^m})^*} q'$), such that $q'$ is a stable state (i.e., $\stable(q')$).
\end{lemma}
We define a notion of equivalence between states of the transformed model and states of the initial system.
\begin{definition}[Equivalent states]
\label{Equivalent states}
Let $q^r=(q^r_1,\cdots,q^r_n,q^r_{n+1})\in Q^r$ be a state in the transformed model where $q^r_{n+1}$ is the state of component $\RGT$, function $\equ: Q^r \longrightarrow Q^\perp$ is defined as follows: $\equ(q^r)=q$, where  $q=(q_1,\cdots,q_n)$, $(\forall i\in[1 \upto n]  \quant  q^r_i=q_i) \wedge \stable(q^r_{n+1})$.
\end{definition}
A state in the initial model is said to be equivalent to a state in the transformed model if the state of each component in the initial model is equal to the state of the corresponding component in transformed model and the state of component $\RGT$ is stable.

The following lemma is a direct consequence of Definition~\ref{Equivalent states}.
The lemma states that, if an interaction is enabled in the transformed model, then the corresponding interaction is enabled in the initial model when the states of two models are equivalent. 
\begin{lemma}
\label{lemma:enabled-interactions}
For any two equivalent states $q \in Q^\bot$ and $q^r \in Q^r$ (i.e., $\equ(q^r)=q$), if interaction $ a \in \gamma^\perp $ is enabled in state $q$, then $a^r \in \gamma^r$ is enabled at state $q^r$.
\end{lemma}
Based on the above lemmas, we can now state the correctness of our transformations.
\begin{theorem}[Transformation Correctness ]
\label{theorem_Correctness}
$\gamma^\perp(B^\perp_{1},\ldots,B^\perp_{n})\sim \gamma^r(B_1^r,\ldots,B_n^r,RGT)$.
\end{theorem}
Theorem~\ref{theorem_Correctness} states that the initial model and the transformed model are observationally equivalent.
\begin{proof}
The proof relies on exhibiting a bi-simulation relation between the set of states of $B^r= \gamma^r(B_1^r,\ldots,B_n^r,RGT)$, that is $Q^r$, and the set of states of $B= \gamma^\perp(B^\perp_{1},\ldots,B^\perp_{n})$, that is $Q^\perp$. The proof is given in Appendix~\ref{proof:transfomation_correcness} (p.~\pageref{proof:transfomation_correcness}).
\end{proof}

Combined together, Theorem~\ref{theorem_Correctness} and Lemma~\ref{lemma:enabled-interactions} imply that, for each state in the initial system, there exists an equivalent state in the transformed system in which all enabled interactions in the initial system are also enabled in the transformed system.
Hence, we can conclude that the transformed system is as concurrent as the initial system.

Consequently, we can substantiate our claims stated in the introduction about the transformations: instrumenting atomic components and adding component $\RGT$ (i) preserves the semantics and concurrency of the initial model, and (ii) verdicts are sound and complete. 
\begin{remark}[Alternative $\RGT$ atoms]
In the definition of atom $\RGT$ (Definition~\ref{def:RGT-atom}), one can observe  that whenever component $\RGT$ has reconstructed global states to deliver, the system cannot proceed and must wait until all the reconstructed global states are sent (because of the guards of transitions $T_{\rm{upd}}$ and $T_{\rm{new}}$).
This gives precedence to monitoring rather than to the evolution of the system.   

Three alternative definitions of $\RGT$ can be considered by changing the guards of the transitions in $T_{\rm{new}}$ and $T_{\rm{upd}}$.
For both transitions, by suppressing the guards, one gives less precedence to the transmission of reconstructed global states.
By suppressing the guards in transitions in $T_{\rm{new}}$, we let the system starting a new interaction while there may be still some reconstructed global states for $\RGT$ to deliver.
By suppressing the guards in transitions in $T_{\rm{upd}}$, we let the system  execute $\beta$-transitions while there may be still some reconstructed global states for $\RGT$ to deliver.

Suppressing these guards favors the performance of the system but may delay the transmission of global states to the monitor and thus it may also delay the emission of verdicts.
There is thus a tradeoff between the performance of the system and the emission of verdicts. 
\end{remark}
%
\subsection{Monitoring}
%
As it is shown in \mfigref{fig:abstract}, one can reuse the results in~\cite{FalconeJNBB15} to monitor a system with partial-state semantics.
One just has to transform this system with the previous transformations and plug a monitor for a property on the global-states of the system to component $\RGT$ through the dedicated ports.
At runtime, such monitor will (i) receive the sequence of reconstructed global states corresponding to the witness trace, (ii) preserve the concurrency of the system, and iii) state verdicts on the witness trace.
\begin{figure}
    \centering
   \scalebox{0.9} {  
    \begin{subfigure}[b]{0.42\textwidth}   
    \centering
\begin{tikzpicture}[->,>=stealth',shorten >=1pt,auto,node distance=5cm, semithick]
  \tikzstyle{every state}=[fill=none,draw=black,text=black]  

\filldraw[fill=none,draw=black](0,0)--(0,2)--(1,2)--(1,0)--cycle;
\filldraw[fill=none,draw=black,xshift=3cm](0,0)--(0,2)--(1,2)--(1,0)--cycle;
\node at (0.5,1) [rotate=90,text width=2cm,text centered]  {{\footnotesize \redd{Single-Threaded}} CBS};
\node at (3.5,1) [rotate=90,text width=2cm,text centered]  {Monitor \cite{FalconeJNBB15}};
\draw[-] (1,0.25)--(3,0.25);\fill (1,0.25) circle (0.08cm);\fill (3,0.25) circle (0.08cm);
\draw[-] (1,1.25)--(3,1.25);\fill (1,1.25) circle (0.08cm);\fill (3,1.25) circle (0.08cm);
\draw[-] (1,1.5)--(3,1.5);\fill (1,1.5) circle (0.08cm);\fill (3,1.5) circle (0.08cm);
\draw[-] (1,1.75)--(3,1.75);\fill (1,1.75) circle (0.08cm);\fill (3,1.75) circle (0.08cm); \node at (2,2) [text centered]  {$\forall a\in \gamma$};
\fill (2,0.5) circle (0.05cm);
\fill (2,0.75) circle (0.05cm);
\fill (2,1) circle (0.05cm);
\end{tikzpicture}
\caption{Monitoring of single-threaded CBSs}
 \end{subfigure}
     \begin{subfigure}[b]{0.50\textwidth}   
    \centering
\begin{tikzpicture}[->,>=stealth',shorten >=1pt,auto,node distance=5cm, semithick]
  \tikzstyle{every state}=[fill=none,draw=black,text=black]  

\filldraw[fill=none,draw=black](0,0)--(0,2)--(1,2)--(1,0)--cycle;
\filldraw[fill=none,draw=black,xshift=3cm](0,0)--(0,2)--(1,2)--(1,0)--cycle;
\filldraw[fill=none,draw=black,xshift=6cm](0,0)--(0,2)--(1,2)--(1,0)--cycle;
\node at (0.5,1) [rotate=90,text width=2cm,text centered]  {{\footnotesize \redd{Multi-Threaded}} CBS};
\node at (3.5,1) [rotate=90,text width=2cm,text centered]  {$\RGT$};
\node at (6.5,1) [rotate=90,text width=2cm,text centered]  {Monitor \cite{FalconeJNBB15}};
\draw[-] (1,0.25)--(3,0.25);\fill (1,0.25) circle (0.08cm);\fill (3,0.25) circle (0.08cm);
\draw[-] (1,1.25)--(3,1.25);\fill (1,1.25) circle (0.08cm);\fill (3,1.25) circle (0.08cm);
\draw[-] (1,1.5)--(3,1.5);\fill (1,1.5) circle (0.08cm);\fill (3,1.5) circle (0.08cm);
\draw[-] (1,1.75)--(3,1.75);\fill (1,1.75) circle (0.08cm);\fill (3,1.75) circle (0.08cm); \node at (2,2) [text centered]  {$\forall a\in \gamma^\bot$};
\fill (2,0.5) circle (0.05cm);
\fill (2,0.75) circle (0.05cm);
\fill (2,1) circle (0.05cm);
\draw[-,xshift=3cm] (1,0.25)--(3,0.25);\fill [xshift=3cm](1,0.25) circle (0.08cm);\fill[xshift=3cm] (3,0.25) circle (0.08cm);
\draw[-,xshift=3cm] (1,1.25)--(3,1.25);\fill [xshift=3cm](1,1.25) circle (0.08cm);\fill[xshift=3cm] (3,1.25) circle (0.08cm);
\draw[-,xshift=3cm] (1,1.5)--(3,1.5);\fill[xshift=3cm] (1,1.5) circle (0.08cm);\fill[xshift=3cm] (3,1.5) circle (0.08cm);
\draw[-,xshift=3cm] (1,1.75)--(3,1.75);\fill[xshift=3cm] (1,1.75) circle (0.08cm);\fill[xshift=3cm] (3,1.75) circle (0.08cm); \node at (2,2) [text centered,xshift=3cm]  {$\forall a\in \gamma$};
\fill [xshift=3cm](2,0.5) circle (0.05cm);
\fill [xshift=3cm](2,0.75) circle (0.05cm);
\fill [xshift=3cm](2,1) circle (0.05cm);
\end{tikzpicture}
\caption{Monitoring of multi-threaded CBSs}
 \end{subfigure}
}
\caption{Abstract view of runtime monitoring of single-threaded vs. multi-threaded CBSs} 
\label{fig:abstract}
\end{figure}
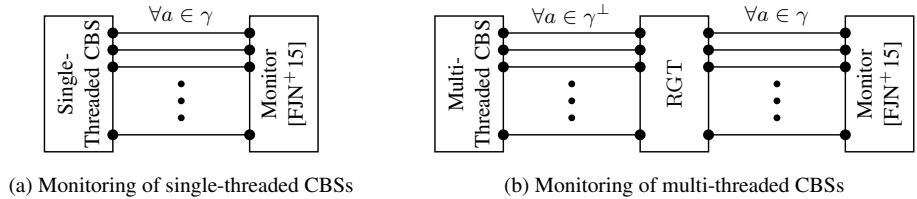
\begin{example}[Monitoring system Task]
Figure~\ref{fig:running_monitor} depicts the transformed system Task with a monitor (for the homogeneous distribution of the tasks among the workers) where $e_1$, $e_2$, and $e_3$ are events related to the pairwise comparison of the number of executed tasks by $\worker$s.
For $i \in [1 \upto 3]$, event $e_i$ evaluates to true whenever $| x_{{(i\ \mathrm{mod}\ 3}) + 1} - x_{i}|$ is lower than 3 (for this example).
Component $\monitor$ evaluates $(e_1 \wedge e_2 \wedge e_3)$ upon the reception of a new global state from $\RGT$ and emits the associated verdict till reaching bad state $\bot$.
The global trace $(\free,\free,\free,\ready)\cdot \ex_{12} \cdot (\done,\done,\free,\delivered)\cdot n_t$ (see Table~\ref{Table:RGT}) is sent by component $\RGT$ to the monitor which in turn produces the sequence of verdicts $\top_c \cdot \top_c$ (where $\top_c$ is verdict ``currently good", see~\cite{BauerLS10,FalconeFM12}).
\begin{figure}[t]
    \centering
   \scalebox{0.8} {   
\begin{tikzpicture}[->,>=stealth',shorten >=1pt,auto,node distance=5cm, semithick]
  \tikzstyle{every state}=[fill=none,draw=black,text=black]  
      \draw[
        -triangle 90,
        line width=2mm,
        color=yellow!80,
        postaction={ draw, line width=10cm, shorten >=1cm, -}](1.9,8.5)--(1.9,9.5);
    
    \draw[-,line width=1.5em,yellow!80] (-5.3,8.5)--(7.3,8.5);
    \draw[-,line width=2.5em,yellow!80] (1.9,8.4)--(1.9,9);
    \node [blue] at (1.9,8.45) {\textbf{$\RGT.V$}}; 
\draw[-] (0.75,2)--(0.25,1.7)--(-2.75,1.7)--(-2.25,2);
\draw[-] (3.75,2)--(4.25,1.7)--(1.25,1.7)--(0.75,2);  
\draw[-,gray] (3.75,2)--(3.75,1.4)--(-2.25,1.4)--(-2.25,2);
\draw[-,gray] (0.75,1)--(0.75,1.43);\node  at (0.75,1.56) {\textbf{$ex^r_{13}$}};
\draw[-] (0.75,1)--(2.25,1.71);\node  at (2.3,1.86) {\textbf{$ex^r_{23}$}};
\draw[-] (0.75,1)--(-1,1.71);  \node  at (-0.7,1.86) {\textbf{$ex^r_{12}$}};
\draw[-] (0.75,-0.5)--(0.75,-1)--(-5,-1)--(-5,4.5);\node  at (0.4,-0.8) {\textbf{$n^r_t$}};
\draw[-] (0.7,3.5)--(0.7,4.5); \node  at (1,3.9) {\textbf{$f^r_2$}};
\draw[-] (-0.1,3.5)--(-0.1,4.5); \node  at (-0.4,3.9) {\textbf{$r^r_2$}};
\draw[-] (3.7,3.5)--(3.7,4.5); \node  at (4,3.9) {\textbf{$f^r_3$}};
\draw[-] (2.9,3.5)--(2.9,4.5); \node  at (2.6,3.9) {\textbf{$r^r_3$}};
\draw[-] (-2.3,3.5)--(-2.3,4.5); \node  at (-2,3.9) {\textbf{$f^r_1$}};
\draw[-] (-3.1,3.5)--(-3.1,4.5); \node  at (-3.4,3.9) {\textbf{$r^r_1$}};
\draw[-] (1.6,3.5)--(1.6,4.5); \node  at (1.9,3.9) {\textbf{$\beta^r_2$}};
\draw[-] (-1.4,3.5)--(-1.4,4.5); \node  at (-1.1,3.9) {\textbf{$\beta^r_1$}};
\draw[-] (4.6,3.5)--(4.6,4.5); \node  at (4.9,3.9) {\textbf{$\beta^r_3$}};
\draw[-] (1.6,-0.5)--(1.6,-1)--(7,-1)--(7,4.5); \node  at (1.9,-0.8) {\textbf{$\beta^r_4$}};
\draw[-] (-2.75,1.7)--(-4.25,1.7)--(-4.25,4.5);
\draw[-] (4.25,1.7)--(5.5,1.7)--(5.5,4.5);
\draw[-,gray] (3.75,1.4)--(6.25,1.4)--(6.25,4.5);
  \filldraw[fill=none,draw=black](-0.7,-0.5)--(-0.7,1)--(2.2,1)--(2.2,-0.5)--(-0.7,-0.5)--cycle; 
  \node at (0.75,0.25)   {$\mathit{Generator^r}$};   
  \fill (0.75,1) circle (0.1cm);  \node at (0.75,0.8) {{\scriptsize $\mathit{deliver}$}};
  \fill (0.75,-0.5) circle (0.1cm); \node at (0.75,-0.3) {{\scriptsize $\mathit{newtask}$}}; 
  \fill (1.6,-0.5) circle (0.1cm); \node at (1.6,-0.3)  {{\scriptsize $\beta_4$}}; 
  \filldraw[fill=none,draw=black](-0.5,2)--(-0.5,3.5)--(2,3.5)--(2,2)--(-0.5,2)--cycle;  
  \node at (0.75,2.75)   {$\mathit{Worker^r_2}$};
  \fill (0.75,2) circle (0.1cm);  \node at (0.75,2.2) {{\scriptsize $\mathit{exec_2}$}}; 
  \fill (0.7,3.5) circle (0.1cm); \node at (0.9,3.2)  {{\scriptsize $\mathit{finish_2}$}};
  \fill (-0.1,3.5) circle (0.1cm);\node at (0,3.2){{\scriptsize $\mathit{\maintenance_2}$}};
  \fill (1.6,3.5) circle (0.1cm); \node at (1.7,3.2)  {{\scriptsize $\beta_2$}};  
  \filldraw[fill=none,draw=black,xshift=-3cm](-0.5,2)--(-0.5,3.5)--(2,3.5)--(2,2)--(-0.5,2)--cycle;  
  \node [xshift=-3cm] at (0.75,2.75)   {$\mathit{Worker^r_1}$};
  \fill [xshift=-3cm](0.75,2) circle (0.1cm);
  \fill [xshift=-3cm](0.7,3.5) circle (0.1cm); 
  \fill [xshift=-3cm](-0.1,3.5) circle (0.1cm); 
  \fill [xshift=-3cm](1.6,3.5) circle (0.1cm); 
  
  \node [xshift=-3cm]at (0.75,2.2) {{\scriptsize $\mathit{exec_1}$}}; 
  \node [xshift=-3cm]at (0,3.2)  {{\scriptsize $\mathit{\maintenance_1}$}};
  \node [xshift=-3cm]at (0.9,3.2)  {{\scriptsize $\mathit{finish_1}$}};
  \node [xshift=-3cm]at (1.7,3.2)  {{\scriptsize $\beta_1$}};
  \filldraw[fill=none,draw=black,xshift=3cm](-0.5,2)--(-0.5,3.5)--(2,3.5)--(2,2)--(-0.5,2)--cycle;   
  \node [xshift=3cm] at (0.75,2.75)   {$\mathit{Worker^r_3}$};
  \fill [xshift=3cm](0.75,2) circle (0.1cm);
  \fill [xshift=3cm](0.7,3.5) circle (0.1cm); 
  \fill [xshift=3cm](-0.1,3.5) circle (0.1cm);
  \fill [xshift=3cm](1.6,3.5) circle (0.1cm);   
  
  \node [xshift=3cm]at (0.75,2.2) {{\scriptsize $\mathit{exec_3}$}}; 
  \node [xshift=3cm]at (0,3.2)  {{\scriptsize $\mathit{\maintenance_3}$}};
  \node [xshift=3cm]at (0.9,3.2)  {{\scriptsize $\mathit{finish_3}$}};
  \node [xshift=3cm]at (1.7,3.2)  {{\scriptsize $\beta_3$}};
 \filldraw[fill=none,draw=black](-5.5,4.5)--(-5.5,8)--(7.5,8)--(7.5,4.5)--(-5,4.5)--cycle;  
  \node  at (5,6.75)   {$\mathit{RGT}$};
  
  \fill (7,4.5) circle (0.1cm);
  \fill (6.25,4.5) circle (0.1cm);
  \fill (5.5,4.5) circle (0.1cm);
  \fill (4.6,4.5) circle (0.1cm);
  \fill (3.7,4.5) circle (0.1cm);  
  \fill (2.9,4.5) circle (0.1cm);
  \fill (1.6,4.5) circle (0.1cm);  
  \fill (0.7,4.5) circle (0.1cm);
  \fill (-0.1,4.5) circle (0.1cm); 
  \fill (-1.4,4.5) circle (0.1cm);
  \fill (-2.3,4.5) circle (0.1cm);
  \fill (-3.1,4.5) circle (0.1cm);
  \fill (-4.25,4.5) circle (0.1cm);
  \fill (-5,4.5) circle (0.1cm);

  \fill (6.25,8) circle (0.1cm);\draw[-] (6.25,8)--(6.25,9.5); \node at (6.7,8.5) {\textbf{$ex'_{13}$}};  
  \fill (5.5,8) circle (0.1cm);\draw[-] (5.5,8)--(5.5,9.5); \node at (5.1,8.5) {\textbf{$ex'_{23}$}};
  \fill (3.7,8) circle (0.1cm);\draw[-] (3.7,8)--(3.7,9.5); \node at (3.95,8.5) {\textbf{$f'_3$}};
  \fill (2.9,8) circle (0.1cm);\draw[-] (2.9,8)--(2.9,9.5); \node at (3.15,8.5) {\textbf{$r'_3$}};
  \fill (0.7,8) circle (0.1cm);\draw[-] (0.7,8)--(0.7,9.5); \node at (0.45,8.5) {\textbf{$f'_2$}};
  \fill (-0.1,8) circle (0.1cm);\draw[-] (-0.1,8)--(-0.1,9.5); \node at (-0.35,8.5) {\textbf{$r'_2$}};
  \fill (-2.3,8) circle (0.1cm);\draw[-] (-2.3,8)--(-2.3,9.5); \node at (-2.1,8.5) {\textbf{$f'_1$}};
  \fill (-3.1,8) circle (0.1cm);\draw[-] (-3.1,8)--(-3.1,9.5); \node at (-2.9,8.5) {\textbf{$r'_1$}};
  \fill (-4.25,8) circle (0.1cm);\draw[-] (-4.25,8)--(-4.25,9.5); \node at (-3.85,8.5) {\textbf{$ex'_{12}$}};
  \fill (-5,8) circle (0.1cm);\draw[-] (-5,8)--(-5,9.5); \node at (-4.75,8.5) {\textbf{$n'_t$}};
     
\node at (-5,4.8) {$p_{n_t}$};
\node at (-4.15,4.8) {$p_{ex_{12}}$};
\node at (-3.1,4.8) {$p_{r_1}$}; 
\node at (-2.3,4.8) {$p_{f_1}$}; 
\node at (-1.4,4.8) {$\beta_1$};
\node at (-0.1,4.8) {$p_{r_2}$};
\node at (0.7,4.8) {$p_{f_2}$};
\node at (1.6,4.8) {$\beta_2$};
\node at (2.9,4.8) {$p_{r_3}$};
\node at (3.7,4.8) {$p_{f_3}$};  
\node at (4.6,4.8) {$\beta_3$};
\node at (5.5,4.8) {$p_{ex_{23}}$};
\node at (6.35,4.8) {$p_{ex_{13}}$};
\node at (7,4.8) {$\beta_4$};  
      
\node at (-5,7.65) {$p'_{n_t}$};
\node at (-4.15,7.65) {$p'_{ex_{12}}$};
\node at (-3.1,7.65) {$p'_{r_1}$}; 
\node at (-2.3,7.65) {$p'_{f_1}$}; 
\node at (-0.1,7.65) {$p'_{r_2}$};
\node at (0.7,7.65) {$p'_{f_2}$};
\node at (2.9,7.65) {$p'_{r_3}$};
\node at (3.7,7.65) {$p'_{f_3}$};
\node at (5.5,7.65) {$p'_{ex_{23}}$};
\node at (6.35,7.65) {$p'_{ex_{13}}$};

 \node at (0,6.25) [state] (A)        [minimum size=0.4cm]              {$l$};
 \path [out=-100,in=-20] (A) edge [loop]   node[right,text width=4cm,xshift=0.2cm,yshift=0cm] {$p_a$, $\mathtt{new}(p_a)$\\ $(\bigwedge_{a \in \gamma} (\neg \mathit{gs}_a))$} ();
 \path [out=20,in=100] (A) edge [loop]   node[text centered,right,text width=3cm,xshift=-0.2cm,yshift=-0.3cm] {$p'_a$, $\mathtt{get}$ $(\mathit{gs}_a == \true)$} ();
 \path [out=140,in=220] (A) edge [loop]   node[text centered,left,text width=5cm,xshift=1.2cm] {$\beta_i$, $\mathtt{upd}(i)$ \\ for $i\in[1 \upto 4]$\\ $(\bigwedge_{a \in \gamma} (\neg \mathit{gs}_a))$} ();
 \node at (5.15,6) {$a \in \gamma=\{\mathit{ex_{12}},\mathit{ex_{13}},\mathit{ex_{23}},$};   
 \node at (5.15,5.5) {$\mathit{r_1},\mathit{r_2},\mathit{r_3},\mathit{f_1},\mathit{f_2},\mathit{f_3},\mathit{n_t}\}$}; 
  \filldraw[fill=none,draw=black,yshift=5cm](-5.5,4.5)--(-5.5,7.5)--(7.5,7.5)--(7.5,4.5)--(-5,4.5)--cycle;
  \node  at (3.4,12.2)   {$\mathit{Monitor}$};   
  
  \fill (6.25,9.5) circle (0.1cm);
  \fill (5.5,9.5) circle (0.1cm);
  \fill (3.7,9.5) circle (0.1cm);  
  \fill (2.9,9.5) circle (0.1cm); 
  \fill (0.7,9.5) circle (0.1cm);
  \fill (-0.1,9.5) circle (0.1cm); 
  \fill (-2.3,9.5) circle (0.1cm);
  \fill (-3.1,9.5) circle (0.1cm);
  \fill (-4.25,9.5) circle (0.1cm);
  \fill (-5,9.5) circle (0.1cm);
  \fill (-4.25,12.5) circle (0.1cm);
  
  \node at (-5,9.8) {$p_{n_t}$};
  \node at (-4.15,9.8) {$p_{ex_{12}}$};
  \node at (-3.1,9.8) {$p_{r_1}$}; 
  \node at (-2.3,9.8) {$p_{f_1}$}; 
  \node at (-0.1,9.8) {$p_{r_2}$};
  \node at (0.7,9.8) {$p_{f_2}$};
  \node at (2.9,9.8) {$p_{r_3}$};
  \node at (3.7,9.8) {$p_{f_3}$};  
  \node at (5.5,9.8) {$p_{ex_{23}}$};
  \node at (6.35,9.8) {$p_{ex_{13}}$}; 
  \node at (-4.15,12.8) {$p_{\mathit{intern}}$};
    
  \node[state,initial,minimum size=0.4cm] at (-3.7,11)(A)  {$\top_c$};
  \node[state]         (B) [right of=A,node distance=5cm,minimum size=0.4cm] {$~~~~~$};
  \node[state]         (C) [right of=B,node distance=4.5cm,minimum size=0.4cm] {$\bot$};
 
  \path (A) edge  [bend left]    node[text centered,above,text width=4cm,yshift=-0.5cm] {{\footnotesize $p_a$, $\mathtt{calculate}\ e_1, e_2, e_3$\\ for $a\in\gamma$}} (B);
  \path (B) edge            node[text centered,below,text width=6.4cm] {{\footnotesize $p_{\mathit{intern}}$, $[$print "currently good"$]$\\ $(e_1\wedge e_2\wedge e_3)$ }} (A);
  
  \path (B) edge   node[text centered,above,text width=4.2cm] {$p_{\mathit{intern}}$, $[$print "bad"$]$ $(\neg e_1\vee \neg e_2\vee \neg e_3)$ } (C);
  
  \path (C) edge [loop above] node[text centered,above right,xshift=-0.4cm] {$p_a$,  for $a\in\gamma$ } ();

\end{tikzpicture}}
\caption{Monitored version of system Task} 
\label{fig:running_monitor}
\end{figure}
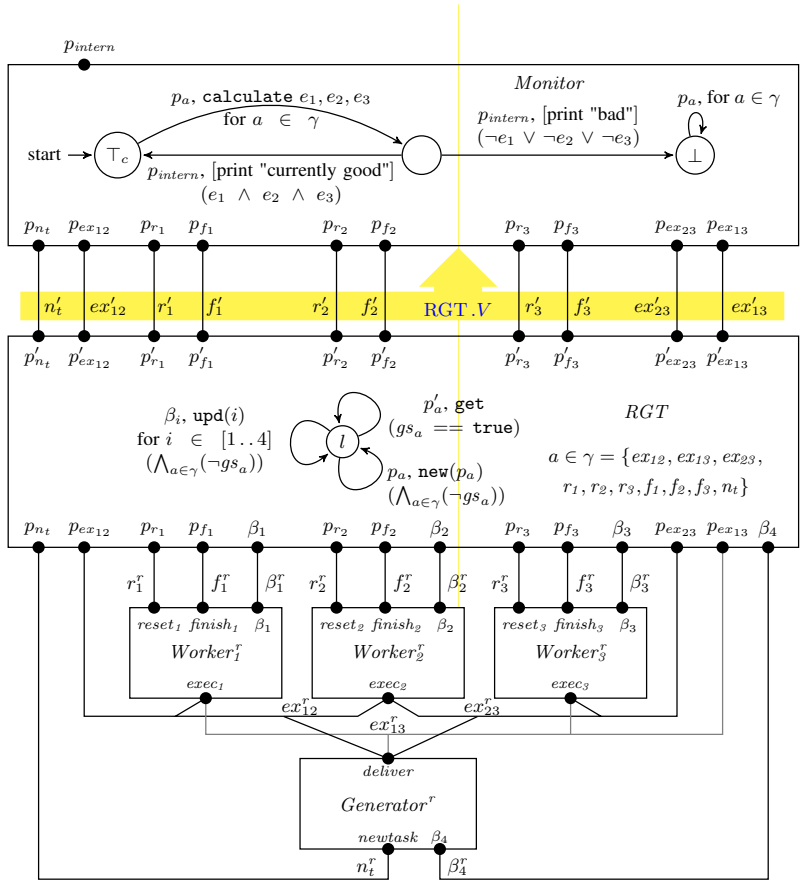 
\end{example}

%
\section{Implementation and Performance Evaluation}
\label{sec:implem}
%
We present an evaluation of our monitoring approach implemented in a tool called RVMT-BIP.
RVMT-BIP is a prototype tool implementing the algorithms presented in \secref{sec:inst}.

This section is organized as follows.
In \secref{subsec:archi}, we present the architecture of RVMT-BIP.
In \secref{subsec:case_study}, we present the systems and properties used in our case studies.
We experiment with RVMT-BIP on four systems where each system is monitored against dedicated properties.
In \secref{subsec:evaluation}, we present the evaluation principles.
In \secref{subsec:results}, we present the experimental results and discuss the performance of RVMT-BIP.
%
\subsection{Architecture of RVMT-BIP}
\label{subsec:archi}
%
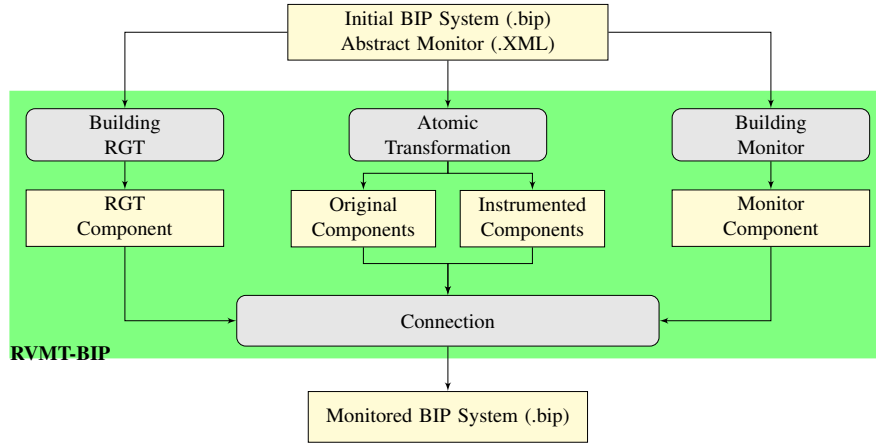
\begin{figure}
    \centering
   \scalebox{.85} {             
\tikzstyle{block} = [rectangle, draw, fill=gray!20, 
    text width=5em, text centered, rounded corners, minimum height=2.5em]   
\tikzstyle{block2} = [rectangle, draw, fill=yellow!20, 
    text width=5em, text centered, minimum height=2.5em]           
\tikzstyle{line} = [draw, -latex']  
\def\myarm{1cm}
\def\myangle{0}
\tikzset{
  arm/.default=1cm,
  arm/.code={\def\myarm{#1}},
  angle/.default=0,
  angle/.code={\def\myangle{#1}} 
}   
\tikzset{
    myncbar/.style = {to path={
        let
            \p1=($(\tikztotarget)+(\myangle:\myarm)$)
        in
            -- ++(\myangle:\myarm) coordinate (tmp)
            -- ($(\tikztotarget)!(tmp)!(\p1)$)
            -- (\tikztotarget)\tikztonodes
    }}
} 
\begin{tikzpicture}[node distance = 1.5cm, auto]   
\node [block2, text width=15em]  (BIP) {Initial BIP System (.bip)\\ Abstract Monitor (.XML)};
\node [block, below of=BIP,text width=9em,node distance = 1.6cm](AT){Atomic \\ Transformation};
\node [block2, below right of=AT,text width=6.4em,node distance = 1.85cm](AT2){Instrumented \\ Components};
\node [block2, below left of=AT,text width=6.4em,node distance = 1.85cm](AT3){Original \\ Components};
\node [block, left of=AT,text width=9em,node distance = 5cm](RGT){Building \\ RGT}; 
\node [block2, below of=RGT,text width=9em,node distance = 1.3cm](RGT2){RGT \\ Component};
\node [block, right of=AT,text width=9em,node distance = 5cm](M){Building \\ Monitor}; 
\node [block2, below of=M,text width=9em,node distance = 1.3cm](M2){Monitor \\ Component};
\node [block,below of=BIP,text width=20em,node distance = 4.5cm](C){Connection}; 
\node [block2,below of=BIP,text width=13em,node distance = 6cm](O){Monitored BIP System (.bip)};
     \path [line] (BIP) -| (M); 
     \path [line] (BIP)  -- (AT);
     \path [line] (BIP)  -| (RGT);
     \path [line] (RGT)  -- (RGT2);
     \draw [line] (AT)   to [myncbar,angle=-90,arm=0.6] (AT2);
     \draw [line] (AT)   to [myncbar,angle=-90,arm=0.6](AT3);
     \path [line] (M)    -- (M2);
     \path [line] (RGT2)  |- (C);
     \draw [line] (AT2)  to [myncbar,angle=-90,arm=0.7] (C);
     \draw [line] (AT3)  to [myncbar,angle=-90,arm=0.7](C);
     \path [line] (M2)    |- (C);
     \path [line] (C)    -- (O);
   \begin{pgfonlayer}{background}
    \draw[line width=13.2em,green!50,cap=rect]
   (-4.7,-3)--(4.7,-3);
   \node at (-6,-5)   {\textbf{RVMT-BIP}};
\end{pgfonlayer}
\end{tikzpicture}
}
      \setlength{\abovecaptionskip}{0.1cm}
\caption{Overview of RVMT-BIP work-flow}
\label{fig:workflow}
\end{figure}
RVMT-BIP (Runtime Verification of Multi-Threaded BIP) is a Java implementation of ca. 2,200 LOC.
RVMT-BIP is integrated in the BIP tool suite~\cite{Basu06modelingheterogeneous}.
The BIP (Behavior, Interaction, Priority) framework is a powerful and expressive framework for the formal construction of heterogeneous systems.
RVMT-BIP takes as input a BIP CBS and a monitor description for a property, and outputs a new BIP system whose behavior is monitored against the property while running concurrently.
RVMT-BIP uses the following modules:

\begin{itemize}
	\item
	Module \emph{Atomic Transformation} takes as input the initial BIP system and a monitor description.
From the input abstract monitor description, it extracts the list of components, and the set of their states and variables that influence the truth-value of the property and are used by the monitor. 
Then, this module instruments the atomic components in the extracted list so as to observe their states and the values of the variables.
Finally, the transformed components and the original version of the components that do not influence the property are returned as output.
	\item
Module \emph{Building $\RGT$} takes as input the initial BIP system and a monitor description and produces component $\RGT$ (Reconstructor of Global Trace) which reconstructs and accumulates global states at runtime to produce ``on-the-fly" the global trace.
	\item
	Module \emph{Building $\monitor$} takes as input the initial BIP system and a monitor description and then outputs the atomic component implementing the monitor (following~\cite{FalconeJNBB15}).
	Component $\monitor$ receives and consumes the reconstructed global trace generated by component $\RGT$ at runtime and emits verdicts.
	\item
	Module \emph{Connections} constructs the new composite and monitored component.
	The module takes as input the output of the \textit{Atomic Transformation}, \textit{Building $\RGT$} and \emph{Building $\monitor$} modules and then outputs a new composite component with new connections.
	The new connections are purposed to synchronize instrumented components and component $\RGT$ in order to transfer updated states of the components to $\RGT$.
	Instrumented components interact with $\RGT$ independently and concurrently.
\end{itemize}
%
\subsection{Case Studies}
\label{subsec:case_study}
%
We present some case studies on executable BIP systems conducted with RVMT-BIP.
%
%
\subsubsection{Process Completion of System Demosaicing}
\begin{figure}
    \centering
    \begin{subfigure}[b]{0.2\textwidth}
    \centering
 	\scalebox{.7} {
	\begin{tikzpicture}[->,>=stealth',shorten >=1pt,auto,node distance=2.8cm, semithick,baseline=-0.3cm]
  	\node[state,initial left] (s1){$s_1$};
  	\node[state](s2)[right of=s1,node distance=2.3cm] {$s_2$};  
  	\path (s1) edge     node {$e_2$} (s2); 
  	\path (s2) edge  [bend left]   node {$e_1$} (s1);  
  	\path (s1) edge [loop above]   node {$\neg e_2$} (); 
	\end{tikzpicture}}
    \caption{$\varphi_1$}
    \label{fig:phi1}
    \end{subfigure}
    \hfill    
    \begin{subfigure}[b]{0.7\textwidth}
       \centering
\scalebox{.7} {
\begin{tikzpicture}[->,>=stealth',shorten >=1pt,auto,node distance=1.8cm, semithick,baseline=-0.3cm]
  \node[state,initial below,minimum size=0.1cm] (s1)   {$s_1$};
  \node[state]         (s2) [right of=s1,node distance=1.7cm,minimum size=0.1cm] {$s_2$};
  \node[state]         (s3) [right of=s2,node distance=3cm,minimum size=0.1cm] {$s_3$};
  \node[state]         (s4) [right of=s3,node distance=3cm,minimum size=0.1cm] {$s_4$};
  \node[state]         (s12)[right of=s4,node distance=4cm,minimum size=0.1cm] {$s_{12}$};
  \node[state]         (s5) [below of=s4,node distance=0.7cm,minimum size=0.1cm] {$s_5$};
  \node[state]         (s6) [below of=s3,node distance=1.5cm,minimum size=0.1cm] {$s_6$};
  \node[state]         (s7) [right of=s6,node distance=3cm,minimum size=0.1cm] {$s_7$};  
  \node[state]         (s8) [below of=s7,node distance=0.7cm,minimum size=0.1cm] {$s_8$};
  \node[state]         (s9) [below of=s6,node distance=1.5cm,minimum size=0.1cm] {$s_9$};
  \node[state]         (s10)[right of=s9,node distance=3cm,minimum size=0.1cm] {$s_{10}$};  
  \node[state]         (s11)[below of=s10,node distance=0.8cm,minimum size=0.1cm] {$s_{11}$};

  \path (s1) edge     node {$e_1$} (s2); 
  \path (s2) edge     node {$e_2$} (s3); 
  \path (s3) edge     node {$e_3$} (s4); 
  \path (s2) edge  [bend right=10]   node {$e_3$} (s6); 
  \path (s2) edge  [bend right=10]   node {$e_4$} (s9);  
  \path (s3) edge  [bend right=10]   node {$e_4$} (s5);
  \path (s6) edge     node {$e_2$} (s7);
  \path (s6) edge   [bend right=10]  node {$e_4$} (s8);
  \path (s9) edge     node {$e_3$} (s10);
  \path (s9) edge [bend right=10]    node {$e_2$} (s11);
  \path (s4) edge     node {$e_4$} (s12);  
  \path (s5) edge [bend right=10]   node {$e_3$} (s12);
  \path (s7) edge [bend right=15]   node {$e_4$} (s12);
  \path (s8) edge [bend right=25]   node {$e_2$} (s12);
  \path (s10)edge [bend right=30]   node {$e_2$} (s12);
  \path (s11)edge [bend right=40]   node {$e_3$} (s12);
  \path (s12)edge [bend right=20]   node [above]{$e_1$} (s2);                 
\end{tikzpicture}
}
        \caption{$\varphi_2$}
        \label{fig:phi2}
    \end{subfigure}
    \caption{Automata of properties of demosaicing}
    \label{fig:automata demosaicing}
\end{figure}
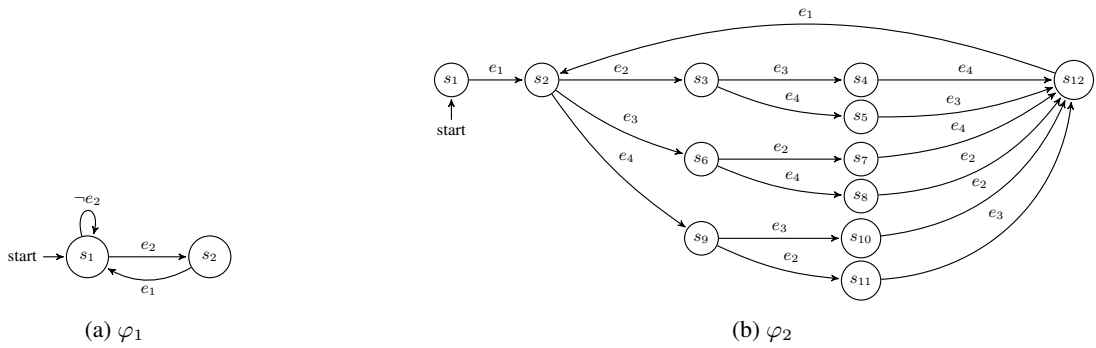

%
Demosaicing is an algorithm for digital image processing used to reconstruct a full color image from the incomplete color samples output from an image sensor.
\figref{fig:Demosaicing} shows a simplified version of the the processing network of Demosaicing.
Demosaicing contains a $\mathit{Splitter}$ and a $\mathit{Joiner}$ process, a pre-demosaicing ($\mathit{Demopre}$) and a post-demosaicing ($\mathit{Demopost}$) process and three internal demosaicing $\mathit{Demo}$ processes that run in parallel.
The real model contains ca. 1,000 lines of code, consists of 26 atomic components interacting through 35 interactions. 
We consider two specifications related to process completion:
\begin{figure}
    \centering
   \scalebox{.75} {     
        
\tikzstyle{block} = [rectangle, draw, fill=gray!20, 
    text width=5em, text centered, rounded corners, minimum height=2.5em]   
     
\tikzstyle{line} = [draw, -latex'] 
\def\myarm{1cm}
\def\myangle{0}
\tikzset{
  arm/.default=1cm,
  arm/.code={\def\myarm{#1}},
  angle/.default=0,
  angle/.code={\def\myangle{#1}} 
}   
\tikzset{
    myncbar/.style = {to path={
        let
            \p1=($(\tikztotarget)+(\myangle:\myarm)$)
        in
            -- ++(\myangle:\myarm) coordinate (tmp)
            -- ($(\tikztotarget)!(tmp)!(\p1)$)
            -- (\tikztotarget)\tikztonodes
    }}
}  
    
\begin{tikzpicture}[node distance = 1.5cm, auto]
   
 \node [block]                         (splitter) {$\mathit{Splitter}$};
 \node [block,right of=splitter,node distance = 3cm]  (demopre)  {$\mathit{Demopre}$}; 
  \node [block,right of=demopre,node distance = 3cm] (demo2)          {$\mathit{Demo_2}$};
 \node [block,above  of=demo2,node distance = 1cm]  (demo1)  {$\mathit{Demo_1}$};
 \node [block,below of=demo2,node distance = 1cm] (demo3)   {$\mathit{Demo_3}$};
 \node [block,right of=demo2,node distance = 3cm]  (demopost)  {$\mathit{Demopost}$};
 \node [block,right of=demopost,node distance = 3cm]  (joiner)  {$\mathit{Joiner}$};

     \path [line] (splitter)  -- (demopre);
     \draw [line] (demopre) to [myncbar,angle=0,arm=1.5] (demo1); 
     \draw [line] (demopre) to [myncbar,angle=0,arm=1.5] (demo3);
     \path [line] (demopre)  -- (demo2);
     \draw [line] (demo1) to [myncbar,angle=0,arm=1.5] (demopost);
     \draw [line] (demo3) to [myncbar,angle=0,arm=1.5] (demopost);
     \path [line] (demo2)  -- (demopost);
     \path [line] (demopost)  -- (joiner);

\end{tikzpicture}}
\caption{Processing network of system Demosaicing}
\label{fig:Demosaicing} 
\end{figure}
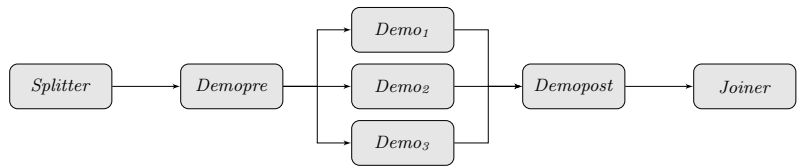
\begin{enumerate}
\item[$\varphi_1$:] It is necessary that all the internal demosaicing units finish their process before the post-demosaicing unit starts processing. 
The post-demosaicing unit receives the output results of internal demosaicing units through port $\mathit{getimg}$.
We add variable $\mathit{port}$ to record the last executed port.
Each demosaicing unit has a boolean variable $\mathit{done}$ which is set to $\true$ whenever the demosaicing process completes.
This requirement is formalized as property $\varphi_1$ defined by the automaton depicted in \mfigref{fig:phi1} where the events are $e_1:\mathit{Demopost.port == getimg}$ and $e_2:(\mathit{Demo_1.done}\wedge\mathit{Demo_2.done}\wedge \mathit{Demo_3.done})$.
From the initial state $s_1$, the automaton moves to state $s_2$ when all the internal demosaicing units finish their process.
Receiving the processed images by post-demosaicing causes a move from state $s_2$ to $s_1$.
\item[$\varphi_2$:]
Moreover, internal demosaicing units ($\mathit{Demo_1}$, $\mathit{Demo_2}$, $\mathit{Demo_3}$) should not start the demosaicing process until the pre-demosaicing unit finishes its process.
The pre-demosaicing unit sends its output to the internal demosaicing units through port $\mathit{transmit}$ and each internal demosaicing unit starts the demosaicing process by executing a transition labeled by port $\mathit{start}$.
This requirement is formalized as property $\varphi_2$ which is defined by the automaton depicted in \mfigref{fig:phi2} where $e_1 : \mathit{Demopre.port} == $ $\mathit{transmit}$, $e_2 : \mathit{Demo_1.port == start}$, $e_3 : \mathit{Demo_2.port == start}$ and $e_4 : \mathit{Demo_3.port == start}$.
From the initial state $s_1$, whenever the pre-demosaicing unit transmits its processed output to the internal demosaicing units, the automaton moves to state $s_2$.
Internal demosaicing units can start in different order.
Moreover, all demosaicing units must eventually start their internal process and the automaton reaches state $s_{12}$. From state $s_{12}$, the automaton moves back to state $s_2$ whenever the pre-demosaicing unit sends the next processed data to the internal demosaicing units.
\end{enumerate}
%

\subsubsection{Data-freshness of System Reader-WriterV1}
%
System Reader-WriterV1 (ca. 130 LOC) consists of a set of independent composite components.
Each composite component consists of four components: a $\mathit{Reader}$, a $\mathit{Writer}$, a $\mathit{Clock}$ and a $\mathit{Poster}$. (in total, 12 components and 9 interactions).
$\mathit{Reader}$ and $\mathit{Writer}$ communicate with each other through the $\mathit{Poster}$.
The data generated by $\mathit{Writer}$ is written in a $\mathit{Poster}$ that can be accessed by $\mathit{Reader}$.
The Reader-Writer model is presented in \mfigref{fig:RW}.
We consider a specifications related to data freshness:
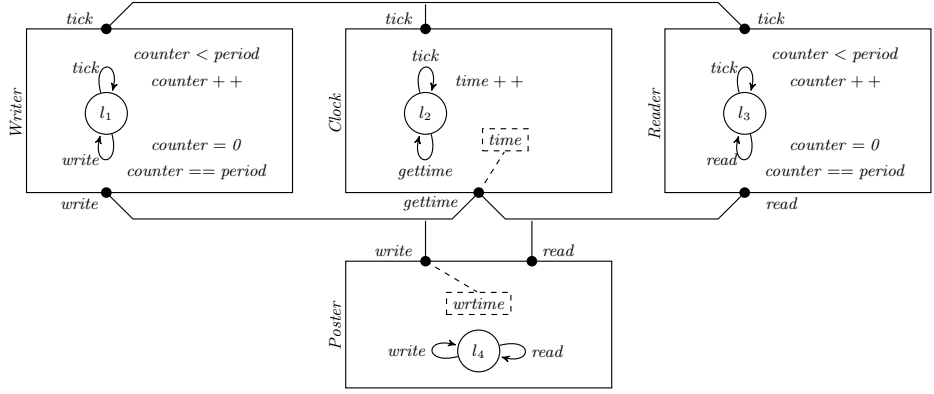
\begin{figure}[t]
\centering
\scalebox{.7} {
\begin{tikzpicture}[->,>=stealth',shorten >=1pt,auto,node distance=2.8cm, semithick]
  \tikzstyle{every state}=[fill=none,draw=black,text=black]
 
  \node[state] (A)  {$l_1$};
  \node[state]         (B) [right of=A,node distance=6cm] {$l_2$};
  \node[state]         (C) [right of=B,node distance=6cm] {$l_3$};
  \node[state]         (D) [below of=B,node distance=4cm,xshift=1cm,yshift=-0.5cm] {$l_4$};
  
  \filldraw[fill=none,draw=black](-1.5,-1.5)--(-1.5,1.6)--(3.5,1.6)--(3.5,-1.5)--(-1.5,-1.5)--cycle;
  \fill (0,1.6) circle (0.1cm);\node at (-0.5,1.8){$\mathit{tick}$};
  \fill (0,-1.5) circle (0.1cm);\node at (-0.5,-1.7){$\mathit{write}$};
  
  \filldraw[fill=none,draw=black,xshift=6cm](-1.5,-1.5)--(-1.5,1.6)--(3.5,1.6)--(3.5,-1.5)--(-1.5,-1.5)--cycle;
  \fill [xshift=6cm](0,1.6) circle (0.1cm);\node [xshift=6cm] at (-0.5,1.8){$\mathit{tick}$};
  \fill [xshift=6cm](1,-1.5) circle (0.1cm);\node [xshift=6cm] at (0.1,-1.7){$\mathit{gettime}$};
  
  \filldraw[fill=none,draw=black,xshift=12cm](-1.5,-1.5)--(-1.5,1.6)--(3.5,1.6)--(3.5,-1.5)--(-1.5,-1.5)--cycle;
  \fill [xshift=12cm](0,1.6)circle (0.1cm);\node[xshift=12cm] at (0.5,1.8){$\mathit{tick}$};
  \fill [xshift=12cm](0,-1.5)circle (0.1cm);\node[xshift=12cm]at (0.7,-1.7){$\mathit{read}$};
  
  \filldraw[fill=none,draw=black,xshift=6cm,yshift=-4cm](-1.5,-1.2)--(-1.5,1.2)--(3.5,1.2)--(3.5,-1.2)--(-1.5,-1.2)--cycle;
  \fill [xshift=6cm,yshift=-4cm](0,1.2) circle (0.1cm);\node [xshift=6cm,yshift=-4cm] at (-0.6,1.4){$\mathit{write}$};
  \fill [xshift=6cm,yshift=-4cm](2,1.2) circle (0.1cm);\node [xshift=6cm,yshift=-4cm] at (2.5,1.4){$\mathit{read}$};
  
  \node at (-1.7,0) {\rotatebox{90} {$\mathit{Writer}$}}; 
  \node at (4.3,0) {\rotatebox{90} {$\mathit{Clock}$}};
  \node at (10.3,0) {\rotatebox{90} {$\mathit{Reader}$}};
  \node at (4.3,-4) {\rotatebox{90} {$\mathit{Poster}$}};
  
  \draw[-] (0,-1.5)--(0.5,-2)--(6.5,-2)--(7,-1.5);  
  \draw[-] (7,-1.5)--(7.5,-2)--(11.5,-2)--(12,-1.5); 
  \draw[-] (6,-2.8)--(6,-2); 
  \draw[-] (8,-2.8)--(8,-2); 
  \draw[-] (0,1.6)--(0.5,2.1)--(11.5,2.1)--(12,1.6);
  \draw[-] (6,1.6)--(6,2.1);
  
  \path (A) edge [loop above]              node [left]{$\mathit{tick}$} ();  
  \path (A) edge [loop below]              node [left]{$\mathit{write}$} ();
  \path (B) edge [loop above]              node {$\mathit{tick}$} ();  
  \path (B) edge [loop below]              node {$\mathit{gettime}$} ();
  \path (C) edge [loop above]              node [left]{$\mathit{tick}$} ();  
  \path (C) edge [loop below]              node [left]{$\mathit{read}$} ();
  \path (D) edge [loop left]               node {$\mathit{write}$} ();  
  \path (D) edge [loop right]              node {$\mathit{read}$} ();

  \node [draw,dashed] at (7.5,-0.5) {$\mathit{time}$};  
  \node [draw,dashed] at (7,-3.6) {$\mathit{wrtime}$};
  \draw[dashed,-] (7,-1.5)--(7.5,-0.7);
  \draw[dashed,-] (6,-2.8)--(7,-3.4);
  
  \node at (1.7,1.1) {$\mathit{counter<period}$};
  \node at (1.7,0.6) {$\mathit{counter++}$};
  \node at (1.7,-1.1) {$\mathit{counter==period}$};
  \node at (1.7,-0.6) {$\mathit{counter=0}$};
  
  \node [xshift=6cm] at (1.2,0.6) {$\mathit{time++}$};
  
  \node [xshift=12cm] at (1.7,1.1) {$\mathit{counter<period}$};
  \node [xshift=12cm] at (1.7,0.6) {$\mathit{counter++}$};
  \node [xshift=12cm] at (1.7,-1.1) {$\mathit{counter==period}$};
  \node [xshift=12cm] at (1.7,-0.6) {$\mathit{counter=0}$};
           
\end{tikzpicture}
}
\caption{Model of system Reader-Writer}
\label{fig:RW} 
\end{figure}
%
%
\begin{itemize}
\item[$\varphi_3$:]

It is necessary that the data is up-to-date: the data read by component $\mathit{Reader}$ must be fresh enough compared to the moment it has been written by $\mathit{Writer}$.
If $t_1$ and $t_2$ are the moments of reading and writing actions respectively, then the difference  between $t_2$ and $t_1$ must be less than a specific duration $\delta$, i.e., $(t_2 - t_1) \leq\delta$.
In the model, the time counter is implemented by a component $\mathit{Clock}$, and the $\mathit{tick}$ transition occurs every second.
This requirement is formalized as property $\varphi_3$ which is defined by the automaton depicted in \mfigref{fig:phi3}, where $\delta=2$,  $e_1:\mathit{Writer.port == write}$, $e_2:\mathit{Clock.port == tick}$ and $e_3:\mathit{Reader.port == read}$.
Whenever $\mathit{Writer}$ writes into $\mathit{Poster}$, the automaton moves from the initial state $s_1$ to $s_2$.
When $\mathit{Reader}$ reads $\mathit{Poster}$, the automaton moves from $s_2$ to $s_1$.
$\mathit{Reader}$ is allowed to read $\mathit{Poster}$ after one $\mathit{tick}$ transition.
In this case, the automaton moves from $s_2$ to $s_3$ after the $\mathit{tick}$, and then moves from $s_3$ to  $s_1$ after reading $\mathit{Poster}$.
$\varphi_3$ also allows to read $\mathit{Poster}$ after two $\mathit{tick}$ transitions.
In this case, the automaton moves from $s_2$ to $s_4$ after the first $\mathit{tick}$, then moves from $s_4$ to  $s_3$ on the second $\mathit{tick}$, and finally moves from $s_3$ to $s_1$ after reading $\mathit{Poster}$.
\end{itemize}
%
%
\subsubsection{Execution Order of System Reader-WriterV2}

\begin{figure}[t]
    \centering
    \begin{subfigure}[b]{0.4\textwidth}
    \centering
\scalebox{.7} {
\begin{tikzpicture}[->,>=stealth',shorten >=1pt,auto,node distance=2cm, semithick,baseline=-0.3cm]
  \node[state,initial] (A)  {$s_1$};
  \node[state]         (B) [below right of=A,node distance=1.5cm,xshift=0.5cm] {$s_2$};
  \node[state]         (C) [above right of=A,node distance=1.5cm,xshift=0.5cm] {$s_3$};
  \node[state]         (D) [right of=A,node distance=3cm] {$s_4$};
  \path (A) edge     node {$e_1$} (B); 
  \path (B) edge      [bend left]   node {$e_3$} (A); 
  \path (D) edge                    node [above]{$e_3$} (A); 
  \path (B) edge      [bend right]  node [below right] {$e_2$} (D); 
  \path (D) edge      [bend right]  node [above right] {$e_2$} (C); 
  \path (C) edge      [bend right]  node [above left]{$e_3$} (A);                    
\end{tikzpicture}
}
        \caption{$\varphi_3$}
        \label{fig:phi3}
    \end{subfigure}
    \begin{subfigure}[b]{0.4\textwidth}
        \centering
        \scalebox{.7} {\begin{tikzpicture}[->,>=stealth',shorten >=1pt,auto,node distance=2.8cm, semithick,baseline=-0.3cm]
  \node[state,initial] (A)  {$s_1$};
  \node[state]         (B) [above right of=A,node distance=5cm,yshift=-2.8cm] {$s_2$};
  \node[state]         (C) [below right of=A,node distance=5cm,yshift=2.8cm] {$s_3$};
  \path (A) edge [bend left]               node {$e_1$} (B); 
  \path (B) edge [bend left]               node {$e_2$} (C);
  \path (C) edge [bend left]               node {$e_3$} (A);
  \path (A) edge [loop right]       node {$\neg e_1\wedge\neg e_2\wedge\neg e_3$} ();
  \path (B) edge [loop right]       node {$\neg e_1\wedge\neg e_2\wedge\neg e_3$} ();
  \path (C) edge [loop right]       node {$\neg e_1\wedge\neg e_2\wedge\neg e_3$} ();
  
\end{tikzpicture}}
        \caption{$\varphi_4$}
        \label{fig:phi4}
    \end{subfigure}
    \caption{Automata of the properties of system Reader-Writer}
    \label{fig:automata RW}
\end{figure}
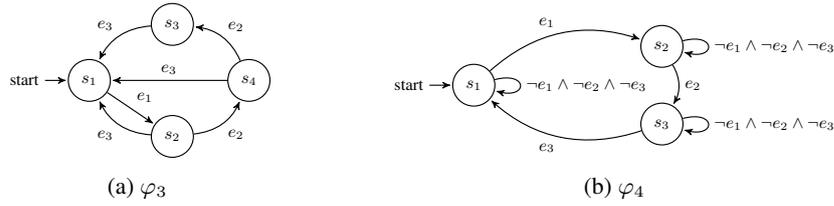

System Reader-WriterV2 (ca. 150 LOC) is a more complex version of Reader-WriterV1 and involves several writers. 
This system has six components: $\mathit{Reader}$, $\mathit{Writer_1}$, $\mathit{Writer_2}$, $\mathit{Writer_3}$, $\mathit{Clock}$ and $\mathit{Poster}$.
The Writers are synchronized together. $\mathit{Reader}$ and Writers communicate with each other through $\mathit{Poster}$. 
The data generated by each writer is written to $\mathit{Poster}$ and can then be accessed by $\mathit{Reader}$.
Having several writers, a more complex specification on the execution order can be defined.
%
%
%
%
%
We consider a specifications related to execution order:
%
%
%
\begin{itemize}
\item[$\varphi_4$:]
The writers should periodically write data to a poster in a specific order.
The specification concerns 3 writers: $\mathit{Writer_1}$, $\mathit{Writer_2}$ and $\mathit{Writer_3}$.
During each period , the writing order must be as follows: $\mathit{Writer_1}$ writes to the poster first, then $\mathit{Writer_2}$ can write only when $\mathit{Writer_1}$ finishes writing to the poster, $\mathit{Writer_3}$ can write only when $\mathit{Writer_2}$ finishes writing to the poster, and the same goes on for the next periods. 
To do so, each writer is assigned a unique id that is passed to the poster when it starts using the poster.
This id is then used to determine the last writer that used the poster.
For example, when $\mathit{Writer_2}$ wants to access the poster, it has to check whether the id stored in the poster corresponds to $\mathit{Writer_1}$ or not.

This requirement is formalized as property $\varphi_4$ which is defined by the automaton depicted in \mfigref{fig:phi4} where:
\begin{itemize}
\item
$e_1:(\mathit{Writer_1.port == write \wedge Poster.port == write \wedge}  \mathit{Clock.port == getTime})$,
\item
$e_2:(\mathit{Writer_2.port == write \wedge Poster.port == write \wedge}  \mathit{Clock.port == getTime})$,
\item
$e_3:(\mathit{Writer_3.port == write \wedge Poster.port == write \wedge}  \mathit{Clock.port == getTime})$.
\end{itemize}
When $\mathit{Writer_1}$ writes to the poster, the automaton moves from initial state $s_1$ to state $s_2$.
From state $s_2$, the automaton moves to state $s_3$ when $\mathit{Writer_2}$ writes to the poster.
From state $s_3$, the automaton moves to the initial state $s_1$ when $\mathit{Writer_3}$ writes to the poster.
This writing order must always be followed.
\end{itemize}

\subsubsection{Distribution of Tasks in System Task}
\begin{figure}[t]
    \centering
\scalebox{.8} {
\begin{tikzpicture}[->,>=stealth',shorten >=1pt,auto,node distance=2.8cm, semithick,baseline=-0.3cm]
  \node[state,initial] (A)  {$s_1$};
   \path (A) edge [loop right]       node {$e_1\wedge e_2\wedge e_3$} ();
\end{tikzpicture}
}       
    \caption{Automaton of the property of system Task}
    \label{fig:automata task system}
\end{figure}
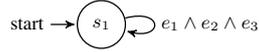

%
We consider our running example system Task and a specification of the homogeneous distribution of the tasks among the workers:
\begin{itemize}
\item[$\varphi_5$:]
The satisfaction of this specification depends on the execution time of each worker.
Different tasks may have different execution times for different workers.
Obviously, the faster a worker completes each task, the higher is the number of its accomplished tasks.
After executing a task, the value of the variable $x$ of a worker is increased by one.
Moreover, the absolute difference between the values of variable $x$ of any two workers must always be less than a specific integer value (which is 3 for this case study).
This requirement is formalized as property $\varphi_5$ which is defined by the automaton depicted in \mfigref{fig:automata task system} where $e_1:|\mathit{worker_1.x-worker_2.x|<3}$ , $e_2:|\mathit{worker_2.x}$ $\mathit{-worker_3.x|<3}$ and $e_3:|\mathit{worker_1.x}$ $\mathit{-worker_3.x|<3}$.
The property holds as long as $e_1$, $e_2$ and $e_3$ hold.
\end{itemize}

%
\subsection{Evaluation Principles}
\label{subsec:evaluation}
%
For each system and all its properties, we synthesized a BIP monitor following~\cite{FalconeJNBB11,FalconeJNBB15} and combined it with the CBS output from RVMT-BIP.
We obtain a new CBS with corresponding $\RGT$ and monitor components.
We run each system by using various numbers of threads and observe the execution time.
Executing these systems with a multi-threaded controller results in a faster run because the systems benefit from the parallel threads.
Additional steps are introduced in the concurrent transitions of the system.
Note, these are asynchronous with the existing interactions and can be executed in parallel. 
These systems can also execute with a single-threaded controller which forces them to run sequentially. 
Varying the number of threads allows us to assess the performance of the (monitored) system under different degrees of parallelism.
In particular, we expected the induced overhead to be insensitive to the degree of parallelism.
For instance, an undesirable behavior would have been to observe a performance degradation (and an overhead increase) which would mean either that the monitor sequentializes the execution or that the monitoring infrastructure is not suitable for multi-threaded systems.
We also extensively tested the functional correctness of RVMT-BIP, that is whether the verdicts of the monitors are sound and complete.
%
\subsection{Results and Conclusions}
\label{subsec:results}
%
\renewcommand{\arraystretch}{1.2}
\begin{table}[t]
\centering
\caption{Results of monitoring Demosaicing, Reader-WriterV1, Reader-WriterV2, and Task with RVMT-BIP}
\resizebox{\textwidth}{!}{
\begin{tabular}{|c|c|c|c|c|c|c|c|c|c|c|c|c|c|}
\cline{1-14}
\multirow{2}{*}{system} & \multirow{2}{*}{\begin{tabular}[c]{@{}c@{}}\# executed\\ interactions\end{tabular}} & \multicolumn{10}{c|}{execution time and overhead according to the number of threads} & \multirow{2}{*}{\# events} & \multirow{2}{*}{\begin{tabular}[c]{@{}c@{}}\# extra executed\\ interactions\end{tabular}} \\ \cline{3-12}
 &  & 1 & 2 & 3 & 4 & 5 & 6 & 7 & 8 & 9 & 10 &  &  \\ \cline{1-14}
 Demosaicing (26,35) & 1,300 & 18.98 & 10.24 & 7.75 & 6.85 & 6.58 & 6.09 & 6.33 & 6.45 & 6.29 & 6.27 & n/a & n/a \\ \cline{1-14}
 \multirow{2}{*}{\textit{\begin{tabular}[c]{@{}c@{}}Demosaicing (27,69)\\ $\varphi_1$ (11)\end{tabular}}} & \multirow{2}{*}{3,051} & 19.02 & 11.53 & 8.17 & 7.43 & 6.68 & 6.50 & 6.27 & 6.05 & 6.03 & 6.18 & \multirow{2}{*}{1,300} & \multirow{2}{*}{1,751} \\ \cline{3-12}
 &  & 0.1\% & 12.6\% & 5.4\% & 8.5\% & 4.3\% & 6.6\% & \lowoh & \lowoh & \lowoh & \lowoh &  &  \\ \cline{1-14}
 \multirow{2}{*}{\textit{\begin{tabular}[c]{@{}c@{}}Demosaicing (27,46)\\ $\varphi_2$ (4)\end{tabular}}} & \multirow{2}{*}{1,850} & 18.68 & 11.05 & 7.65 & 7.80 & 6.77 & 6.38 & 6.22 & 6.45 & 6.17 & 6.35 & \multirow{2}{*}{400} & \multirow{2}{*}{550} \\ \cline{3-12}
 &  & \lowoh & 7.9\% & \lowoh & 13.8\% & 2.8\% & 4.8\% & \lowoh & \lowoh & \lowoh & \lowoh &  &  \\ \cline{1-14}
Reader-WriterV1 (12,9) & 120,000 & 61.48 & 29.67 & 20.03 & 20.00 & 20.05 & 20.21 & 20.60 & 21.54 & 21.92 & 22.13 & n/a & n/a \\ \cline{1-14}
\multirow{2}{*}{\textit{\begin{tabular}[c]{@{}c@{}}ReaderWriterV1 (13,12)\\ $\varphi_3$ (3)\end{tabular}}} & \multirow{2}{*}{200,000} & 62.53 & 38.29 & 21.96 & 22.28 & 22.62 & 22.71 & 22.88 & 23.48 & 24.15 & 24.47 & \multirow{2}{*}{40,000} & \multirow{2}{*}{80,000} \\ \cline{3-12}
 &  & 1.6\% & 27.7\% & 9.6\% & 11.4\% & 12.8\% & 12.4\% & 11.0\% & 9.0\% & 10.1\% & 10.5\% &  &  \\ \cline{1-14}
Reader-WriterV2 (6,7) & 20,000 & 32.06 & 21.45 & 12.04 & 11.37 & 11.33 & 11.37 & 11.44 & 11.49 & 11.53 & 11.58 & n/a & n/a \\ \cline{1-14}
\multirow{2}{*}{\textit{\begin{tabular}[c]{@{}c@{}}ReaderWriterV2 (7,12)\\ $\varphi_4$ (5)\end{tabular}}} & \multirow{2}{*}{85,000} & 33.92 & 22.72 & 13.90 & 13.77 & 14.09 & 14.36 & 14.83 & 15.18 & 15.41 & 15.57 & \multirow{2}{*}{20,000} & \multirow{2}{*}{65,000} \\ \cline{3-12}
 &  & 5.8\% & 5.9\% & 15.4\% & 21.1\% & 24.3\% & 26.2\% & 29.6\% & 32.1\% & 33.5\% & 34.4\% &  &  \\ \cline{1-14}
Task (4,10) & 399,999 & 117.28 & 70.18 & 60.91 & 60.06 & 58.98 & 60.01 & 60.93 & 61.77 & 63.13 & 65.45 & n/a & n/a \\ \cline{1-14}
\multirow{2}{*}{\textit{\begin{tabular}[c]{@{}c@{}}Task (5,16)\\ $\varphi_5$ (3)\end{tabular}}} & \multirow{2}{*}{600,197} & 123.98 & 71.73 & 62.28 & 63.26 & 62.79 & 62.78 & 63.35 & 64.57 & 65.61 & 66.27 & \multirow{2}{*}{100,198} & \multirow{2}{*}{200,198} \\ \cline{3-12}
 &  & 5.7\% & 2.2\% & 2.2\% & 5.3\% & 6.4\% & 4.4\% & 3.9\% & 4.5\% & 3.9\% & 1.2\% &  &  \\ \cline{1-14}
\end{tabular}}
\label{Table:result}
\end{table}

\paragraph{Performance evaluation.}
Tables~\ref{Table:result} and \ref{Table:result_RV-BIP} report the timings obtained when checking the following specifications: \textit{complete process} property on Demosaicing, \textit{data freshness} and \textit{execution ordering} property on Reader-Writer systems, and \textit{task distribution} property on Task, with RVMT-BIP and RV-BIP respectively.
Each measurement is an average value obtained over 100 executions of these systems.
In Tables~\ref{Table:result} and \ref{Table:result_RV-BIP}, the columns have the following meanings:
\begin{itemize}
	\item
	Column \textit{system} indicates the systems.
	System in \textit{italic} format represents the monitored version of the initial system.
	Moreover, $(x,y)$ in front of the system name means that $x$ (resp. $y$) is the number of components (resp. interactions) of the system.
	The monitored property is written below each monitored system name with a value $(z)$ which indicates that $z$ components have variables influencing the truth-value of the property (and were thus instrumented by RVMT-BIP or RV-BIP).
	\item Column \textit{\# executed interactions} indicates the number of interactions executed by the engine which also represents the number of functional steps of the system. 
	\item Columns \textit{execution time and overhead according to the number of threads} report (i) the execution time of the systems when varying the number of threads and (ii) the overhead induced by monitoring (for monitored systems).
	\item Column \textit{events} indicates the number of reconstructed global states (events sent to the associated monitor). 
	\item Column \textit{extra executed interactions} reports the number of additional interactions (\ie execution of interactions which are added into the initial system for monitoring purposes).	
\end{itemize}
As shown in Table~\ref{Table:result}, using more threads reduces significantly the execution time in both the initial and transformed systems.
Comparing the overheads according to the number of threads shows that the proposed monitoring technique (i) does not restrict the performance of parallel execution and (ii) scales up well with the number of threads.
\paragraph{Performance comparison of RV-BIP and RVMT-BIP.}
\begin{table}[t]
\centering
\caption{Results of monitoring Demosaicing, Reader-WriterV1, Reader-WriterV2, and Task with RV-BIP}
\resizebox{\textwidth}{!}{
\begin{tabular}{|c|c|c|c|c|c|c|c|c|c|c|c|c|c|}
\cline{1-14}
\multirow{2}{*}{system} & \multirow{2}{*}{\begin{tabular}[c]{@{}c@{}}\# executed\\ interactions\end{tabular}} & \multicolumn{10}{c|}{execution time and overhead w.r.t. different number of threads} & \multirow{2}{*}{\# events} & \multirow{2}{*}{\begin{tabular}[c]{@{}c@{}}\# extra executed\\ interactions\end{tabular}} \\ \cline{3-12}
 &  & 1 & 2 & 3 & 4 & 5 & 6 & 7 & 8 & 9 & 10 &  &  \\ \cline{1-14}
Demosaicing (26,35) & 1,300 & 18.98 & 10.24 & 7.75 & 6.85 & 6.58 & 6.09 & 6.33 & 6.45 & 6.29 & 6.27 & n/a & n/a \\ \cline{1-14}
\multirow{2}{*}{\textit{\begin{tabular}[c]{@{}c@{}}Demosaicing (27,37)\\ $\varphi_1$ (11)\end{tabular}}} & \multirow{2}{*}{2,450} & 19.66 & 27.34 & 32.28 & 32.61 & 33.03 & 32.23 & 31.17 & 31.24 & 31.22 & 31.81 & \multirow{2}{*}{1,300} & \multirow{2}{*}{1,300} \\ \cline{3-12}
 &  & 3.5\% & 167\% & 316\% & 376\% & 402\% & 429\% & 392\% & 384\% & 369\% & 407\% &  &  \\ \cline{1-14}
\multirow{2}{*}{\textit{\begin{tabular}[c]{@{}c@{}}Demosaicing (27,37)\\ $\varphi_2$ (11)\end{tabular}}} & \multirow{2}{*}{1,700} & 19.50 & 14.79 & 13.87 & 13.11 & 13.13 & 12.75 & 11.18 & 11.34 & 11.19 & 11.16 & \multirow{2}{*}{400} & \multirow{2}{*}{400} \\ \cline{3-12}
 &  & 2.7\% & 44.4\% & 78.8\% & 91.4\% & 99.7\% & 109\% & 76.5\% & 75.7\% & 78.0\% & 78.0\% &  &  \\ \cline{1-14}
Reader-WriterV1 (12,9) & 120,000 & 61.48 & 29.67 & 20.03 & 20.00 & 20.05 & 20.21 & 20.60 & 21.54 & 21.92 & 22.13 & n/a & n/a \\ \cline{1-14}
\multirow{2}{*}{\textit{\begin{tabular}[c]{@{}c@{}}Reader-WriterV1 (13,11)\\ $\varphi_3$ (3)\end{tabular}}} & \multirow{2}{*}{1600,000} & 61.97 & 37.77 & 21.94 & 22.13 & 22.62 & 23.14 & 25.09 & 26.21 & 26.73 & 27.18 & \multirow{2}{*}{40,000} & \multirow{2}{*}{40,000} \\ \cline{3-12}
 &  & 0.8\% & 26.0\% & 9.5\% & 10.6\% & 12.8\% & 14.5\% & 21.8\% & 21.7\% & 21.9\% & 22.7\% &  &  \\ \cline{1-14}
Reader-WriterV2 (6,7) & 20,000 & 32.06 & 21.45 & 12.04 & 11.37 & 11.33 & 11.37 & 11.44 & 11.49 & 11.53 & 11.58 & n/a & n/a \\ \cline{1-14}
\multirow{2}{*}{\textit{\begin{tabular}[c]{@{}c@{}}Reader-WriterV2 (7,9)\\ $\varphi_4$ (5)\end{tabular}}} & \multirow{2}{*}{40,000} & 33.11 & 23.80 & 13.31 & 13.32 & 13.37 & 13.82 & 14.28 & 14.35 & 14.79 & 14.96 & \multirow{2}{*}{20,000} & \multirow{2}{*}{20,000} \\ \cline{3-12}
 &  & 3.2\% & 10.9\% & 10.5\% & 17.1\% & 18.0\% & 21.5\% & 24.8\% & 24.8\% & 28.2\% & 29.2\% &  &  \\ \cline{1-14}
Task (4,10) & 399,999 & 117.28 & 70.18 & 60.91 & 60.06 & 58.98 & 60.01 & 60.93 & 61.77 & 63.13 & 65.45 & n/a & n/a \\ \cline{1-14}
\multirow{2}{*}{\textit{\begin{tabular}[c]{@{}c@{}}Task (5,12)\\ $\varphi_5$ (3)\end{tabular}}} & \multirow{2}{*}{500,197} & 121.61 & 70.12 & 72.25 & 75.11 & 75.66 & 80.54 & 81.62 & 84.58 & 89.65 & 90.21 & \multirow{2}{*}{100,198} & \multirow{2}{*}{100,198} \\ \cline{3-12}
 &  & 3.6\% & \lowoh & 18.6\% & 25.0\% & 28.2\% & 34.0\% & 33.9\% & 36.9\% & 42.01\% & 37.8\% &  &  \\ \cline{1-14}
\end{tabular}
}
\label{Table:result_RV-BIP}
\end{table}

To illustrate the advantages of monitoring multi-threaded systems with RVMT-BIP, we compared the performance of RVMT-BIP and RV-BIP (\cite{FalconeJNBB15}); see Tables~\ref{Table:result} and \ref{Table:result_RV-BIP} for the results.
Monitoring with RV-BIP amounts to use a standard runtime verification technique, i.e., not tailored to multi-threaded systems.
At runtime, the RV-BIP monitor consumes the global trace (\ie sequence of global states) of the system (where global snapshots are obtained by synchronization among the components) and yields verdicts regarding property satisfaction.
It has been shown in \cite{FalconeJNBB15} that RV-BIP efficiently handles CBSs with sequential executions.

In the following, we highlight some of the main observations and draw conclusions:
\begin{enumerate} 
	\item
	Fixing a system and a property, the number of events received by the monitors of RV-BIP and RVMT-BIP are similar, because both techniques produce monitored systems that are observationally equivalent to the initial ones~\cite{FalconeJNBB15,NazarpourFBBC16}.
	 Moreover, increasing the number of threads does not change the global behavior of the system, therefore the number of events is not affected by the number of threads.
	\item
	Fixing a system and a property, the number of extra interactions imposed by RVMT-BIP is greater than the one imposed by RV-BIP.
	In the monitored system obtained with RVMT-BIP, after the execution of an interaction, the components that are involved in the interaction and influencing the truth-value of the property independently send their updated state to component $\RGT$ (whenever their internal computation is finished).
	In the monitored system obtained with RV-BIP, after the execution of an interaction influencing the truth value of the property, all the updated states will be sent at once (synchronously) to the component monitor. 
	Hence, the evaluation of an event in RV-BIP is done in one step and the number of extra interactions imposed by RV-BIP is the same as the number of monitored events (see Table~\ref{Table:result_RV-BIP}). 
	\item
	In spite of the higher number of extra interactions imposed by RVMT-BIP, during a multi-threaded execution, the fewer synchronous interactions of monitored components imposed by RV-BIP induces a significant overhead.
	This phenomenon is especially visible for the two most concurrent systems: Demosaicing and Task. 
	\item
	\emph{On the independence of components}: Consider systems Demosaicing and Task, which consist of independent components with low-level synchronization and high degree of parallelism, and for which the monitored property requires the states of these independent components.
	On the one hand, at runtime, RV-BIP imposes synchronization among the components whose execution influences the truth value of the property and the component monitor.
	It results in a loss of the performance when executing with multiple threads.
	On the other hand, RVMT-BIP collects updated states of the components independently right after their state update.
	Consequently, with RVMT-BIP, the system performance in the multi-threaded setting is preserved (systems Demosaicing and Task) as a negligible overhead is observed.
	 This is a usual and complex problem which depends on many factors such as platform, model, external codes, compiler, etc.
	 This renders the computation of the number of threads leading to peak performance complex.
	\item
	\emph{Synchronization of independent components}: In RV-BIP, the thread synchronizations and the synchronization of components with the monitor induce a huge overhead especially when concurrent component are concerned with the desired property (system Demosaicing and property $\varphi_1$).
	\item
	\emph{Synchronized components}:
	We observe that, for system ReaderWriterV2, the overhead obtained with RVMT-BIP monitor is slightly higher than the one obtained with RV-BIP monitors.
	Indeed, system ReaderWriterV2 consists of 3 writers synchronized by a clock component.
	 Moreover, property $\varphi_4$ is defined over the states of all the writers.
	 As a matter of fact, if one of the writers needs to communicate with component $\RGT$, then all the other writers need to wait until the communication ends.
	 That is, when the concurrency of the monitored system is limited by internal synchronizations, the global-state reconstruction performed by RVMT-BIP is less effective than the technique used by RV-BIP from a performance point of view.
	\item
	\emph{Synchronized components in independent composite components}: If the initial system (i) consists of independent composite components working concurrently, (ii) the components in each composite are highly synchronized (low degree of parallelism in each composition) and (iii) the desired property is defined over the states of the components of a specific composite component, then RVMT-BIP performs similarly to RV-BIP.
	Indeed, in the monitored system, the independent entities (i.e., composite component) are able to run as concurrently as in the initial system and the overhead is caused by the synchronized components.
	However, by increasing the number of threads, RVMT-BIP monitors offer better performance (system Reader-WriterV1). 
\end{enumerate}
%
\section{Related Work}
\label{sec:rw}
%
Several approaches are related to the one in this paper, as they either target CBSs or address the problem of concurrently runtime verifying systems.
%
\subsection{Runtime Verification of Single-threaded
 CBSs}
%
Dormoy et al. proposed an approach to runtime check the correct reconfiguration of components at runtime~\cite{DormoyKL10}.
They propose to check configurations over a variant of RV-LTL where the usual notion of state is replaced by the notion of component configuration.
RV-LTL is a 4-valued variant of LTL dedicated to runtime verification introduced in~\cite{BauerLS10} and used in~\cite{DBLP:conf/rv/FalconeFM09}.
Our approach offers several advantages compared to the approach in~\cite{DormoyKL10}.
First, our approach is not bound to temporal logic since it only requires a monitor written as a finite-state machine. This state-machine can be then generated by several already existing tools (e.g., Java-MOP) since it uses a generic format to express monitors. Thus, existing monitor synthesis algorithms from various specification formalisms can be re-used, up to a syntactic adaptation layer. Second, the instrumentation of the initial system and the addition of the monitor is formally defined, contrarily to~\cite{DormoyKL10} where the process is only overviewed. Moreover, the whole approach leverages the formal semantics of BIP allowing us to provide a formal proof of the correctness of the proposed approach. All these features confers to our approach a higher-level of confidence.

In~\cite{FalconeJNBB11,FalconeJNBB15}, we proposed a first approach for the runtime verification of CBSs.
The approach in \cite{FalconeJNBB11,FalconeJNBB15} takes a CBS and a regular property as input and generates a monitor implemented as a component.
Then, the monitor component is integrated within an existing CBS.
At runtime, the monitor consumes the global trace (i.e., sequence of global states) of the system and yields verdicts regarding property satisfaction.
The technique in~\cite{FalconeJNBB11,FalconeJNBB15} only efficiently handles CBSs with sequential executions: if applied to a multi-threaded CBS, the monitor would sequentialize completely the execution.
Hence, the approach proposed in this paper can be used in conjunction with the approach in~\cite{FalconeJNBB11,FalconeJNBB15} when dealing with multi-threaded CBSs: (only) the monitor-synthesis algorithm in~\cite{FalconeJNBB11,FalconeJNBB15} can be used to obtained a monitor that can be plugged to the $\RGT$ component (defined in this paper) reconstructing the global states of the system.
%
\subsection{Synthesizing Correct Concurrent Runtime Monitors}
%
In~\cite{FrancalanzaS15}, the authors investigate the synthesis of correct monitors in a concurrent setting, whereby (i) the system being verified executes concurrently with the synthesized monitor (ii) the system and the monitor themselves consist of concurrent sub-components.
Authors have constructed a formally-specified tool that automatically synthesizes monitors from sHML (adaptation of SafeHML (SHML) a sub-logic of the Hennessy-Milner Logic) formulas so as to asynchronously detect property violation by Erlang programs at runtime.
SHML syntactically limits specifications to safety properties which can be monitored at runtime.
Our approach is not bounded to any particular logic.
Moreover, properties in our approach are not restricted to safety properties but can encompass co-safety, and properties that are neither safety nor co-safety properties.
Moreover, the monitored properties can express the desired behavior not only on the internal states of components but they also on the states of external interactions.
%
\subsection{Decentralized Runtime Verification}
%
The approaches in~\cite{bauer2012decentralised,falcone2014efficient,BauerF16} decentralize monitors for linear-time specifications on a system made of synchronous black-box components that cannot be executed concurrently.
Moreover, monitors only observe the outside visible behavior of components to evaluate the formulas at hand.
The decentralized monitor evaluates the global trace by considering the locally-observed traces obtained by local monitors.
To locally detect global violations and satisfactions, local monitors need to communicate, because their traces are only partial w.r.t. the global behavior of the system. 
In~\cite{bauer2012decentralised,falcone2014efficient,BauerF16}, multiple components in a system each observe a subset of some global event trace.
Given an LTL property $\varphi$, the objective is to create sound formula derived from $\varphi$ that can be monitored on each local trace, while minimizing inter-component communication. 
However, they assume that the projection of the global trace upon each component is well-defined and known in advance.
Moreover, all components consume events from the trace synchronously.

Inspired by the decentralized monitoring approach to LTL properties in~\cite{bauer2012decentralised}, Kouchnarenko and Weber~\cite{kouchnarenko14} defines a progressive FTPL semantics allowing a decentralized evaluation of FTPL formula over component-based systems.
Complementarily, Kouchnarenko and Weber~\cite{kouchnarenko13} propose the use of temporal logics to integrate temporal requirements to adaptation policies in the context of Fractal components~\cite{bruneton04}.
The policies are used for specifying reflection or enforcement mechanisms, which refers respectively to corrective reconfiguration triggered by unwanted behaviors, and avoidance of reconfiguration leading to unwanted states.
However, the approaches in~\cite{kouchnarenko13,kouchnarenko14} fundamentally differs from ours because (i) they target architectural invariants and (ii) our approach is specific to CBSs that can be executed in a multi-threaded fashion.
The components in~\cite{kouchnarenko13,kouchnarenko14} are seen as black boxes and the interaction model considers only unidirectional connections.
On the contrary, our approach leverages the internal behavior of components and their interactions for the instrumentation and global-state reconstruction.
%
\subsection{Monitoring Safety Properties in Concurrent Systems}
%
The approach in~\cite{SenVAR06} addresses the monitoring of asynchronous multi-threaded systems against temporal logic formulas expressed in M{\small T}TL.
M{\small T}TL augments LTL with modalities related to the distributed/multi-threaded nature of the system.
The monitoring procedure in~\cite{SenVAR06} takes as input a safety formula and a partially-ordered execution of a parallel asynchronous system, and then predicts a potential property violation on one of the causally-consistent interleavings of the observed execution.
Our approach mainly differs from~\cite{SenVAR06} in that we target CBSs.
Moreover, we assume a central scheduler and we only need to monitor the unique causally-consistent global trace with the observed partial trace. 
Also, we do not place any expressiveness restriction on the formalism used to express properties.
%
\subsection{Parallel Runtime Verification of Sequential Programs}
%
Berkovich et al.~\cite{BerkovichBF15} introduce parallel algorithms to speed up the runtime verification of sequential programs against complex LTL formulas using a graphics processing unit (GPU).
Berkovich et al. consider two levels of parallelism: the monitor (i) works along with the program in parallel, and (ii) evaluates a set of properties in a parallel fashion.
Monitoring threads are added to the program and directly execute on the GPU.
The approach in~\cite{BerkovichBF15} is not tailored to CBSs and is a complementary technique that adds significant computing power to the system to handle the monitoring overhead.
Note that, as shown by our experiments, our approach preserves the performance of the monitored system.
Finally, our approach is not bound to any particular logic, and allows for Turing-complete monitors.
%
%
\section{Conclusions and Future Work}
\label{sec:conc}
%
We draw conclusions and outline avenues for future work.
%
\subsection{Conclusions}

This paper introduces runtime verification for component-based systems that execute concurrently on several threads.
Our approach considers an input system with partial-state semantics  and transforms it to integrate a global-state reconstructor, i.e., a component that produces the witness trace at runtime.
The witness trace is the sequence of global states that could be observed if the system was not multi-threaded and which contain the global information gathered from the partial-states actually traversed by the system at runtime. 
A runtime monitor can be then plugged to the global state reconstructor to monitor the system against properties referring to the global state of the system, while preserving the performance and benefits from concurrency.
We implemented the model transformation in a prototype tool RVMT-BIP.
We evaluated the performance and functional correctness of RVMT-BIP against three case studies and our running examples.
Our experimental results show the effectiveness of our approach and that monitoring with RVMT-BIP induces a cheap overhead at runtime.
%
%
%
\subsection{Future Work}
%
Several research perspectives can be considered.

A first direction is to consider monitoring for fully decentralized and completely distributed models where a central controller does not exist.
For this purpose, we intend to make controllers collaborating in order to resolve conflicts in a distributed fashion.
This setting should rely on the distributed semantics of CBSs as presented in~\cite{bonakdarpour2012framework} and study the influence of the organization of decentralized monitors~\cite{BauerF16} as done for black box components with a global clock in~\cite{ColomboF16}.

Moreover, much work has been done in order to monitor properties on a distributed (monolithic) systems; such as~\cite{SenG07} for online monitoring of CTL properties,~\cite{MostafaB15} for online monitoring of LTL properties, \cite{SenG03} for offline monitoring of properties expressed in a variant of CTL, and~\cite{TomlinsonG97} for online monitoring of global-state predicates.
In the future, we plan to adapt these approaches to the context of CBSs.

Another possible direction is to extend the proposed framework to runtime verify~\cite{BauerLS11} and enforce~\cite{FalconeJMP16} timed specifications on timed components~\cite{Basu06modelingheterogeneous}.
%
%
%
\bibliographystyle{alpha}
\bibliography{main}

\newcommand{\etalchar}[1]{$^{#1}$}
\begin{thebibliography}{NFB{\etalchar{+}}16}

\bibitem[BBBS08]{basu2008distributed}
Ananda Basu, Philippe Bidinger, Marius Bozga, and Joseph Sifakis.
\newblock Distributed semantics and implementation for systems with interaction
  and priority.
\newblock In Kenji Suzuki, Teruo Higashino, Keiichi Yasumoto, and Khaled
  El{-}Fakih, editors, {\em Formal Techniques for Networked and Distributed
  Systems - {FORTE} 2008, 28th {IFIP} {WG} 6.1 International Conference, Tokyo,
  Japan, June 10-13, 2008, Proceedings}, volume 5048 of {\em Lecture Notes in
  Computer Science}, pages 116--133. Springer, 2008.

\bibitem[BBF15]{BerkovichBF15}
Shay Berkovich, Borzoo Bonakdarpour, and Sebastian Fischmeister.
\newblock Runtime verification with minimal intrusion through parallelism.
\newblock {\em Formal Methods in System Design}, 46(3):317--348, 2015.

\bibitem[BBJ{\etalchar{+}}12]{bonakdarpour2012framework}
Borzoo Bonakdarpour, Marius Bozga, Mohamad Jaber, Jean Quilbeuf, and Joseph
  Sifakis.
\newblock A framework for automated distributed implementation of
  component-based models.
\newblock {\em Distributed Computing}, 25(5):383--409, 2012.

\bibitem[BBS06]{Basu06modelingheterogeneous}
Ananda Basu, Marius Bozga, and Joseph Sifakis.
\newblock Modeling heterogeneous real-time components in {BIP}.
\newblock In {\em Fourth {IEEE} International Conference on Software
  Engineering and Formal Methods {(SEFM} 2006), 11-15 September 2006, Pune,
  India}, pages 3--12. {IEEE} Computer Society, 2006.

\bibitem[BCL{\etalchar{+}}04]{bruneton04}
Eric Bruneton, Thierry Coupaye, Matthieu Leclercq, Vivien Qu{\'e}ma, and
  Jean-Bernard Stefani.
\newblock An open component model and its support in java.
\newblock In {\em International Symposium on Component-based Software
  Engineering}, pages 7--22. Springer, 2004.

\bibitem[BF12]{bauer2012decentralised}
Andreas~Klaus Bauer and Yli{\`{e}}s Falcone.
\newblock Decentralised {LTL} monitoring.
\newblock In Dimitra Giannakopoulou and Dominique M{\'{e}}ry, editors, {\em
  {FM} 2012: Formal Methods - 18th International Symposium, Paris, France,
  August 27-31, 2012. Proceedings}, volume 7436 of {\em Lecture Notes in
  Computer Science}, pages 85--100. Springer, 2012.

\bibitem[BF16]{BauerF16}
Andreas Bauer and Yli{\`{e}}s Falcone.
\newblock Decentralised {LTL} monitoring.
\newblock {\em Formal Methods in System Design}, 48(1-2):46--93, 2016.

\bibitem[BLS10]{BauerLS10}
Andreas Bauer, Martin Leucker, and Christian Schallhart.
\newblock Comparing {LTL} semantics for runtime verification.
\newblock {\em Journal of Logic and Computation}, 20(3):651--674, 2010.

\bibitem[BS07]{bliudze07}
Simon Bliudze and Joseph Sifakis.
\newblock The algebra of connectors: structuring interaction in bip.
\newblock In {\em Proceedings of the 7th ACM \& IEEE international conference
  on Embedded software}, pages 11--20. ACM, 2007.

\bibitem[CF16]{ColomboF16}
Christian Colombo and Yli{\`{e}}s Falcone.
\newblock Organising {LTL} monitors over distributed systems with a global
  clock.
\newblock {\em Formal Methods in System Design}, 49(1-2):109--158, 2016.

\bibitem[DKL10]{DormoyKL10}
Julien Dormoy, Olga Kouchnarenko, and Arnaud Lanoix.
\newblock Using temporal logic for dynamic reconfigurations of components.
\newblock In Lu\'{\i}s~Soares Barbosa and Markus Lumpe, editors, {\em
  Proceedings of the 7th International Workshop on Formal Aspects of Component
  Software (FACS 2010)}, volume 6921 of {\em LNCS}, pages 200--217. Springer,
  2010.

\bibitem[FCF14]{falcone2014efficient}
Yli{\`{e}}s Falcone, Tom Cornebize, and Jean{-}Claude Fernandez.
\newblock Efficient and generalized decentralized monitoring of regular
  languages.
\newblock In Erika {\'{A}}brah{\'{a}}m and Catuscia Palamidessi, editors, {\em
  Formal Techniques for Distributed Objects, Components, and Systems - 34th
  {IFIP} {WG} 6.1 International Conference, {FORTE} 2014, Held as Part of the
  9th International Federated Conference on Distributed Computing Techniques,
  DisCoTec 2014, Berlin, Germany, June 3-5, 2014. Proceedings}, volume 8461 of
  {\em Lecture Notes in Computer Science}, pages 66--83. Springer, 2014.

\bibitem[FFM09]{DBLP:conf/rv/FalconeFM09}
Yli{\`e}s Falcone, Jean-Claude Fernandez, and Laurent Mounier.
\newblock Runtime verification of safety-progress properties.
\newblock In Saddek Bensalem and Doron Peled, editors, {\em Proceedings of the
  9th International Workshop on Runtime Verification (RV 2009), Selected
  Papers}, volume 5779 of {\em LNCS}, pages 40--59. Springer, 2009.

\bibitem[FFM12]{FalconeFM12}
Yli{\`{e}}s Falcone, Jean{-}Claude Fernandez, and Laurent Mounier.
\newblock What can you verify and enforce at runtime?
\newblock {\em {STTT}}, 14(3):349--382, 2012.

\bibitem[FJN{\etalchar{+}}11]{FalconeJNBB11}
Yli{\`{e}}s Falcone, Mohamad Jaber, Thanh{-}Hung Nguyen, Marius Bozga, and
  Saddek Bensalem.
\newblock Runtime verification of component-based systems.
\newblock In {\em {SEFM} 2011}, pages 204--220, 2011.

\bibitem[FJN{\etalchar{+}}15]{FalconeJNBB15}
Yli{\`{e}}s Falcone, Mohamad Jaber, Thanh{-}Hung Nguyen, Marius Bozga, and
  Saddek Bensalem.
\newblock Runtime verification of component-based systems in the {BIP}
  framework with formally-proved sound and complete instrumentation.
\newblock {\em Software and System Modeling}, 14(1):173--199, 2015.

\bibitem[FS15]{FrancalanzaS15}
Adrian Francalanza and Aldrin Seychell.
\newblock Synthesising correct concurrent runtime monitors.
\newblock {\em Formal Methods in System Design}, 46(3):226--261, 2015.

\bibitem[Hoa78]{hoare78}
Charles Antony~Richard Hoare.
\newblock Communicating sequential processes.
\newblock In {\em The origin of concurrent programming}, pages 413--443.
  Springer, 1978.

\bibitem[KW13]{kouchnarenko13}
Olga Kouchnarenko and Jean-Fran{\c{c}}ois Weber.
\newblock Adapting component-based systems at runtime via policies with
  temporal patterns.
\newblock In {\em International Workshop on Formal Aspects of Component
  Software}, pages 234--253. Springer, 2013.

\bibitem[KW14]{kouchnarenko14}
Olga Kouchnarenko and Jean-Francois Weber.
\newblock Decentralised evaluation of temporal patterns over component-based
  systems at runtime.
\newblock In {\em International Workshop on Formal Aspects of Component
  Software}, pages 108--126. Springer, 2014.

\bibitem[MB15]{MostafaB15}
Menna Mostafa and Borzoo Bonakdarpour.
\newblock Decentralized runtime verification of {LTL} specifications in
  distributed systems.
\newblock In {\em 2015 {IEEE} International Parallel and Distributed Processing
  Symposium, {IPDPS} 2015, Hyderabad, India, May 25-29, 2015}, pages 494--503.
  {IEEE} Computer Society, 2015.

\bibitem[Mil95]{mil95}
R.~Milner.
\newblock {\em Communication and concurrency}.
\newblock Prentice Hall International (UK) Ltd., Hertfordshire, UK, 1995.

\bibitem[Naz]{rvmt}
Hosein Nazarpour.
\newblock Website of {RVMT-BIP}, a tool for the {Runtime Verification of
  Multi-Threaded BIP} systems.
\newblock \url{http://www-verimag.imag.fr/\textasciitilde nazarpou/rvmt.html}.

\bibitem[NFB{\etalchar{+}}16]{NazarpourFBBC16}
Hosein Nazarpour, Yli\`es Falcone, Saddek Bensalem, Marius Bozga, and Jacques
  Combaz.
\newblock Monitoring multi-threaded component-based systems.
\newblock In Erika Abraham and Marieke Huisman, editors, {\em Proceedings of
  the 12th International Conference on integrated Formal Methods}, LNCS, 2016.

\bibitem[SG03]{SenG03}
Alper Sen and Vijay~K. Garg.
\newblock Detecting temporal logic predicates in distributed programs using
  computation slicing.
\newblock In Marina Papatriantafilou and Philippe Hunel, editors, {\em
  Principles of Distributed Systems, 7th International Conference, {OPODIS}
  2003 La Martinique, French West Indies, December 10-13, 2003 Revised Selected
  Papers}, volume 3144 of {\em Lecture Notes in Computer Science}, pages
  171--183. Springer, 2003.

\bibitem[SG07]{SenG07}
Alper Sen and Vijay~K. Garg.
\newblock Formal verification of simulation traces using computation slicing.
\newblock {\em {IEEE} Trans. Computers}, 56(4):511--527, 2007.

\bibitem[SVAR06]{SenVAR06}
Koushik Sen, Abhay Vardhan, Gul Agha, and Grigore Rosu.
\newblock Decentralized runtime analysis of multithreaded applications.
\newblock In {\em 20th International Parallel and Distributed Processing
  Symposium {(IPDPS} 2006), Proceedings, 25-29 April 2006, Rhodes Island,
  Greece}. {IEEE}, 2006.

\bibitem[TG97]{TomlinsonG97}
Alexander~I Tomlinson and Vijay~K Garg.
\newblock Monitoring functions on global states of distributed programs.
\newblock {\em Journal of Parallel and Distributed Computing}, 41(2):173--189,
  1997.

\bibitem[vGV97]{van97}
Rob van Glabbeek and Frits Vaandrager.
\newblock The difference between splitting semantics.
\newblock {\em Information and Computation}, 136(2):109--142, 1997.

\end{thebibliography}
\appendix
%
\section{Correctness Proof of the Approach}
\label{sec:proofs}
%
Before tackling the proof of correctness of our approach, we provide an intuitive description of the proof content.
The correctness of our approach relies on three results.

The first result concerns the witness trace.
Given a CBS $B$ whose semantics is described as per \secref{sec:cbs}, that is the general semantics of CBS.
One can build $B^\perp$, a transformed version of $B$ that can execute concurrently and which is bi-similar to $B$.
$B^\perp$ executes following the partial-state semantics described in \secref{sec:partial_state_sem}.
Any trace of an execution of $B^\perp$ can be related to the trace of a unique execution of $B$, i.e., its witness.
Property~\ref{property:same_sequence} states that any witness trace corresponds to the execution in global-state semantics that has the same sequence of interaction executions, i.e., that the witness relation captures the abovementioned relation between a system in global-state semantics and the corresponding system in partial-state semantics.
Property~\ref{property:unique_witness} states that from any execution in partial-state semantics, the witness exists and is unique.

The second result states that function $\RGT$ builds the witness trace from a trace in partial-state semantics in an online fashion.
Theorem~\ref{theorem_witness} states the correctness of this function.

The third result states that the transformed components, the synthesized components, and their connection are correct.
That is, the obtained system (i) computes the witness and implements function $\RGT$ (Proposition~\ref{proposition_RGT}), and (ii) is bisimilar to the initial system (Theorem~\ref{theorem_Correctness}).
\paragraph{Proof outline.}
The following proofs are organized as follows.
The proof of Property~\ref{property:same_sequence} is in Appendix~\ref{proof:same_sequence}.
The proof of Property~\ref{property:unique_witness} is in Appendix~\ref{proof:unique_witness}.
Some intermediate lemmas with their proofs are introduced in Appendix~\ref{Intermediate Lemmas} in order to prove Theorem~\ref{theorem_witness} in Appendix~\ref{proof:RGT-witness}.
The proofs of Propositions~\ref{proposition:length-discriminant} and~\ref{proposition_RGT} are respectively given in Appendices~\ref{proof:RGT-witness} and \ref{proof:RGT_correcness}.
Some intermediate definitions and lemmas with their proofs are given in Appendix~\ref{Intermediate Definitions and Lemmas} in order to prove Theorem~\ref{theorem_Correctness} in Appendix~\ref{proof:transfomation_correcness}.
%
\subsection{Proof of Property~\ref{property:same_sequence} (p.~\pageref{property:same_sequence})}
\label{proof:same_sequence}
%
We shall prove that:
\[
\forall (\sigma_1,\sigma_2)\in \W  \quant  \interactions(\sigma_1) = \interactions (\sigma_2),
\]
where $W$ is the witness relation defined in Definition~\ref{def:witness} (using a bi-simulation relation $R$), and $\interactions(\sigma)$ is the sequence of interactions of trace $\sigma$. 
\begin{proof}
The proof is done by structural induction on $W$.
\begin{itemize}
\item Base case. By definition of $W$, $(\mathit{Init},\mathit{Init}) \in W$ and $\interactions(\mathit{Init}) = \epsilon$.
\item
Induction case.
Let us consider $(\sigma_1,\sigma_2) \in W$ and suppose that $\interactions (\sigma_1) = \interactions(\sigma_2)$.
According to the definition of $W$, there are two rules for constructing a new element in $W$.
\begin{itemize}
\item
Consider $(\sigma_1 \cdot a\cdot q_1, \sigma_2\cdot a \cdot q_2)\in \W$ such that $a\in \gamma$ and $(q_1,q_2)\in R$.
We have $\interactions(\sigma_1 \cdot a\cdot q_1) = \interactions(\sigma_1) \cdot a$ and $\interactions(\sigma_2\cdot a \cdot q_2) = \interactions(\sigma_2) \cdot a$, and thus the expected result using the induction hypothesis.
\item 
Consider $(\sigma_1,\sigma_2\cdot\beta\cdot q_2)\in \W$ such that $(\last(\sigma_1),q_2) \in R$.
We have $\interactions(\sigma_2\cdot\beta\cdot q_2) = \interactions(\sigma_2)$ and thus the expected result using the induction hypothesis.
\end{itemize}
\end{itemize}
\end{proof}
%
\subsection{Proof of Property~\ref{property:unique_witness} (p.~\pageref{property:unique_witness})}
\label{proof:unique_witness}
%
We shall prove that:
\[
\forall \sigma_2\in \Tr(B^\perp), \exists!\sigma_1\in \Tr(B) \quant  (\sigma_1, \sigma_2)\in \W,
\]
where $B$ is a component-based system (with set of traces $\Tr(B)$) and $B^\perp$ is the corresponding component-based system with partial-state semantics (with set of traces $\Tr(B^\perp)$). 
\begin{proof}

First, let us note that from the weak bi-simulation of a global-state semantics model with its corresponding partial-state semantics model~\cite{Basu06modelingheterogeneous}, we can conclude that, for any trace in the partial-state semantics model, there exists a corresponding trace in the global-state semantics model. 
We prove that the witness trace is unique by contradiction.

Let us assume that for a trace in partial-state semantics $\sigma_2\in \Tr(B^\perp)$, there exist two witness traces $\sigma'_1,\sigma_1\in \Tr(B)$ such that $(\sigma_1,\sigma_2),(\sigma'_1,\sigma_2)\in W$ and $\sigma_1 \neq \sigma'_1$.
From Property~\ref{property:same_sequence}, $\interactions(\sigma_1) = \interactions(\sigma_2)$ and $\interactions(\sigma'_1) = \interactions(\sigma_2)$, therefore $\interactions(\sigma_1) = \interactions(\sigma'_1)$.
Moreover, $\sigma_1$ and $\sigma_1'$ have the same initial state because of the definition of $W$ and $(\sigma_1,\sigma_2)$, $(\sigma'_1,\sigma_2)\in W$.
From the semantics of composite components, a sequence of interactions is associated to a unique trace (from a unique initial state).
This is thus a contradiction.
\end{proof}
\subsection{Intermediate Lemmas}
\label{Intermediate Lemmas}
We prove the intermediate lemmas that are needed to prove Theorem~\ref{theorem_witness}.
\paragraph{Proof of Lemma~\ref{lemma:length-acc} (p.~\pageref{lemma:length-acc}).}
We shall prove 
$\forall (\sigma_1,\sigma_2)\in W \quant  |\Acc(\sigma_2)|=|\sigma_1|=2\times s+1$, where $s=|\interactions (\sigma_1)|$,
where $\Acc$ is the accumulator used in the definition of function $\RGT$ (Definition~\ref{def:rgt}), and function $\interactions$ (defined in \secref{sec:partial_state_sem}) returns the sequence of interactions in a trace (removing $\beta$).
\begin{proof}
The proof is done by structural induction on $W$.

\begin{itemize}
\item Base case. By definition of $W$, $(\mathit{Init},\mathit{Init}) \in W$ and we have $\Acc(\mathit{Init}) = \mathit{Init}$, $|\mathit{Init}| = 1$ and $|\interactions(\mathit{Init})| = |\epsilon| = 0$.

\item Induction case. Let us consider $(\sigma_1,\sigma_2)\in W$ such that $\interactions(\sigma_2) = s$ and suppose that Lemma~\ref{lemma:length-acc} holds for $(\sigma_1,\sigma_2)$.
According to the definition of $W$, there are two rules for constructing a new element in $W$.
\begin{itemize}
\item
Consider $(\sigma_1 \cdot a \cdot q_1, \sigma_2 \cdot a \cdot q_2) \in \W$ such that $a\in \gamma$ and $(q_1,q_2)\in R$.
According to Definition~\ref{def:rgt}, $\Acc(\sigma_2 \cdot a \cdot q_2) = \Acc(\sigma_2) \cdot a \cdot q_2$.
Using the induction hypothesis, $|\Acc(\sigma_2)| = | \sigma_1 |$.
Hence $|\Acc(\sigma_2 \cdot a \cdot q_2)| = | \sigma_2 | + 2 =  | \sigma_1 | + 2 = |\sigma_1 \cdot a \cdot q_1 |$.
\item 
Consider $(\sigma_1,\sigma_2\cdot\beta\cdot q_2)\in \W$ such that $(\last(\sigma_1),q_2) \in R$.
According to Definition~\ref{def:rgt} and using the definition of operator $\Map$, we have $| \Acc(\sigma_2 \cdot \beta \cdot q_2) | = | \Map \ [ x \mapsto \upd(q,x)] \ (\Acc(\sigma_2)) | = | \Acc(\sigma_2) |$, and thus we obtain the expected result using the induction hypothesis.
\end{itemize}
\end{itemize}
\end{proof}
\paragraph{Proof of Lemma~\ref{lemma:acc} (p.~\pageref{lemma:acc}).}
We shall prove that:
$\forall \sigma\in \Tr(B^\perp), \exists k\in[1 \upto s] \quant  q_k\in Q\implies\forall z\in[1 \upto k] \quant  q_z\in Q$ , $q_{z-1} \stackrel{a_z}{\longrightarrow}q_z$ where $(q_0\cdot a_1 \cdot q_1 \cdots a_s\cdot q_s)=\Acc(\sigma)$. 
\begin{proof}
According to Lemma~\ref{lemma:length-acc} and the definition of function $\Acc$ (see Definition~\ref{def:rgt}), a state is generated and added to sequence $\Acc(\sigma)$ just after the execution of an interaction $a\in\gamma$.
This state is obtained from the last state in $\Acc(\sigma)$, say $q$, such that the new state has state information about less components than $q$ because the states of all components involved in $a$ are undetermined and the states of all other components are identical.
Since after any busy transition, function $\upd$ (see Definition~\ref{def:rgt}) updates all the generated partial states that do not have the state information regarding the components that performed a busy transition, the completion of each partial state guarantees the completion of previously generated states.
Therefore, if there exists a global state (possibly completed through function $\upd$) in trace $\Acc(\sigma)$, then all the previously generated states are global states.

Moreover, the sequence of reconstructed global states follow the global-state semantics.
This results stems from two facts.
First, according to the definition of function $\upd$, whenever function $\upd$ completes a partial state in the trace by adding the state of a component for which the last state in the trace is undetermined, it uses the next state reached by this component according to partial-state semantics.
Second, according to Definition~\ref{def:atomic-partial}, the transformation of a component to make it compatible with partial-state semantics is such that an intermediate busy state, say $\perp$, is added between the starting state $q$ and arriving state $q'$ of any transition $(q, p, q')$.
Moreover, the transitions $(q, p, \perp)$ and $(\perp, \beta, q')$ in the partial-state semantics replace the previous transition $(q, p, q')$ in the global-state semantics.
Hence, whenever a component in partial-state semantics is in a busy state $\perp$, the next state that it reaches is necessarily the same state as the one it would have reached in the global-state semantics.
\end{proof}
\paragraph{Proof of Proposition~\ref{proposition:length-discriminant} (p.~\pageref{proposition:length-discriminant}).}
We shall prove that
$\forall \sigma\in \Tr(B^\perp) \quant $
\[
\begin{array}{l}
\quad |\D(\Acc(\sigma))|\leq |\Acc(\sigma)| \\
\wedge \D(\Acc(\sigma))=q_0\cdot a_1 \cdot q_1 \cdots a_d\cdot q_d \implies \forall i\in[1 \upto d] \quant q_{i-1} \stackrel{a_i}{\longrightarrow}q_i,
\end{array}
\]
where $\Acc$ is the accumulator function and $\D$ is the discriminant function used in the definition of function $\RGT$ (Definition~\ref{def:rgt}) such that $\RGT(\sigma)=\D(\Acc(\sigma))$.
\begin{proof}
The proof directly follows from the definitions of functions $\Acc$ and $\D$, and Lemma~\ref{lemma:acc}.
Let us consider $\sigma\in \Tr(B^\perp)$.

Regarding the first conjunct, according to the definition of function $\D$, $\D(\Acc(\sigma))$ is the longest prefix of $\Acc(\sigma)$ such that the last state of $\D(\Acc(\sigma))$ is a global state.
Thus, the length of sequence $\D(\Acc(\sigma))$ is always lesser than or equal to the length of sequence $\Acc(\sigma)$.

Regarding the second conjunct, according to Lemma~\ref{lemma:acc}, all the states of $\D(\Acc(\sigma))$ are global states and follow the global-state semantics.
Moreover, one can note that function $\D$ removes the longest suffix made of partial states output by function $\Acc$ and function $\Acc$ only updates partial states.
\end{proof}
\paragraph{Proof of Lemma~\ref{lemma:last-acc} (p.~\pageref{lemma:last-acc}).}
We shall prove that:
$\forall \sigma\in \Tr(B^\perp) \quant  \last(\Acc(\sigma))=\last(\sigma)$.
\begin{proof}
The proof is done by induction on the length of the trace in partial-state semantics, \ie $\sigma\in \Tr(B^\perp)$.
\begin{itemize}
\item 
Base case: The property holds for the initial state. 
Indeed, in this case $\sigma=\mathit{Init}$ and according to the definition of function $\Acc$ (see Definition~\ref{def:rgt}) $\last(\Acc(\mathit{Init}))=\mathit{Init}$.
\item 
Induction case: Let us  assume that $\sigma=q_0\cdot a_1\cdot q_1 \cdots a_m\cdot q_m$ is a trace in partial-state semantics and $\Acc(\sigma)=q'_0\cdot a'_1\cdot q'_1 \cdots a'_s\cdot q'_s$ such that $q_m=q'_s$.
We have two cases according to whether the next move of the  partial-state semantics model is an interaction or a busy transition:
\begin{itemize}
\item 
If $a_{m+1}\in \gamma$, then according to the definition of the function $\Acc$, we have: $\last(\Acc(\sigma\cdot a_{m+1} \cdot q_{m+1}))=q_{m+1}$. 
\item 
If $a_{m+1}\in\{\beta_i\}^n_{i=1}$, then according to the definition of function $\Acc$, we have: $\last(\Acc(\sigma\cdot a_{m+1} \cdot q_{m+1}))=\upd(q_{m+1},q'_s)$.
From the induction hypothesis: $\upd(q_{m+1},q'_s)=\upd(q_{m+1},$ $q_m)$ and from the fact that the only difference between state $q_m$ and state $q_{m+1}$ is that in state $q_m$ the state of the component that executed $a_{m+1}$ is a busy state, while in state $q_{m+1}$ it is not a busy state.
From the definition of function $\upd$ (Definition~\ref{def:rgt}), we can  conclude that $\upd(q_{m+1},q_m)=q_{m+1}$.
\end{itemize}
In both cases, $\last(\Acc(\sigma))=\last(\sigma)$.
\end{itemize}

\end{proof}
\paragraph{Proof of Lemma~\ref{lemma:interactions-acc} (p.~\pageref{lemma:interactions-acc}).}
We shall prove that 
$\forall \sigma \in \Tr(B^\perp) \quant  \interactions(\Acc(\sigma))=\interactions(\sigma)$.
\begin{proof}
By an easy induction on the length of $\sigma$ and case analysis on the definition of function $\Acc$ (Definition~\ref{def:rgt}).
\end{proof}
%
\subsection{Proof of Theorem~\ref{theorem_witness} (p.~\pageref{theorem_witness})}
\label{proof:RGT-witness}
%
We shall prove that, for a given CBS $B=\gamma(B_1,\ldots,B_n)$ with set of traces $\Tr(B)$ and $B^\perp$, the following holds on the set of traces $\Tr(B^\perp)$ of the corresponding CBS with partial-state semantics:
\[
\begin{array}{l}
\forall \sigma\in \Tr(B^\perp) \quant \\
\qquad \quad \last(\sigma) \in Q \implies \RGT(\sigma)=W(\sigma)\\
\qquad \wedge \last(\sigma) \notin Q \implies
\RGT(\sigma)=W(\sigma') \cdot a, \text{with}\\
\sigma' = \min_{\preceq} \{ \sigma_p \in \Tr(B^\perp) \mid \exists a\in\gamma, \exists \sigma'' \in \Tr(B^\perp)  \quant  \sigma=\sigma_p \cdot a \cdot \sigma''\\
\qquad\qquad\qquad\qquad\qquad\qquad \wedge \exists i\in[1 \upto n] \quant  (B_i.P\cap a \neq \emptyset)\wedge(\forall j\in[1 \upto \length(\sigma'')] \quant \beta_i \neq\sigma''(j))\}
\end{array}
\]
where function $\RGT$ is defined in Definition~\ref{def:rgt} and $W$ is the witness relation defined in Definition~\ref{def:witness}. 
\begin{proof}
For any trace in partial-state semantics $\sigma$, we consider two cases depending on whether the last element of $\sigma$ belongs to $Q$ of not:
\begin{itemize}
\item 
If $\last(\sigma)\in Q$, according to Lemma~\ref{lemma:last-acc}, $\last(\Acc(\sigma))\in Q$ and thus $\RGT(\sigma)=\D(\Acc(\sigma))$ $=\Acc(\sigma)$.
Let us assume that $\Acc(\sigma) = q_0\cdot a_1 \cdot q_1 \cdots a_s \cdot q_s$, with $q_0 = \mathit{Init}$.
According to Lemma~\ref{lemma:acc}, $\forall k\in[1 \upto s] \quant q_{k-1}\stackrel{a_k}{\longrightarrow}q_k \Longrightarrow \Acc(\sigma) \in \Tr(B)$.
Moreover, according to Lemma~\ref{lemma:interactions-acc}, $\interactions(\Acc(\sigma)) = \interactions(\sigma)$.
Furthermore, according to definition of the witness relation (Definition~\ref{def:witness}), from the unique initial state, since $\Acc(\sigma)$ and $\sigma$ have the same sequence of interactions, $(\Acc(\sigma),\sigma)\in W$.
Therefore, $\Acc(\sigma)=\RGT(\sigma)=W(\sigma)$.
\item
If $\last(\sigma) \notin Q$, we treat this case by induction on the length of $\sigma$.
Let us assume that the proposition holds for some $\sigma \in \Tr(B^\perp)$ (induction hypothesis).
Let us consider $\sigma=\sigma' \cdot a'_1 \cdot q'_1 \cdot a'_2 \cdot q'_2 \cdots a'_k \cdot q'_{k}$, with $k > 0$.
Let us assume that the splitting of $\sigma$ is $\sigma' \cdot a'_1 \cdot \sigma''$, where $\sigma'$ is the minimal sequence such that there exists at least one component that is involved in interaction $a'_1 \in \gamma$ and that is still busy.
(We note that in this case $\sigma'$ do exist because $\last(\sigma) \notin Q$ implies that the system has made at least one move.)
Let $i$ be the identifier of this component and $a'_1$ be $s^{th}$ interaction in trace $\sigma$ such that $a'_1=\interactions(\sigma)(s)$.
Let us consider $\sigma \cdot a'_{k+1} \cdot q'_{k+1}$, the trace extending $\sigma$ by one interaction $a'_{k+1}$.
We distinguish again two subcases depending on whether $a'_{k+1} \in \gamma$ or not.
\begin{itemize}
\item
Case $a'_{k+1} \in \gamma$.
We have $\last(\sigma) \notin Q$ and then $\last(\sigma \cdot a'_{k+1} \cdot q'_{k+1}) \notin Q$ (because $a'_{k+1} \in \gamma$, i.e., the system performs an interaction, and the state following an interaction is necessarily a partial state).
Moreover, $\RGT(\sigma) = \RGT(\sigma \cdot a'_{k+1} \cdot q'_{k+1})$, i.e., the reconstructed global state does not change.
Hence, the components which are busy after $a_1'$ are still busy.
Consequently, the splitting of $\sigma$ and $\sigma \cdot a'_{k+1} \cdot q'_{k+1}$ are the same.
Following the induction hypothesis, $\sigma \cdot a'_{k+1} \cdot q'_{k+1}$ has the expected property.
\item 
Case $a'_{k+1} = \beta_j$, for some $j \in [1 \upto n]$.
We distinguish again two subcases.
\begin{itemize}
\item
If $i = j$, that is the busy interaction $\beta_j$ concerns the component(s) for which information was missing in $\sigma''$ (component $i$).
If component $i$ is the only component involved in interaction $a'_1$ for which information is missing in $q'_1 \cdots q'_k$, the reconstruction of the global state corresponding to the execution of $a'_1$ can be done just after receiving the state information of component $i$.
After receiving $q'_{k+1}$, which contains the state information of component $i$, the partial states of $\Acc(\sigma)$ are updated with function $\upd$.
That is, $\RGT(\sigma\cdot a'_{k+1} \cdot q'_{k+1})=\RGT(\sigma) \cdot q''_0 \cdot a''_1 \cdot q''_1 \cdots  q''_{m-1} \cdot a''_m$, where $m > 0$, $q''_0$ is the reconstructed global state associated with interaction $a'_1$, and $a''_m=\interactions(\sigma)(s+m)$ is the first interaction executed after $\sigma$ for which there exists at least one involved component which is still busy.
Indeed, some interactions after $a'_1$ in trace $\sigma$ (\ie $a''_p=\interactions(\sigma)(s+p)$ for $m>p>0$) may exist and be such that component $i$ is the only component involved in them for which information is missing to reconstruct the associated global states.
In this case, updating the partial states of $\Acc(\sigma)$ with the state information of component $i$ yields several global states \ie $q''_1 \cdots q''_{m-1}$.
Then, the splitting of $\sigma$ changes as follows: $\sigma=\sigma'' \cdot a''_m \cdots a'_{k+1} \cdot q'_{k+1}$, where $\sigma''=\sigma' \cdot a'_1 \cdot q'_1 \cdot a'_2 \cdot q'_2 \cdots  q'_t$ and $q'_t$ is the system state before interaction $a''_m$.
Therefore, $\RGT(\sigma \cdot a'_{k+1} \cdot q'_{k+1})=W(\sigma'') \cdot a''_m$ and the property holds again.
\item
If $i \neq j$, we have $\RGT(\sigma) = \RGT(\sigma \cdot a'_{k+1} \cdot q'_{k+1})$. 
Hence, the splitting of $\sigma$ and $\sigma \cdot a'_{k+1} \cdot q'_{k+1}$ are the same.
Following the induction hypothesis, $\sigma \cdot a'_{k+1} \cdot q'_{k+1}$ has the expected property.
\end{itemize}
\end{itemize}  
\end{itemize}
\end{proof}
%
\subsection{Proof of Proposition~\ref{proposition_RGT} (p.~\pageref{proposition_RGT})}
\label{proof:RGT_correcness}
%
Given a CBS $B=\gamma(B_1,\ldots,B_n)$ with corresponding partial-state semantics model $B^\perp=\gamma^\perp(B^\perp_1,\ldots,$ $B^\perp_n)$ and the transformed composite component $B^r=\gamma^r(B_1^r,\ldots,B_n^r,RGT)$ obtained as per Definition~\ref{def:composit_transformation}, we shall prove that for any execution of the system with partial-state semantics with trace $\sigma \in \Tr(B^\perp)$, component $\RGT$ (Definition~\ref{def:RGT-atom}) implements function $\RGT$ (Definition~\ref{def:rgt}), that is $\forall \sigma \in \Tr(B^\perp) \quant  \RGT.V \cong \Acc(\sigma)$.

\begin{proof}
The proof is done by induction on the length of $\sigma \in \Tr(B^\perp)$, i.e., the trace of the system in partial-state semantics.
\begin{itemize}
\item Base case.
By definition of function $\RGT$, at the initial state  $\Acc(\mathit{Init})=\mathit{Init}$.
By definition of component $\RGT$, $V$ is initialized as a tuple representing the initial state of the system.
Therefore, $\RGT.V \cong \Acc(\mathit{Init})$.
\item
Induction case.
Let us suppose that the proposition holds for a trace $\sigma \in \Tr(B^\perp)$, that is $\RGT.V \cong \Acc(\sigma)$.
According to the definition of function $\RGT$, $\RGT(\sigma)=\D(\Acc(\sigma))$.
Consequently, there exists $\sigma'\in \Tr(B^\perp)$ of the form $\sigma' = q'_0 \cdot a'_1 \cdot q'_1\cdots q'_{k}$, with $k \geqslant 0$, such that $\Acc(\sigma)= \RGT(\sigma) \cdot \sigma'$. 
We distinguish two cases depending on the action of the system executed after $\sigma$:
\begin{itemize}
\item
The first case occurs when the action is the execution of an interaction $a'_{k+1}$, followed by a partial state $q'_{k+1}$.
On the one hand, we have $\Acc(\sigma \cdot a'_{k+1}\cdot q'_{k+1})=\Acc(\sigma) \cdot a'_{k+1}\cdot q'_{k+1}$.
On the other hand, in component $\RGT$, according to Algorithm~\ref{alg:NEW} (line 6), the corresponding transition $\tau \in T_{\rm{new}}$ extends the sequence of tuples $V$ by a new $(n+1)$-tuple $v$ which consists of the current partial state of the system such that $V=V \cdot v$ and $v \cong q'_{k+1}$.
Therefore, we have $\RGT.V\cong \Acc(\sigma)$ as expected. 
\item
The second case occurs when the next action is the execution of a busy transition.
On the one hand, function $\RGT$ updates all the partial states $q'_0,\ldots,q'_k$.
On the other hand, according to Algorithm~\ref{alg:UPD} (lines 2-6), in component $\RGT$, the corresponding transition $\tau \in T_{\rm{upd}}$ updates the sequence of tuples $V$ such that $\RGT.V \cong \Acc(\sigma)$ hold.

Moreover, function $\RGT$ and component $\RGT$ similarly create new global states from the partial states whenever new global states are computed.
On the one hand, after any update of partial states, through function $\D$, function $\RGT$ outputs the longest prefix of the generated trace which corresponds to the witness trace.
On the other hand, after any update of the sequence of tuples $V$, component $\RGT$ checks for the existence of fully completed tuples in $V$ to deliver them to through the dedicated ports to the runtime monitor.\end{itemize}
\end{itemize}
\end{proof}
\subsection{Proofs of Intermediate Lemmas}
\label{Intermediate Definitions and Lemmas}
%
In the following proofs, we will consider several mathematical objects in order to prove the correctness of our framework:
\begin{itemize}
\item 
a composite component with partial-state semantics $B^\perp=\gamma^\perp(B_{1}^\perp, \ldots,B_{n}^\perp)$ of behavior $(Q^\perp,\gamma^\perp,$ $\goesto)$;
\item 
the transformed composite component $B^r=\gamma^r(B_1^r,\ldots,B_n^r,RGT)$ of behavior $(Q^r,\gamma^r,\goesto_r)$. $B^r$ is obtained from $B^\perp$ by following the transformations described in \secref{sec:inst}.
\end{itemize}
\paragraph{Proof of Lemma~\ref{lemma:reset_befor_set} (p.~\pageref{lemma:reset_befor_set}).}
We shall prove that in any state of the transformed system, if there is a non-empty set $\mathit{GS} \subseteq  \{\RGT.\mathit{gs}_a \mid a \in \gamma\}$ in which all variables are $\true$, the variables in $\{ \RGT.\mathit{gs}_a \mid a \in \gamma \} \setminus \mathit{GS}$ cannot be set to $\true$ until all variables in $\mathit{GS}$ are reset to $\false$ first.
\begin{proof}
According to the definition of component $\RGT$ (Definition~\ref{def:RGT-atom}), on the one hand only the transitions in set $T_{\rm{upd}}$ are able to set the value of the variables in $\{\RGT.\mathit{gs}_a \mid a \in \gamma\}$ to $\true$; on the other hand the transitions in set $T_{\rm{upd}}$ are guarded by $\bigwedge_{a \in \gamma} (\neg \mathit{gs}_a)$ which means that all of the Boolean variables in $\{\RGT.\mathit{gs}_a \mid a \in \gamma\}$ must be $\false$ for one of these transitions to execute. 
Therefore, in any state $q \in Q^r$ such that such a set $\mathit{GS}$ exists, the transitions in $T_{\rm{upd}}$ are not possible.
Moreover, the only possible transitions in state $q$ are the transitions in set $T_{\rm{out}}$ which effect is to reset the value of the variables in $\{ \RGT.\mathit{gs}_a \mid a \in \gamma\}$ to $\false$ using algorithm~$\mathtt{get}$.
\end{proof}
\paragraph{Proof of Lemma~\ref{lemma:stable-state} (p.~\pageref{lemma:stable-state}).}
%
We shall prove that for any state $q \in Q^r$, there exists a state $q'\in Q^r$ reached after interactions in $a^m$ (i.e., $q \xrightarrow{({a^m})^*} q'$), such that $q'$ is a stable state (i.e., $\stable(q')$).
\begin{proof}
Let us consider a non-stable state $q\in \RGT.Q$.
The interactions in $a^m$ involve to execute ports in $\{p_a' \mid a \in \gamma \}$ and transitions in $T_{\rm out}$.
Since $q$ is a non-stable state, at least one of the variables in $\{\RGT.\mathit{gs}_a \mid a \in \gamma\}$ evaluates to $\true$ in $q$ (see Definition~\ref{def:stability}, p.~\pageref{def:stability}).
Such transitions entail to execute algorithm $\mathtt{get}$ (Algorithm~\ref{alg:get}) which resets the Boolean variable to $\false$ by delivering the associated reconstructed global state(s) to the monitor.
After executing algorithm $\mathtt{get}$, if there exists another Boolean variable in $\{\RGT.\mathit{gs}_a \mid a \in \gamma\}$ that evaluates to $\true$, according to Lemma~\ref{lemma:reset_befor_set}, component $\RGT$ returns to a situation where only again algorithm $\mathtt{get}$ can execute (through the interactions in set $a^m$).
The above process executes until the system eventually reaches a state $q'$ where no interaction in $a^m$ is enabled.
Therefore, in $q'$ all Boolean variables in $\{\RGT.\mathit{gs}_a \mid a \in \gamma\}$ evaluate to $\false$, because interactions in $a^m$ are unary interactions, each involving port $\RGT.p'_a$ (Definition~\ref{def:composit_transformation}) guarded by $\bigwedge_{a \in \gamma} (\neg \mathit{gs}_a)$ (Definition~\ref{def:RGT-atom}).
According to Definition~\ref{def:stability}, a state is stable when all Boolean variables in $\{\RGT.\mathit{gs}_a \mid a \in \gamma\}$ evaluate to $\false$.
Thus, $q'$ is stable.
\end{proof}
\paragraph{Proof of Lemma~\ref{lemma:enabled-interactions} (p.~\pageref{lemma:enabled-interactions}).}
Let us consider two states: $q$ of the initial model and $q^r$ its corresponding state in the transformed model such that $\equ(q^r)=q$.
There exists an enabled interaction in the initial model ($a\in \gamma^\perp$) in state $q\in Q^\bot$, if and only if the corresponding interaction in the transformed model ($a^r\in \gamma^r$) is enabled at state $q^r$. 
\begin{proof}
According to the definitions of interaction transformation and atom $\RGT$ (Definition~\ref{def:RGT-atom}), ports $\RGT.p_a$, for $a \in \gamma$, are always enabled.
Since for a given interaction $a$, $a^r$ and $a$ differ only by port $\RGT.p_a$, we can conclude that $a^r\in a^r_\gamma$ is enabled if and only if $a\in \gamma^\perp$ is enabled.
\end{proof}
%
%
\subsection{Proof of Theorem~\ref{theorem_Correctness} (p.~\pageref{theorem_Correctness})}
\label{proof:transfomation_correcness}
%
Before tackling the proof of Theorem~\ref{theorem_Correctness}, we convey a remark preparing the definition of the weak bi-simulation relation defined in the proof.

Following Definition~\ref{def:composit_transformation}, the set of interactions $\gamma^r$ of the instrumented system is partitioned as $\gamma^r = a^r_\gamma \cup a^r_\beta \cup a^m$, where $a^r_\gamma$ is the set of interactions of the initial system augmented by $\RGT$ port, $a^r_\beta$ is a set containing the busy interactions of the initial system (one for each component) augmented by $\RGT$ port, and $a^m$ is a new set of interactions used for monitoring purposes.
First, we note that the set of interactions in the instrumented system $a^r_\gamma$ and $a^r_\beta$ are isomorphic to the sets of interactions $\gamma$ and $\{ \{ \beta_i \} \}_{i=1}^n$ of the initial system because they contain only an additional port to notify component $\RGT$.
We can thus identify these sets of interactions.
Moreover, as usual in monitoring, the actions used for monitoring purposes (i.e., interactions in $a^m$) are considered to be unobservable.
These interactions do not influence the state of the system and execute independently of the interactions in $a^r_\gamma \cup a^r_\beta$; these are interactions occurring between $\RGT$ and the monitor which are components introduced in the instrumentation.
See also~\cite{FalconeJNBB15}, for more arguments along these lines related to the instrumentation of single-threaded CBSs.
\begin{proof}
We exhibit a relation $R \subseteq Q^\bot \times Q^r$ between the set of states of the initial model with partial-state semantics and the set of states of the transformed model.
We define $R=\{(q,q^r)  \mid \exists z^r \in Q^r  \quant   q^r\xrightarrow{\left(a^m\right)^*}_r z^r \wedge \equ(z^r)=q\}$, and 
we shall prove that relation $R$ satisfies the following properties to establish that $R$ is a weak bi-simulation:
\begin{itemize}
\item[(i)]
$\left( (q,q^r)\in R \wedge \exists z^r \in Q^r  \quant q^r\xrightarrow{a^m}_r z^r \right) \implies (q,z^r)\in R$;
\item[(ii)]
$\left( (q,q^r)\in R \wedge \exists z^r \in Q^r  \quant q^r\xrightarrow{a^r_\gamma + a^r_\beta}_r z^r \right) \implies \exists z\in Q  \quant  \left( q \xrightarrow{a} z \wedge (z,z^r)\in R \right)$;
\item[(iii)]
$\left( (q,q^r)\in R \wedge q \xrightarrow{\gamma^\perp} z\right) \implies \exists z^r\in Q^r \quant  \left(q^r\xrightarrow{\left(a^m\right)^*.a^r}_r z^r \wedge (z,z^r)\in R \right)$. 
\end{itemize}
Let us consider $q=(q_1,\cdots,q_n)$ and $q^r=(q^r_1,\cdots,q^r_n, q^r_{n+1})$ such that $(q,q^r)\in R$.

\underline{Proof of (i):}
\newline
Since $(q,q^r) \in R$, there exists a stable state ${q^r}' \in Q^r$ which is reached after unobservable interactions in $a^m$.
After the execution of some unary interaction $\alpha \in a^m$, the corresponding Boolean variable $\RGT.\mathit{gs}_\alpha$ is set to $\false$ (Algorithm~\ref{alg:get}).
Let us consider that the next state after the execution of some interaction $\alpha\in a^m$ is $z^r=(z^r_1,\cdots,z^r_n,z^r_{n+1})$.
If $z^r_{n+1}$ is a stable state then $\equ(z^r)=q$ thus $(q,z^r)\in R$, and if $z^r_{n+1}$ is not a stable state according to Lemma~\ref{lemma:stable-state}, after interaction $\alpha \in a^m$, the state of $\RGT$ (that is $z^r_{n+1}$) will be stable, therefore we conclude that $(q,z^r)\in R$.

\underline{Proof of (ii):}
\newline
Let us consider  $z^r=(z^r_1,\cdots,z^r_n,z^r_{n+1})$ and $z=(z_1,\cdots,z_n)$.
When some $a^r\in (a^r_\gamma \cup a^r_\beta)$ is enabled, from the definition of the semantics of transformed composite component and Lemma~\ref{lemma:enabled-interactions}, we can deduce that the corresponding interaction $a\in\gamma^\perp$ is enabled (recall, that for each interaction $a \in \gamma^\bot$ in the initial model with partial-state semantics there exists a corresponding interaction $a^r$ in the transformed model, as per Definition~\ref{def:composit_transformation}).
Executing the corresponding interactions $a$ and $a^r$ changes the local states $q^r_i$ and $q_i$, for $i\in[1 \upto n]$, to $z^r_i$ and $z_i$ for $i\in[1 \upto n]$ respectively, in such a way that $z^r_i=z_i$, for $i\in[1 \upto n]$, because the transformations do not modify the transitions of the components of the initial model.
After $a^r$, we have two cases depending on whether $z^r_{n+1}$ is stable or not.
\begin{itemize}
\item
If $z^r_{n+1}$ is stable, from the definition of relation $R$, we have $(z,z^r)\in R$.
\item
If $z^r_{n+1}$ is not stable, then according to Lemma~\ref{lemma:stable-state}, $z^r_{n+1}$ will be stable after some interactions $\alpha \in a^m$ (that is $z^r_{n+1}\xrightarrow{\alpha^*}\stable(z^r_{n+1})$). Therefore, $(z,z^r)\in R$.
\end{itemize}

\underline{Proof of (iii):}
\newline
Let us consider $z^r=(z^r_1,\ldots,z^r_n,z^r_{n+1})$.
When $a\in \gamma^\perp$ is enabled in the initial model, 
we can consider two cases depending on whether the corresponding interaction $a^r$ in the transformed model is enabled or not.
\begin{itemize}
\item If $a^r$ is enabled, we have two cases for the next state of component $\RGT$:
\begin{itemize}
\item if $a^r\in a^r_\gamma$, according to the definition of atom $\RGT$, $z^r_{n+1}$ is stable and $(z,z^r)\in R$.
\item if $a^r\in a^r_\beta$, we have two cases:
\begin{itemize}
\item If $\RGT$ has some global states to deliver (that is $z^r_{n+1}$ is not stable), then, according to Lemma~\ref{lemma:stable-state}, $\RGT$ will be stable after some interactions in $a^m$. Hence, $(z,z^r)\in R$.
\item If $\RGT$ has no global state, then atom $\RGT$ is stable and $(z,z^r)\in R$.
\end{itemize}
\end{itemize}
\item If $a^r$ is not enabled, according to the definition of atom $\RGT$, we can conclude that $\RGT$ has some global states to deliver, thus $q^r$ is not stable.
According to Lemma~\ref{lemma:stable-state}, a not stable system  becomes stable after executing some interactions in $a^m$.
Therefore, according to Lemma~\ref{lemma:enabled-interactions}, $a^r$ is necessarily enabled when the system is stable.
Consequently, the same reasoning followed for the previous case can be conducted in which $a^r$ is initially enabled.
Henceforth, $(z,z^r)\in R$. 
\end{itemize} 
\end{proof}
%
\end{document}